\documentclass[]{aastex63}
\usepackage{graphics}
\usepackage{pgfplotstable}
\usepackage{natbib}
\usepackage{float}
\usepackage{textcomp}

\usepackage{dcolumn}% Align table columns on decimal point
\usepackage{bm}% bold math
\usepackage{rotating}
\usepackage{xtab}
\usepackage{booktabs}
\usepackage{rotating}
\usepackage[T1]{fontenc}
\usepackage[utf8]{inputenc}

\usepackage[graphicx]{realboxes}

\let\tablenum\relax
\usepackage{siunitx}
\pgfplotsset{compat=1.16}
\shorttitle{Optical Luminosity-Time Correlation}
\shortauthors{Dainotti et al.}
\graphicspath{{./}{figures/}}
\begin{document}

\title{The Optical Two and Three-Dimensional Fundamental Plane Correlations for Nearly 180 Gamma-Ray Burst Afterglows with \emph{Swift}/UVOT, RATIR, and the SUBARU Telescope}

\correspondingauthor{Maria Giovanna Dainotti$^*$\\
$^*$\href{mailto:maria.dainotti@nao.ac.jp}{maria.dainotti@nao.ac.jp}}

% Dainotti, M. G.; first and second authors contributed equally to this paper.}
\author{M.G. Dainotti}
\affiliation{Division of Science, National Astronomical Observatory of Japan, 2-21-1 Osawa, Mitaka, Tokyo 181-8588, Japan}
\affiliation{The Graduate University for Advanced Studies (SOKENDAI),
2-21-1 Osawa, Mitaka, Tokyo 181-8588, Japan}
\affiliation{Space Science Institute, 4750 Walnut Street, Boulder, CO
80301, USA}
\affiliation{SLAC National Accelerator Laboratory, 2575 Sand Hill Road, Menlo Park, CA 94025, USA}
\author[0000-0002-2488-1899]{S. Young}
\affiliation{Department of Physics and Astronomy, University of Pennsylvania, 209 South 33rd Street, Philadelphia, PA 19104, USA}
\author{L. Li}
\affiliation{ICRANet, Piazza della Repubblica 10, I-65122 Pescara, Italy; liang.li@icranet.org}
\author{D. Levine}
\affiliation{University of Maryland, College Park, US}
\author[0000-0001-9825-7418]{K. K. Kalinowski}
\affiliation{Wydział Fizyki, Astronomii i Informatyki Stosowanej, Uniwersytet Jagielloński, ul. Łojasiewicza 11, 30-348 Kraków, Poland}
\author{D. A. Kann}
\affiliation{Instituto de Astrof\'isica de Andaluc\'ia (IAA-CSIC), Glorieta de la Astronom\'ia s/n, 18008 Granada, Spain}
\author{B. Tran}
\affiliation{Centennial High School, 1820 Rimpau Avenue, Corona, CA 92881, USA}
\author{L. Zambrano-Tapia}
\affiliation{Scientific Caribbean Foundation, 12 Camino Francisco Rivera, San Juan, PR 00926, USA}
%\author{H. T. Tso}
%\affiliation{Jacobs School of Engineering, University of California, San Diego, CA 92093, USA}
\author{A. Zambrano-Tapia}
\affiliation{Scientific Caribbean Foundation, 12 Camino Francisco Rivera, San Juan, PR 00926, USA}
\author{S. B. Cenko}
\affiliation{Astrophysics Science Division, NASA Goddard Space Flight Center, MC 661, Greenbelt, MD 20771, USA}
\affiliation{Space Science Institute, 4765 Walnut Street, Boulder, CO 80301, USA}
\author{M. Fuentes}
\affiliation{Scientific Caribbean Foundation, 12 Camino Francisco Rivera, San Juan, PR 00926, USA}
\author{E. G. Sánchez-Vázquez}
\affiliation{Latino Education Advancement Foundation, 210 E Lexington Street, Baltimore MD 21202, USA}
\author{S. R. Oates}
\affiliation{School of Physics and Astronomy, University of Birmingham, Birmingham B15 2TT, UK}
\author{N. Fraija}
\affiliation{Instituto de Astronomía, Universidad Nacional Autónoma de México, Apartado Postal 70-264, 04510 México, CDMX, Mexico}
%\author{N. Osborn}
%\affiliation{Department of Physics and Astronomy, Purdue University, 525 Northwestern Avenue, West Lafayette, IN 47907, USA}
\author{R. L. Becerra}
\affiliation{Instituto de Ciencias Nucleares, Universidad Nacional Autónoma de México, Apartado Postal 70-264, 04510 México, CDMX, Mexico}
\author{A. M. Watson}
\affiliation{Instituto de Astronomía, Universidad Nacional Autónoma de México, Apartado Postal 70-264, 04510 México, CDMX, Mexico}
\author{N. R. Butler}
\affiliation{School of Earth and Space Exploration, Arizona State University, Tempe, AZ 85287, USA}
%\author{J. S. Bloom}
%\affiliation{Department of Astronomy, University of California, Berkeley, CA 94720-3411, USA}
\author{J. J. González}
\affiliation{Instituto de Astronomía, Universidad Nacional Autónoma de México, Apartado Postal 70-264, 04510 México, CDMX, Mexico}
\author{A. S. Kutyrev}
\affiliation{Department of Astronomy, University of Maryland, College Park, MD 20742-4111, USA}
\affiliation{Astrophysics Science Division, NASA Goddard Space Flight Center, 8800 Greenbelt Road, Greenbelt, MD 20771, USA}
\author{W. H. Lee}
\affiliation{Instituto de Astronomía, Universidad Nacional Autónoma de México, Apartado Postal 70-264, 04510 México, CDMX, Mexico}
\author{J. X. Prochaska}
\affiliation{Department of Astronomy and Astrophysics, UCO/Lick Observatory, University of California, 1156 High Street, Santa Cruz, CA 95064, USA}
\author{E. Ramirez-Ruiz}
\affiliation{Department of Astronomy, University of California, Berkeley, CA 94720-3411, USA}
\author{M. G. Richer}
\affiliation{Instituto de Astronomía, Universidad Nacional Autónoma de México, Unidad Académica en Ensenada, 22860 Ensenada, BC, Mexico}
\author[0000-0003-3609-382X]{S. Zola}
\affiliation{Astronomical Observatory, Jagiellonian University, ul. Orla 271, 30-244 Kraków, Poland}
%\author{T. Sakamoto}
%\affiliation{Aoyama University, Yokohama, Japan}
%\affiliation{Department of Astronomy \& Astrophysics, Las Vegas University,
%---TO BE UPDATED}

%\affiliation{Joint Space-Science Institute, University of Maryland, College Park, MD 20742, USA}
%TC:endignore

\begin{abstract}
Gamma-ray bursts (GRBs) are fascinating events due to their panchromatic nature. We study optical plateaus in GRB afterglows via an extended search into archival data. We comprehensively analyze all published GRBs with known redshifts and optical plateaus observed by many ground-based telescopes (e.g., Subaru Telescope, RATIR) around the world and several space-based observatories such as the \emph{Neil Gehrels Swift Observatory}. We fit 500 optical light curves (LCs), showing the existence of the plateau in 179 cases. This sample is 75\% larger than the previous one \citep{dainotti2021b}, and it is the largest compilation so far of optical plateaus. 
We discover the 3D fundamental plane relation at optical wavelengths using this sample. This correlation is between the rest-frame time at the end of the plateau emission, $T^{*}_{\rm opt}$, its optical luminosity, $L_{\rm opt}$, and the peak in the optical prompt emission, $L_{\rm peak, opt}$, thus resembling the three-dimensional (3D) X-ray fundamental plane relation \citep{dainotti2016}. We correct our sample for redshift evolution and selection effects, discovering that this correlation is indeed intrinsic to GRB physics. We investigate the rest-frame end time distributions in X-rays and optical ($T^{*}_{\rm opt}$, $T^{*}_{\rm X}$), and conclude that the plateau is achromatic only when selection biases are not considered. We also investigate if the 3D optical correlation may be a new discriminant between optical GRB classes and find that there is no significant separation between the classes compared to the Gold sample plane after correcting for evolution.
\end{abstract}

\section{Introduction} \label{sec:intro}
Gamma-Ray Bursts (GRBs), among the most luminous phenomena in the Universe, originate from the deaths of massive stars \citep{1993ApJ...405..273W,1998ApJ...494L..45P, Woosley2006ARA&A,Cano2017} or the merging of two compact objects, like neutron stars \citep[NSs;][]{1992ApJ...392L...9D, 1992Natur.357..472U, 1994MNRAS.270..480T, 2011MNRAS.413.2031M} and black holes \citep[BHs,][]{1992ApJ...395L..83N}. These models can have hyper-accreting BHs or fast-spinning newly born highly-magnetized NSs (magnetars) as central engines.

To distinguish between the different origins, we categorize GRBs according to their phenomenology. 
The GRB prompt emission is observed from hard X-rays to $\ge100$ MeV $\gamma$-rays, and sometimes also in the optical \citep{Vestrand2005Natur,Blake2005Natur,Beskin2010ApJ}. The afterglow \citep[e.g.,][]{1997Natur.387..783C, 1997Natur.386..686V,1998A&A...331L..41P,Gehrels2009ARA&A,Wang2015} is the long-lasting multi-wavelength emission (in X-rays, optical, and sometimes radio) following the prompt. 

GRBs are traditionally further classified as Short (SGRBs) and Long GRBs (LGRBs), depending on their duration: $T_{90}\leq 2\mathrm{\,s}$ or $T_{90} \ge 2\mathrm{\,s}$,\footnote{$T_{90}$ is the time over which a burst emits from $5\%$ to $95\%$ of its prompt emission total measured counts.} respectively \citep{mazets1981catalog, kouveliotou1993identification}.
\cite{zhang2009} proposed a classification based on the GRBs' progenitors, according to which GRBs are divided in Type I/II, see Fig. 8 in \cite{zhang2009} and refer to \cite{kann2011} for a discussion of controversial cases.
Type II GRBs originate from the collapse of massive stars \citep{woosley1993}. These include LGRBs, X-ray flashes (XRFs) with soft spectra and greater fluence in X-rays (2-30 keV) than in $\gamma$-rays (30-400 keV, \citealt{Heise2001grba.conf}), Ultra-Long GRBs (ULGRBs) with $T_{90}>1000\mathrm{\,s}$ \citep{Gendre2013, levan2014, Piro2014, Greiner2015Natur,Kann2018A&A,10.1093/mnras/stz1036}, and GRBs associated with SNe Ic. GRB-SNe Ic are further classified in A, B, C, D, and E classes \citep{hjorth03}. The A, B, and C classes, which are more spectroscopically associated with SNe, are used in this work, hereafter denoted as GRB-SNe ABC.
It is debated if all LGRBs should have a SNe associated \citep{Fynbo2006, 2006Natur.444.1050D}, because there have been cases such as GRB 060614 and GRB 060505 for which the SNe should have been seen and since it was not either the SNe was at least two magnitudes smaller than the other SNe associated or simply was not seen. Due to these observational differences it was suggested \citep{Fynbo2006, 2006Natur.444.1050D} that those GRBs-SNe may form a new population different from the regular LGRBs for which the SNe is not observed. Thus, we choose to keep also separated in our analysis this class of events.
Type I GRBs, resulting from the mergers of compact objects \citep{Abbott2017ApJ1,Abbott2017ApJ2}, include SGRBs, SGRBs with extended emission (SEEs, \citealt{norris2006short,levan2007case,norris2010threshold,dichiara2021,rastinejad2022}) and the intrinsically short (IS), namely Short in the rest frame with $T^*_{90}=T_{90}/(1+z)<2\mathrm{\,s}$ \citep{Levesque2010,Ahumada2021NatAs,Zhang2021NatAs,Rossi2021arXiv}. Here, we use SGRBs, which include both SEEs and ISs as a unique class.

Observations of the X-ray afterglows performed by the {\it Neil Gehrels Swift Observatory} ({\it Swift} hereafter) revealed the presence of an X-ray plateau \citep{o2006early, sakamoto2007evidence,Evans2009,Zhang2006,Nousek2006}. This phase generally lasts from $10^2\mathrm{\,s}$ to $10^5\mathrm{\,s}$ and is followed by a power law (PL) decay phase.
The plateau can be explained with the long-lasting energy injection from the central engine by fallback mass accretion onto a BH 
\citep{Kumar2008,Cannizzo2009, cannizzo2011} or with the energy injection produced by the spin-down luminosity of a highly magnetized millisecond newborn NS, a magnetar \citep[e.g.,][]{1992ApJ...392L...9D, 1992Natur.357..472U, 1994MNRAS.270..480T, Dai1998,Zhang2001, 2007ApJ...665..599T, 2011MNRAS.413.2031M, 2011A&A...526A.121D, Rowlinson2014,Rea2015,Li2018b,Stratta2018, 2019AnPhy.41067923M, Fraija2020}. 
The plateau found in X-rays and optical has been identified as a trait that may standardize GRBs. \cite{dainotti2016,dainotti2017a,dainotti2017c} and \cite{Li2018b} explored the luminosity at the end of the plateau, $L_{X,a}$ vs. its rest-frame time $T^{*}_{X,a}$ (known as the Dainotti relation or 2D L-T relation), with the rest-frame time denoted with an asterisk.
\cite{Rowlinson2014} showed that the Dainotti relation in X-rays is recovered within the magnetar scenario with a slope for $L_a$-$T^{*}_{X, a}$ of $-1$. Within the cosmological context, this correlation has already been applied to construct a GRB Hubble diagram out to $z>8$ \citep{cardone2009revised,cardone2010constraining,postnikov2014nonparametric,dainotti2013}. 

As pointed out in \citet{dainotti2008,dainotti2010,dainotti2016,dainotti2017b,dainotti2017a}, to obtain a class of GRBs that can be well-standardized, we need to select a GRB sub-sample with well-defined properties from a morphological or physical point of view. Thus, we segregate each class in GRBs-SNe Ic, XRFs, X-ray Rich (XRR; an intermediate case between the LGRBs and XRFs), ULGRBs, SGRBs, SEE, IS GRBs, and LGRBs. LGRBs are defined as the total sample from which we remove all other classes.
Regarding the connection between prompt and plateau emission, we annotate the peak prompt luminosity in 1 second, $L_{X, peak}$ vs. $T^{*}_{X, a}$ correlation. A theoretical interpretation of this correlation is within the standard fireball model and with the changing of the microphysical parameters \citep{2014MNRAS.442.3495V,2014MNRAS.445.2414V}.
An extension of the 2D L-T relation has been obtained by adding the peak prompt luminosity, $L_{X, peak}$ leading to the so-called Dainotti 3D relation \citep{dainotti2016,dainotti2017a,dainotti2020arXiv}.
We enhance the previous definition of Gold GRBs from \cite{dainotti2016} with new criteria: identifying the plateaus with fewer gaps in the data points and with less fluctuation in the fluxes. The criteria guarantee a tighter correlation involving the plateau emission to use it as a future cosmological tool and a theoretical model discriminator.

In this work, we investigate (1) the 3D Dainotti relation in the optical; (2) the 3D optical correlation as a discriminant between GRB classes; (3) the 2D optical correlation with our incremented sample size to determine if it can be a discriminant among classes; and (4) if with a larger sample and with the correction for selection biases, whether the plateau is an achromatic feature between X-rays and optical. 
All this analysis has been performed exploring three different types of fitting, to show that the reliability of the results is independent from the particular chosen fitting function. 

In \S \ref{sec:sample selection}, we describe the sample. In \S \ref{sec:methodology}, we detail the methodology showing how we corrected for Galactic and host-galaxy extinction and fitted our sample. In \S \ref{sec:3D correlation}, we present the 2- and 3D optical Dainotti relations, also corrected for selection biases and redshift evolution. We compare our results with previous results in X-ray in \S \ref{sec:comparison}, and we discuss our conclusions in \S \ref{sec:discussion}.

\section{Sample selection}\label{sec:sample selection}
We analyzed 500 GRB optical afterglows with known redshifts, thus building the most comprehensive sample of optical LCs to date by searching the literature for all GRBs detected between May 1997 and May 2021 by several satellites (e.g., the {\it Swift} Ultraviolet/Optical Telescope, UVOT), and ground-based telescopes/detectors, e.g., the Subaru Telescope, Gamma-ray Burst Optical/Near-IR Detector (GROND), Re-ionization and Transients InfraRed camera/telescope (RATIR), the MITSuME \citep{kotani2005}, etc. In our final sample, the redshifts span from $z=0.06$ to $z=8.23$ and the LCs are taken from \citet[][2022a, 2022b in preparation]{Kann2006,Kann2010,kann2011}, \cite{Li2012,Li2015,Li2018a}, \cite{Oates2009,Oates2012}, \cite{Zaninoni2013,Si2018}, the RATIR collaboration, the GCN Circular Archive (GCN)\footnote{https://gcn.gsfc.nasa.gov/}, the {\it Swift} Burst Analyzer \citep{Evans2010}, and other literature. The GRB name, redshift, fitting parameters, and data source of a portion of our data sample is given in Table \ref{table:finaltable} -- the full table with our sample of 179 GRBs is available in the online supplementary material.
Following \cite{dainotti2021b}, we use 69 LCs from \citet[][20221a, 2022b in preparation]{Kann2006,Kann2010,kann2011}, 22 GRBs from \citet[][2022]{Li2012,Li2015,Li2018a}, 3 GRBs from \cite{Oates2009,Oates2012}, 19 GRBs from \cite{Zaninoni2013}, and 16 GRBs from \cite{Si2018} which were in turn taken from \cite{Li2012} and \cite{Kann2006}.
Out of those, we combined the LCs of 6 GRBs from different authors: 3 from \cite{Li2012,Li2015}, 1 from \cite{Kann2010, Zaninoni2013}, and 1 from \cite{Kann2010, Zaninoni2013, Oates2009, Oates2012}. Additionally, we gathered 50 LCs from the GCN, {\it Swift} Burst Analyzer, the RATIR collaboration, and other literature.

We have also investigated GRB data points taken from the SUBARU Telescope, updating previous LCs and improving the fits of the following 18 GRBs: 110422A, 140801A, 140423A, 141121A, 020124A, 110801A, 100513A, 110503A, 980326A, 150413A, 140907A, 180325A, 160131A, 151027A, 160227A, 151029A, 170113A, 140206A.

\section{Methodology}\label{sec:methodology}
\begin{figure*}[!t]
\centering
\includegraphics[width=0.7\textwidth,angle=0,clip]{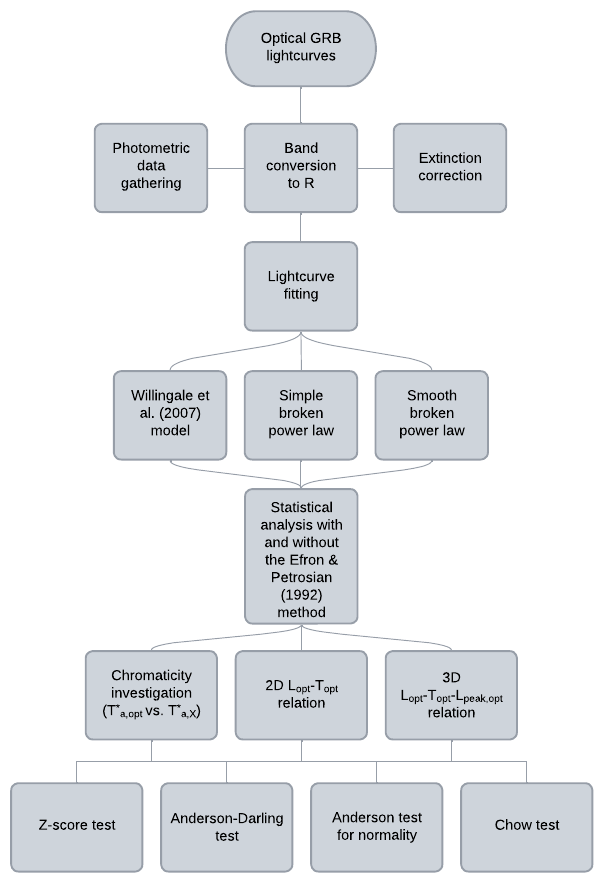}
\caption{Flow chart summarizing all the steps of the analysis.}
\label{workchart}
\end{figure*}

We briefly describe the analysis performed on LCs collected by \cite{Li2012,Li2015,Li2018a}, \cite{Kann2006,Kann2010,kann2011}, \cite{Oates2012}, \cite{Zaninoni2013}, \cite{Si2018}, and the LCs taken from the GCN. A flow chart summarizing all the steps of the analysis can be found in Fig. \ref{workchart}.
All errors in this paper are quoted at the $1\sigma$ level.

\subsection{Correction for Galactic and host-galaxy extinction}

For the GRBs not already corrected for host extinction in the papers cited previously, we computed the extinction factor $10^{-A_{\rm \lambda}/2.5}$ in flux density space. 
For the GRBs taken from \cite{Li2012,Li2015,Li2018a}, we followed \cite{Li2012,Li2015,Li2018a} to correct for Galactic and host-galaxy extinction through the extinction parameter $A_{\rm \lambda}$, assuming $R_{\rm v}=3.1$ and the Small Magellanic Cloud (SMC), Milky Way (MW), or Large Magellanic Cloud (LMC) dust models. The flux contribution coming from the host galaxy at very late times ($\sim 10^{6}\mathrm{\,s}$ after the GRB trigger) for some GRBs has also been subtracted.

For the GRBs taken from \cite{Kann2006, Kann2019} we follow \cite{Kann2006,Kann2019}. More specifically, for each afterglow, the multiband LCs are fit with a single PL, a smoothly broken PL, or a series of these. 
If necessary, a constant host-galaxy component is added, and a special supernova-model fit is applied if such a SN is detected (see \citealt{Kann2019}).  
The LCs are corrected for Galactic extinction, and the spectral energy distribution (SED) is assumed to be constant over the region fit and analyzed to determine the line-of-sight extinction from the host galaxy. 
The SED is then used twofold: first, it allows us (after necessary host- and SN-component removal) to shift other bands to the $R_{\rm C}$ band, for which there are essentially always measurements, creating a compound LC with maximized data density and temporal coverage. 
The LCs gathered by \cite{Oates2009,Oates2012} are corrected for host extinction using the same values as \cite{Oates2012}.
In \cite{Oates2009}, for each GRB, the onset of the prompt $\gamma$-ray emission (the start time of the $T_{90}$ parameter) is equal to the start time of the UVOT LC. However, here we convert it using the \emph{Swift} Burst Alert Telescope (BAT) trigger time as the start time of the UVOT LCs to have a consistent BAT trigger time, as the other LCs in the sample.

For the LCs gathered from \cite{Zaninoni2013}, SEDs are created at early and late times for each GRB, using optical filters for which data were available; spectral index values $\beta_\text{opt}$ are derived from fitting these SEDs, corrected for host and Galactic extinction. 

For the 50 GRBs gathered from GCNs, we correct for Galactic extinction using the reddening maps from \cite{Schlegel1998}, and the $A_b /E(B-V)_{\rm SFD}$ values from \cite{Schlafly2011}. For \emph{Swift} UVOT bandpasses, $A_b/E(B-V)_{\rm SFD}$ values are taken from the York Extinction Solver \citep{McCall2014}. 

\subsection{Magnitude conversions}
We converted magnitudes across 18 bandpasses ($B$, $H$, $I$, $I_C$, $J$, $K$, $K_S$, $R$, $R_C$, $V$, $Z$, $b$, $g$, $i$, $r$, $u$, $v$, and $z$) into energy fluxes (${\rm erg\, cm^{-2}\, s^{-1}}$) to the $R$ band using
a conversion of zero-point flux densities in any given band. We take the optical spectral indices from the literature; when no value has been found, we assume a constant photon index extrapolated from X-ray and taken from \cite{Evans2009} or GCNs.
The formulation for the conversion of flux densities to the R band is the following:
\begin{equation}
f_R = f_X \left(\frac{\lambda_\text{X}}{\lambda_\text{R}}\right)^{-\beta}.
\end{equation}
Then, we use the following equation to convert from magnitude to flux:
\begin{equation}
F_R = \nu_R  f_X \left(\frac{\lambda_X}{\lambda_R}\right)^{-\beta} 10^{-m_X/2.5}
\end{equation}
where, given a band $X$, $\lambda_X$ is the effective wavelength (\AA), $f_X$ the zero-point flux density (erg cm$^{-2}$ s$^{-1}$ Hz$^{-1}$), $\nu_R$ is the effective frequency (Hz) of the R band in the Johnson-Cousins system, $m_X$ is the observed magnitude, and $\beta$ is the optical spectral index either taken from the literature or taken as $\Gamma-1$, where $\Gamma$ is the photon index in X-ray extrapolated from \cite{Evans2009} or GCNs. 
We take effective wavelengths from \cite{Bessel1998} for Johnson-Cousins bands, \cite{Fukugita1996} for SDSS bands, \cite{Poole2008} for {\it Swift} UVOT bands. 

\begin{figure*}[!t]
\centering
\includegraphics[width=0.455\textwidth,angle=0,clip]{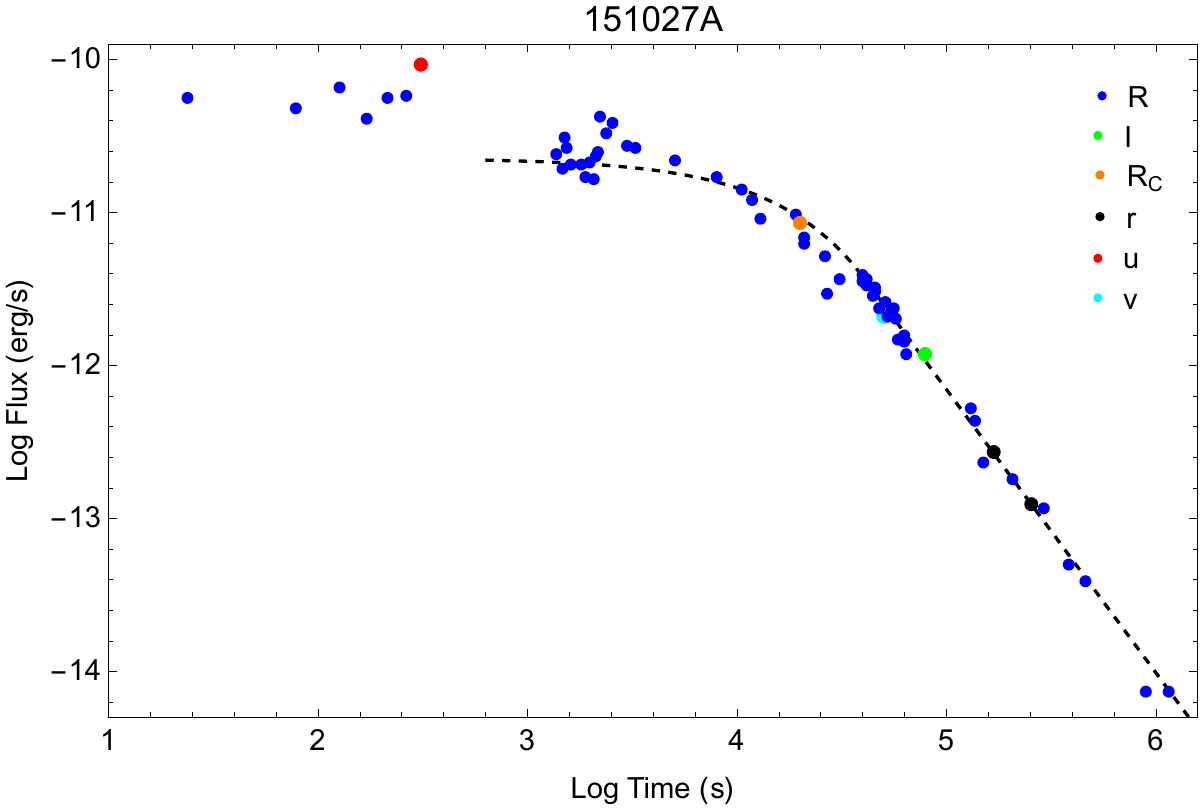}
\includegraphics[width=0.46\textwidth,angle=0,clip]{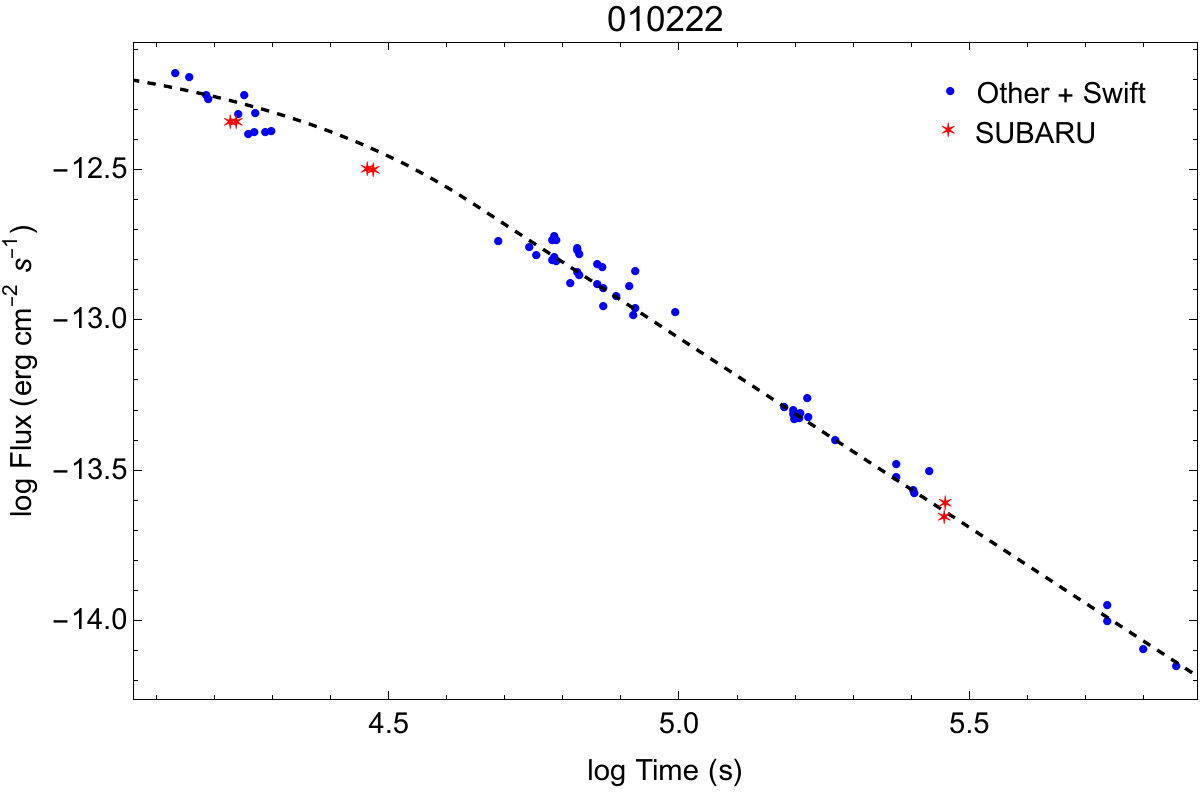}
\includegraphics[width=0.46\textwidth,angle=0,clip]{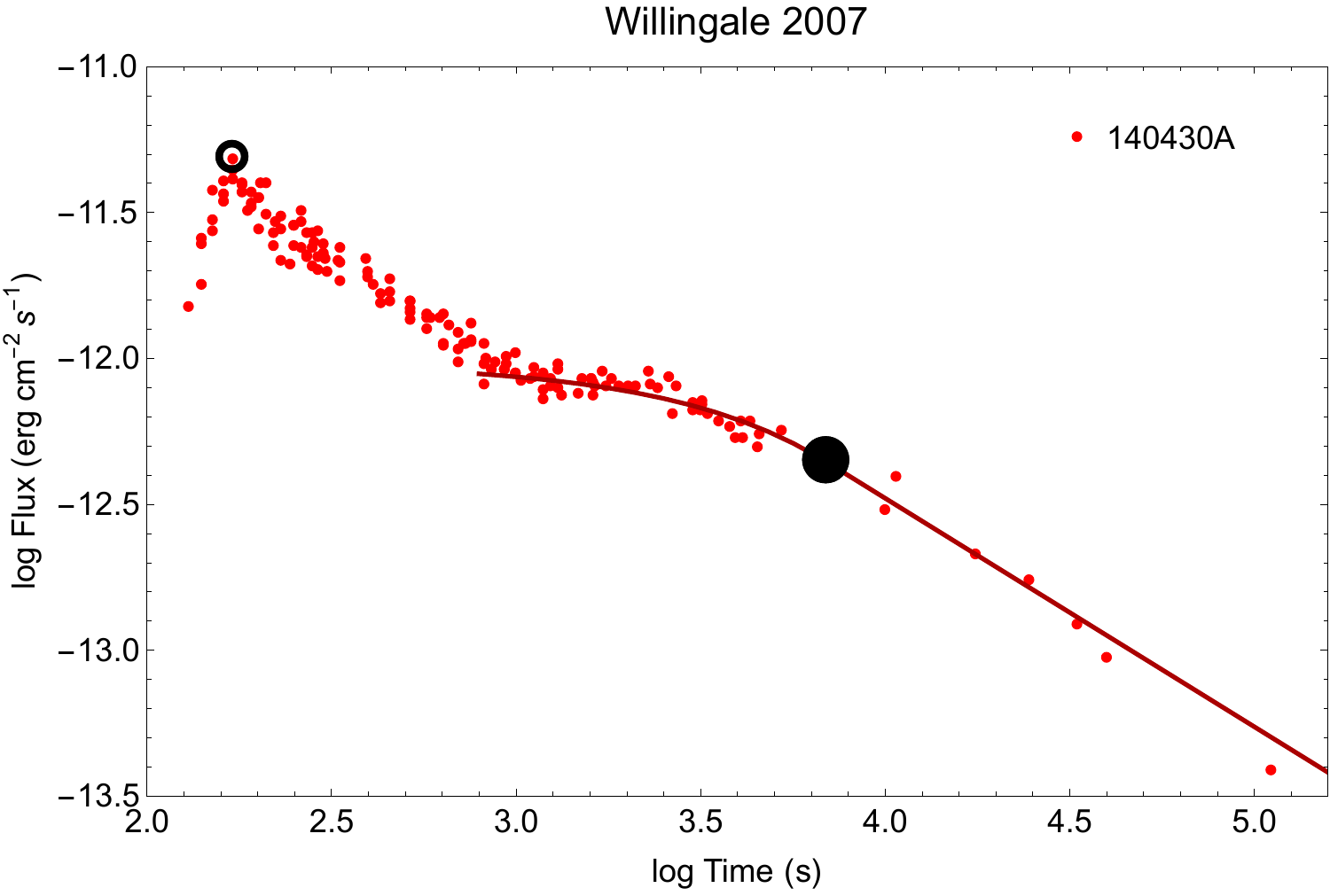}
\includegraphics[width=0.46\textwidth,angle=0,clip]{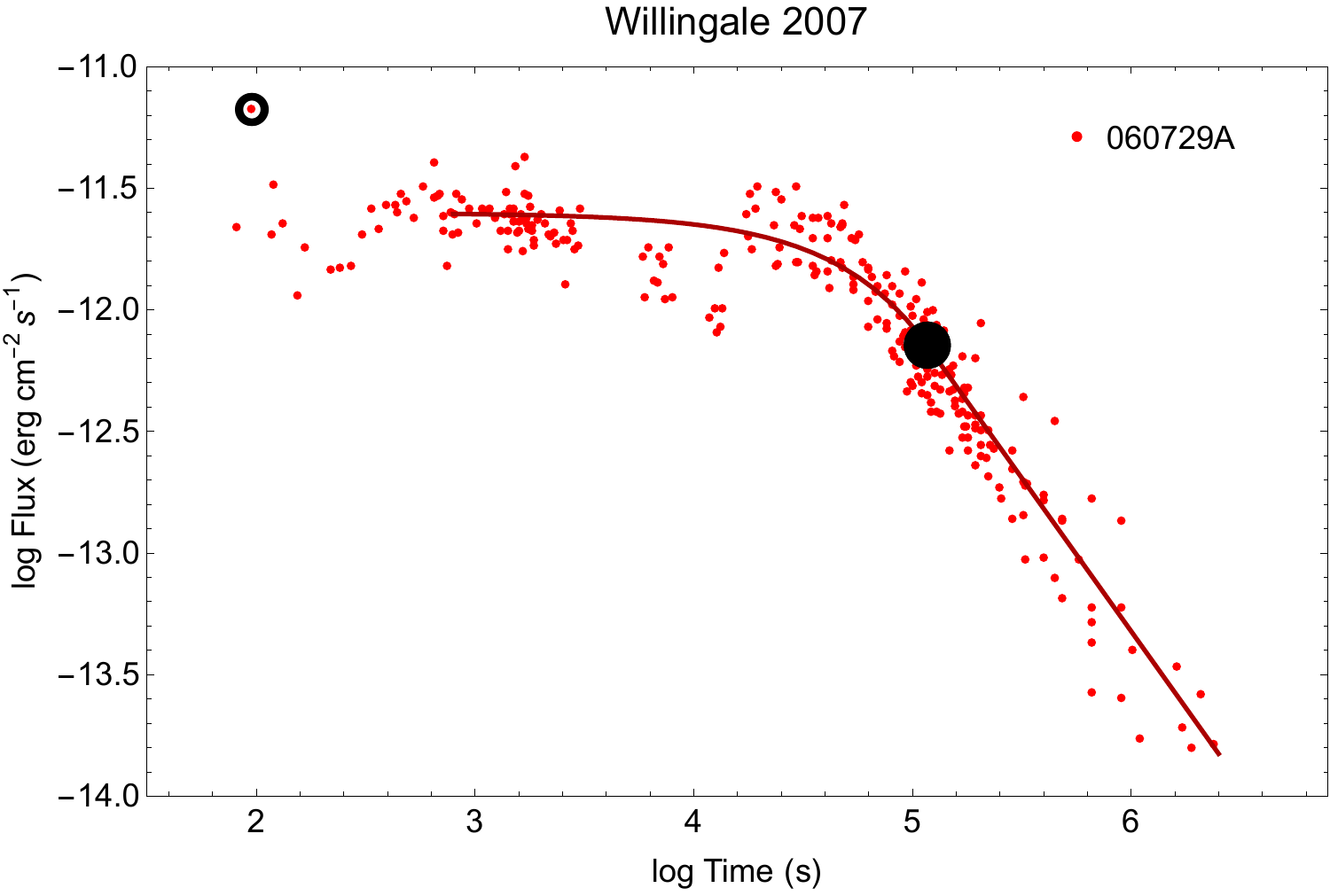}
\includegraphics[width=0.46\textwidth,angle=0,clip]{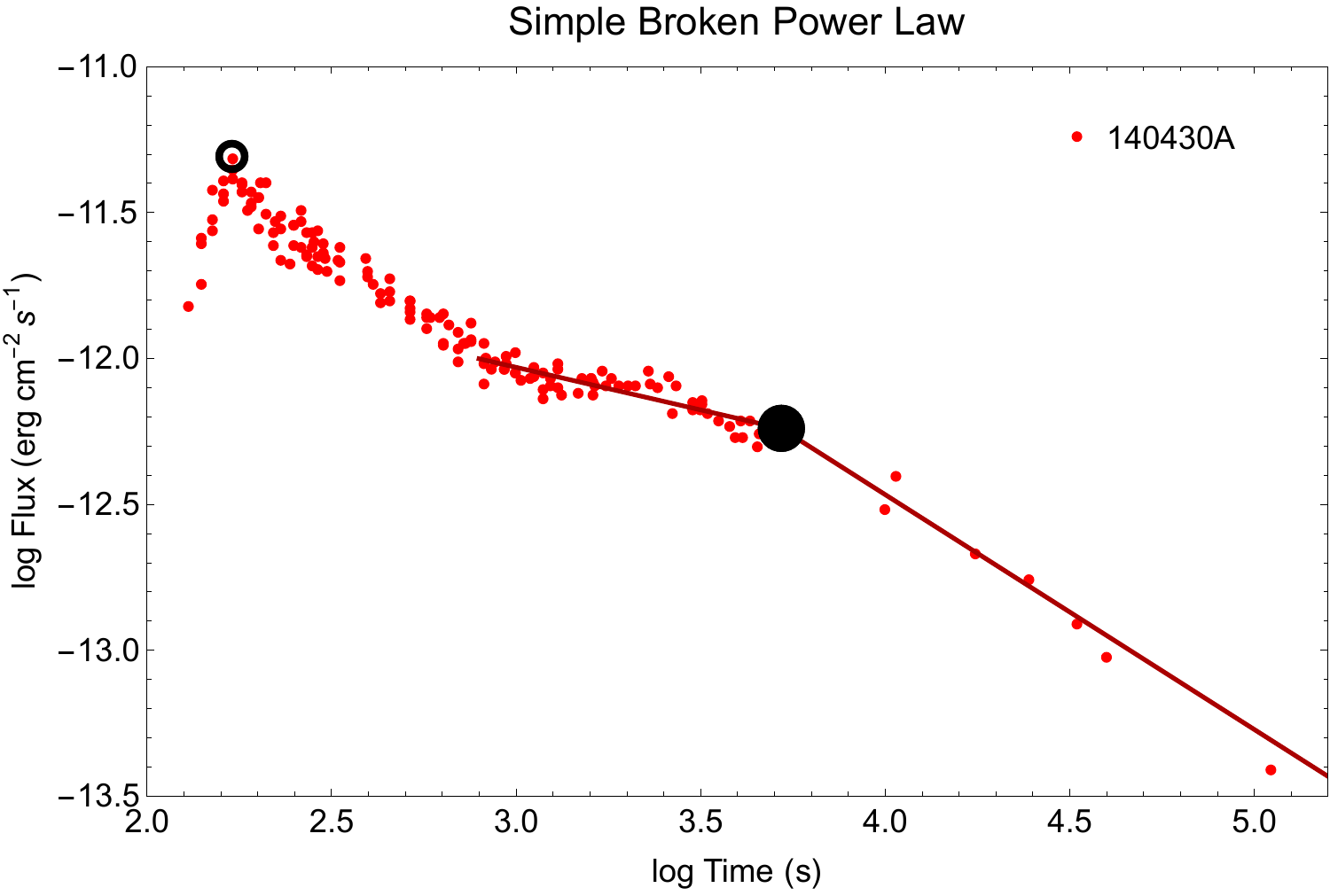}
\includegraphics[width=0.46\textwidth,angle=0,clip]{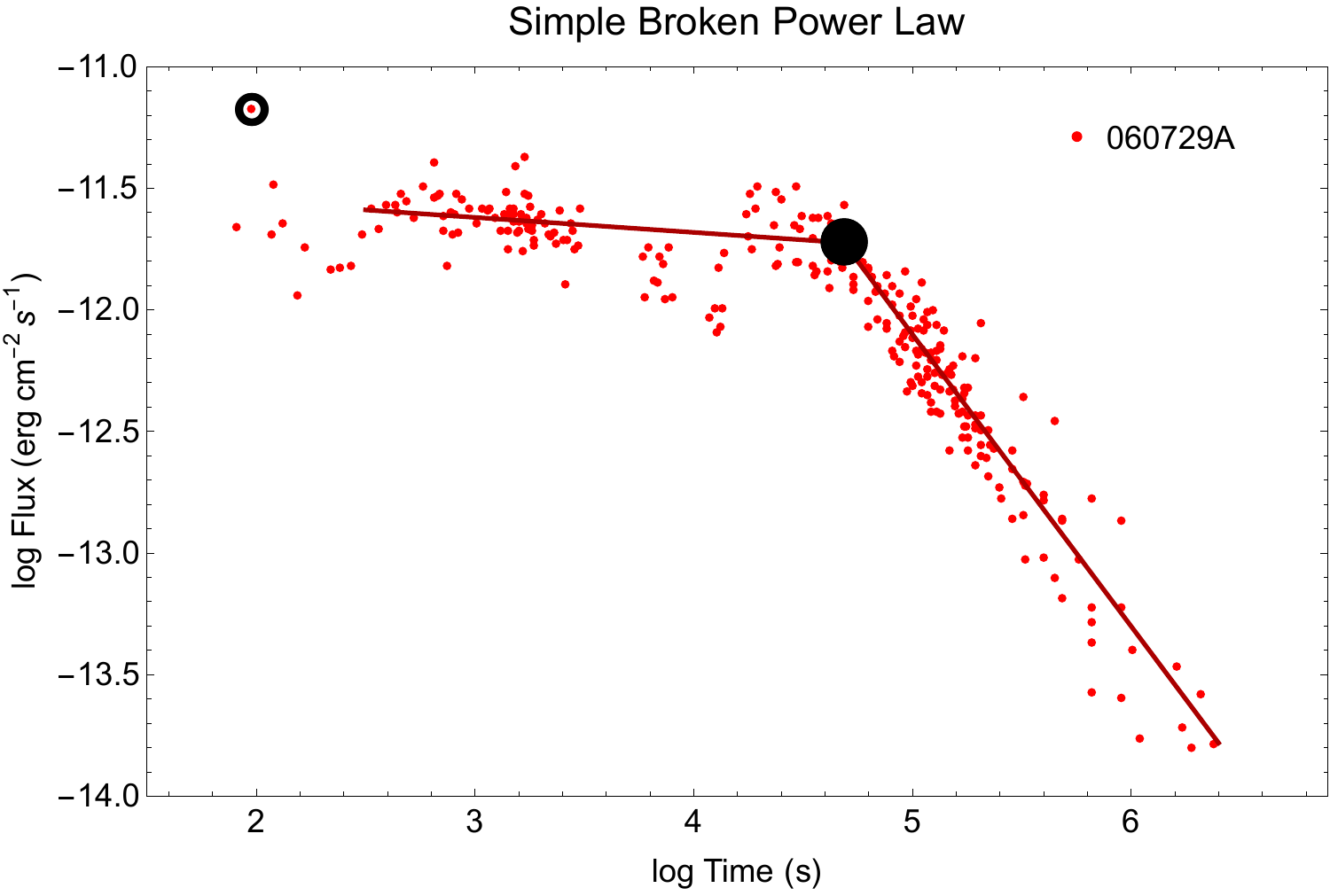}
\includegraphics[width=0.46\textwidth,angle=0,clip]{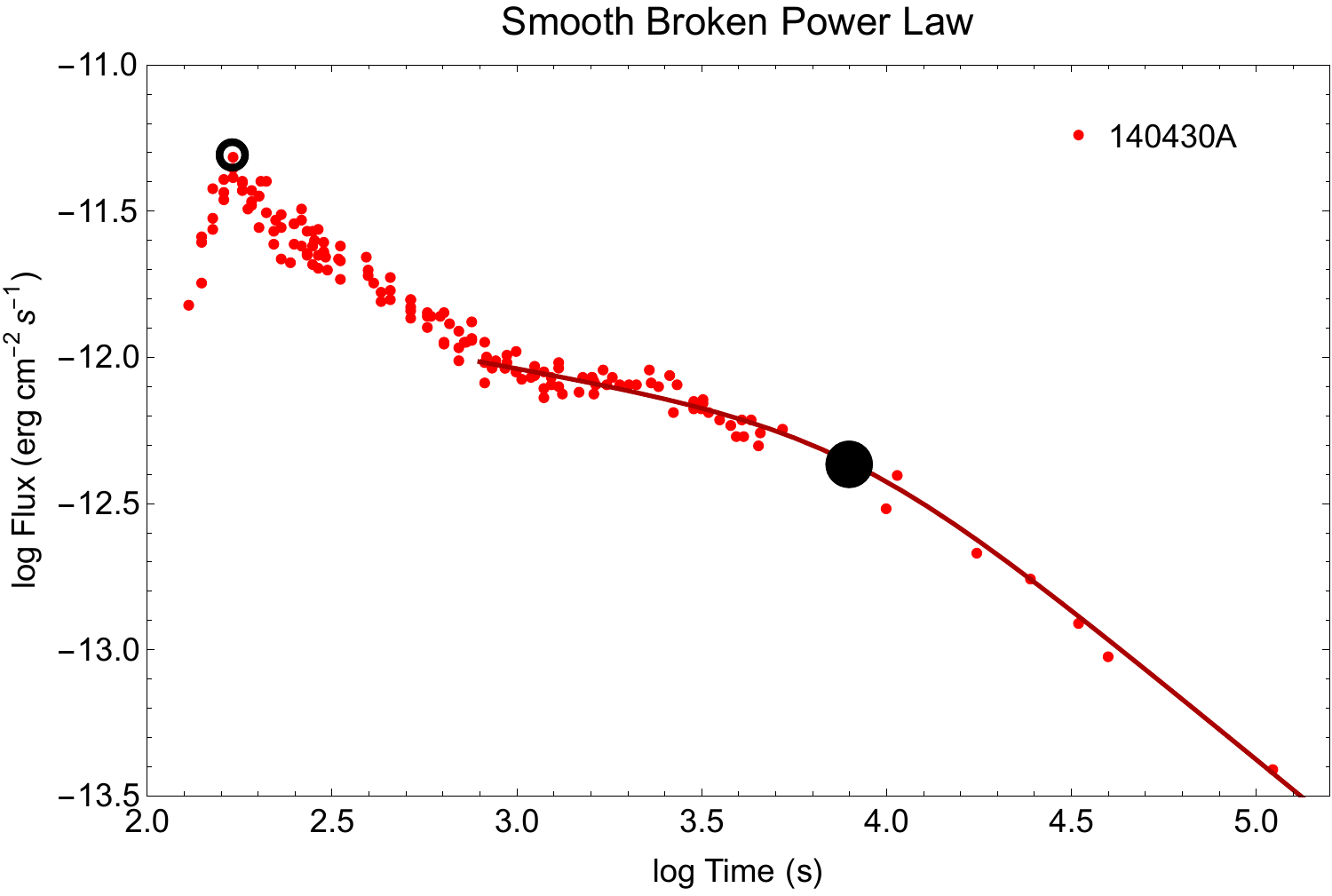}
\includegraphics[width=0.455\textwidth,angle=0,clip]{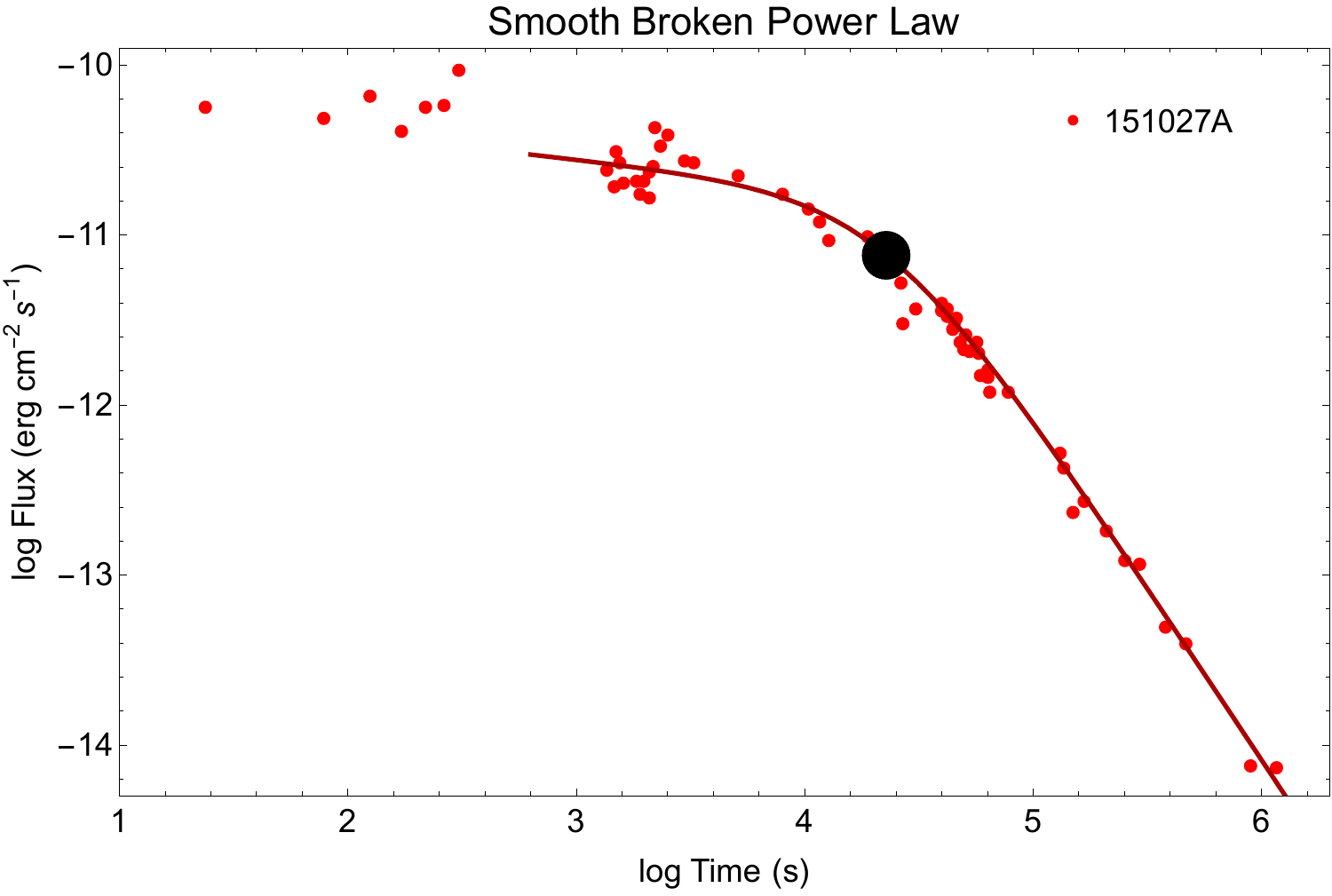}
\caption{The upper left panel shows a LC with multiple bands and the upper right panel shows a LC including the SUBARU data with the W07 function superimposed. The second row of panels show examples of LCs which present a peak of the prompt emission, indicated by an empty black circle and the end of the plateau emission with a filled black circle with the W07 function superimposed with a dark red curve.  The third and fourth rows of panels show the same, but with the simple and smoothly BPL, respectively.}
\label{fig:examples}
\end{figure*}

Examples of two LCs show data across six band-passes (top left panel of Fig. \ref{fig:examples}) and an enhanced coverage with data from the Subaru Telescope (top right panel of Fig. \ref{fig:examples}).

\noindent 

\subsection{Light curve fitting with the Willingale 2007 (W07) model}
We confirm the existence of a plateau by fitting the LCs to the phenomenological \cite{Willingale2007} model (hereafter W07, see the dashed black and solid red lines in Fig. \ref{fig:examples}).\footnote{The W07 model makes no assumptions on the underlying physics.}
We include all 179 GRBs that can be successfully fitted by the W07 model:

\begin{equation}
\label{eq:w07}
f(t) = \left \{
\begin{array}{ll}
\displaystyle{F_i \exp{\left ( \alpha_i \left( 1 - \frac{t}{T_i} \right) \right )} \exp{\left (
- \frac{t_i}{t} \right )}} & {\rm for} \ \ t < T_i \\
~ & ~ \\
\displaystyle{F_i \left ( \frac{t}{T_i} \right )^{-\alpha_i}
\exp{\left ( - \frac{t_i}{t} \right )}} & {\rm for} \ \ t \ge T_i, \\
\end{array}
\right .
\end{equation}

\noindent where the prompt (index `i=\textit{p}') and the afterglow (`i=\textit{a}') could be modelled, in principle. The LC $f_{tot}(t) = f_p(t) + f_a(t)$ contains two sets of four free parameters $(T_{i},F_{i},
\alpha_i,t_i)$, where $T_i$ and $F_i$ are the end times and corresponding fluxes at the plateau end, $\alpha_{i}$ is the temporal PL decay index, and $t_{i}$ is the initial rise timescale, usually fixed at zero. 
We exclude cases when the afterglow fitting procedure fails or the determination of $1\sigma$ confidence intervals does not satisfy the \cite{Avni1976} $\chi^{2}$ rules. 
We show in the middle panel of Fig. \ref{fig:examples} two examples of LCs fitted with the W07 model from a GRB which shows the peak of the prompt emission indicated by an empty black circle. 
Of the 500 LCs fitted, 179 yield a good fitting. We reject 69 for being a PL, 42 for not fulfilling the aforementioned $\Delta \chi ^2$ prescriptions, 43 for being too scattered, 148 for having insufficient data points, and 19 when multiple reasons above are happening at the same time.
We here clarify that the fit shows the presence of the plateau emission, given by several models such as the W07, smoothly BPL, or simple BPL.
For the successfully fitted LCs, we compute the source rest-frame isotropic luminosity $L_\text{opt}$ (erg s$^{-1}$) at the end of the plateau emission and, when possible, the peak prompt luminosity $L_\text{peak, opt}$ (erg s$^{-1}$) following \cite{dainotti2016,dainotti2017a,dainotti2017b,dainotti2020arXiv,dainotti2021b}.
The luminosities are defined as follows:
 \begin{equation}
L_\text{opt}= 4 \pi D_L^2(z) \, F_\text{opt} \textit{K}, \qquad L_\text{peak,opt}= 4 \pi D_L^2(z) \, F_\text{peak,opt} \textit{K},
\label{eq: la}
\end{equation}
where $D_L(z)$ is the luminosity distance assuming a flat $\Lambda$CDM cosmological model with $\Omega_M=0.3$ and $H_0=70$ km s$^{-1}$ Mpc$^{-1}$, $F_\text{opt}$ and $F_{peak, opt}$ are the measured optical energy flux (erg cm$^{-2}$ s$^{-1}$) at time $T_\text{opt}$, the end of the plateau, and in the peak of the prompt emission over a one second interval, respectively. Following \cite{Bloom2001}, we apply the \textit{K}-correction $K=1/(1+z)^{1-\beta_\text{opt}}$, where $\beta$ is the optical spectral index.\footnote{When $\beta_\text{opt}$ is not available, we use the XRT index.} We use the same $\beta_\text{opt}$ for all LCs assuming no spectral evolution. 
Finally, we construct a sub-sample from the 179 GRBs called the new Gold sample via the morphology conditions: 
\begin{itemize}
\item The plateau should not be too steep, with an angle of $<\ang{41}$. The angle of the plateau is defined as $\text{tan}^{-1}\left(\Delta F / \Delta T\right)$, with $\Delta F / \Delta T = (F_i-F_a)/(T_i - T_\text{opt})$, where $i$ denotes the time at the beginning of the plateau.
\item The largest change between times in the first 5 consecutive points in the plateau, normalized to the length of the plateau, should be $\left(\Delta T/(T_\text{opt} - T_{i})\right)_{\rm max} < 0.10$.
\item The largest relative change in flux in the first five consecutive points in the plateau should be $\Delta F_\text{max}/F<0.10$.
\end{itemize}

The definition of the $\ang{41}$ comes from \citet{dainotti2016} in which a Gaussian distribution characterizes the angular distribution; the angles $>$ $\ang{41}$ are the outliers beyond $1\sigma$ from that Gaussian. We show the plot of the Gaussian distribution for the X-ray data in Appendix B (Fig. 10) of \citet{dainotti2017b}. To be consistent with the X-ray sample, we use the same definition.

This data quality criterion defines the new Gold Sample, which includes $12$ GRBs in the W07 fitting ($42\%$ larger than the previous sample of 7). These criteria are chosen such that the definition of the plateau is enhanced, allowing a minimal variation in flux and time while still preserving the existence of the sample; more restrictive criteria would have caused the sample to be smaller, less restrictive criteria would enlarge the sample, but allow larger variation in fluxes and larger gaps in time. Although these criteria may not be the only choices, they safely allow the Gold sample to be reconstructed following these specific criteria. The results of the fitting are stored in Table \ref{table:corr_paramsWL}.

\subsection{Light curve fitting with the simple broken power law and smoothly broken power law models}
To give generality to the analysis and to use a model which is not phenomenological and is driven by the underlying physics of the standard fireball model, we have also attempted to fit the sample of 179 with the simple and smoothly BPL. Specifically, the simple BPL reads as follows:
\begin{equation}
f(t) = \left \{
\begin{array}{ll}
\displaystyle{F_i \left (\frac{t}{T_i} \right)^{-\alpha_1} 
} & {\rm for} \ \ t < T_i \\
\displaystyle{F_i \left ( \frac{t}{T_i} \right )^{-\alpha_2}
} & {\rm for} \ \ t \ge T_i, \\
\end{array}
\right .
\label{eq:simpleBPL}  
\end{equation}
where $T_i$ is the time at the end of the plateau, and $\alpha_1$ and $\alpha_2$ are the slopes of the LC before and after the time $T_i$.
Within this analysis, we consider the cases that have the angle of the plateau $<\ang{41}$ corresponding to $\alpha_1<0.8$ and we removed the cases in which $\sigma_{\alpha_2}/\alpha_2<1$ and $\sigma F/F<1$ and $\sigma_T/T$. We denote with $\sigma$ the error in 1 $\sigma$. 
It's important to clarify here that we did not adopt the condition that the ratio of $\alpha_1$ error to best-fit value $\sigma_{\alpha_1}/\alpha_1<1$ in approving fits because this would naturally introduce a bias against very flat plateaus. Indeed, if a plateau is very close to 0, a small error bar can be larger than 100\%, but the plateau is still flat. To guarantee that the error bars still preserve the condition of flatness, we instead use a criterion in which $\sigma_{\alpha_1}<0.8$.
From the analysis of the 179 LCs, we obtain 99 cases that fulfill these and the $\chi^2$ requirements. 

We also attempt to fit the sample of 179 GRBs with the smoothly BPL: 
\begin{equation}
\label{eq:smoothBPL}
    f(t) = F_i\left(\left(\frac{t}{T_i}\right)^{S\alpha_1} + \left(\frac{t}{T_i}\right)^{S\alpha_2}\right)^{-\frac{1}{S}}
\end{equation}
with all parameters defined as before but with the addition of a smoothing parameter $S$. After the smoothly BPL fitting we are left with 45 GRBs that fulfill the $\chi^2$ requirements. We present the results of the simple and smoothly BPL fittings in Tables \ref{table:corr_params_simpleBPL} and \ref{table:corr_params_smoothBPL}. Two examples of fitting with the simple and smoothly BPL models are shown in the bottom panel of Fig. \ref{table:corr_paramsWL}. As expected, both the end time of the plateau and its corresponding fluxes are compatible within $1\sigma$.

\section{The optical correlations}\label{sec:3D correlation}

We present the results of the fitting with and without evolution for the 2D and 3D correlations using the W07 and the two BPL models. We show the z-score values for all three models calculated as the distance from the Gold fundamental plane to the other classes (see Table \ref{table:corr_paramsWL}, Table \ref{table:corr_params_simpleBPL}, and Table \ref{table:corr_params_smoothBPL}). In all plots, all the logarithmic scales are in the base of 10.

\subsection{The 3D fundamental plane relation with the Willingale (2007) model: $L_{\rm opt}^{(\prime)}$-$T_{\rm opt}^{(*,\prime)}$-$L_{\rm peak,opt}^{(\prime)}$}

We find that 58 out of the total sample of 179 GRBs show a peak in the prompt emission.
To determine the peak flux in these cases, we consider the highest flux before the initial decay phase and when the time is nearly coincident with the peak flux in the X-ray data. Two examples of optical peak fluxes are shown in the bottom panels of Fig. \ref{fig:examples}.
With this information, we build the 3D optical correlation.

The optical fundamental plane relation is defined as:
\begin{equation}
\log L_\text{opt} = C_0 + a \times \log T^{*}_\text{opt} + b \times \log L_\text{peak,opt} \label{the fundamental plane},
\end{equation}
\noindent where $C_0$ is the normalization, $a_\text{opt}$ and $b_\text{opt}$ are the best fit parameters related to $\log T^{*}_\text{opt}$ and $\log L_\text{peak,opt}$, respectively, see the upper left panel of Fig. \ref{fig:3D}.
We also consider the evolutionary effects, and we correct for them following \cite{Efron1992}. 
The fundamental plane corrected for selection biases and redshift evolution is:
\begin{equation}
\log L '_\text{opt} = C '_0 + a ' \times \log T '_\text{opt} + b ' \times \log L '_\text{peak,opt},
\end{equation}

\noindent where $C '_0$, $a '$ and $b '$ are the parameters of the plane corrected for selection biases (upper right panel of Fig. \ref{fig:3D}). The new variables are $L '_\text{opt}=L_\text{opt}/(1+z)^{k_\text{opt,L}}$, $L'_\text{peak,opt}=L_\text{peak,opt}/(1+z)^{k_\text{opt,peak}}$ and $T '_\text{opt}=T_\text{opt}/(1+z)^{k_\text{opt,T}}$ where $k_\text{opt,L}=3.96 \pm 0.43$, $k_\text{opt,peak}=3.10 \pm 1.60$ and $k_\text{opt,T}=-2.09 \pm 0.40$ define the slope of the evolutionary functions. For details about the application of the method, see \cite{dainotti2013, dainotti2015,dainotti2017a,dainotti2020arXiv}.

For the X-ray fundamental plane relation, the correction for evolution is $k_{T_X}=-1.25 \pm 0.28$, for $k_{L_{X,a}}=2.42 \pm 0.58$, $k_{L_{peak}}=2.24 \pm 0.3$. Thus, the evolutionary parameters for the $L_{X, opt}$ and $T_{X, opt}$ are within $2\sigma$, while that for $L_{X, opt peak}$ is within $1\sigma$.

\begin{figure*}[!t]
\centering
\includegraphics[width=0.49\textwidth,angle=0,clip]{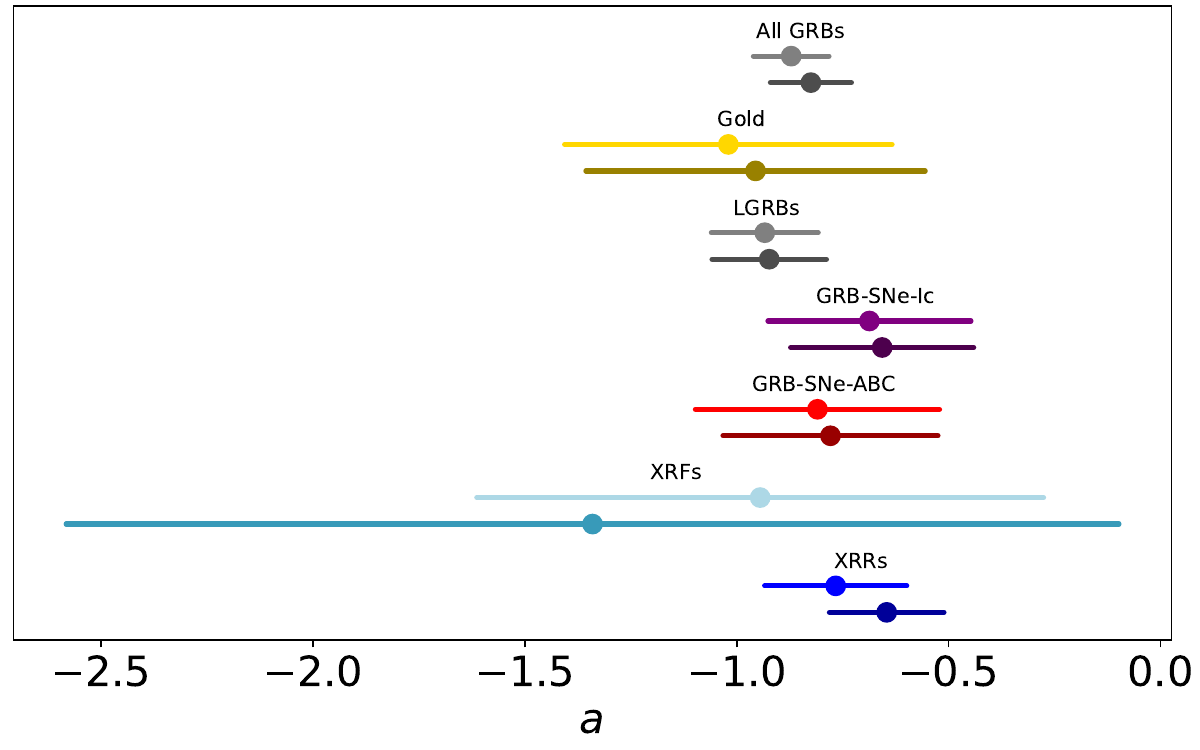}
\includegraphics[width=0.49\textwidth,angle=0,clip]{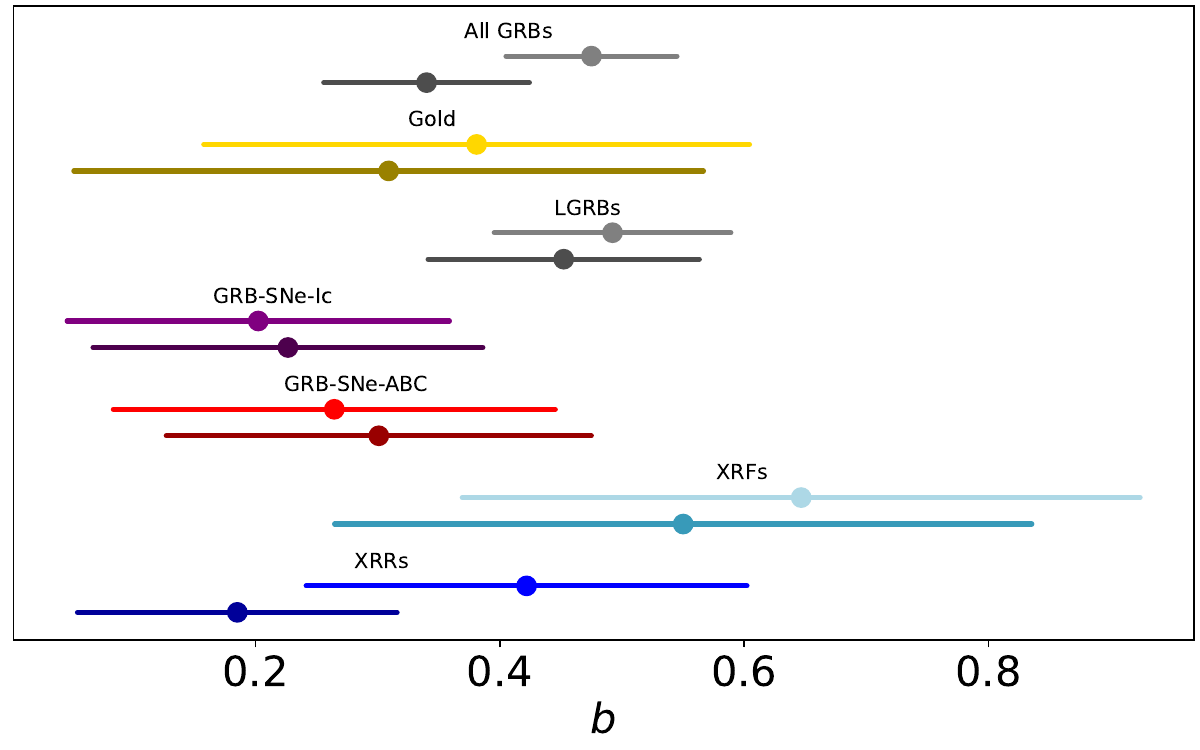}
\includegraphics[width=0.49\textwidth,angle=0,clip]{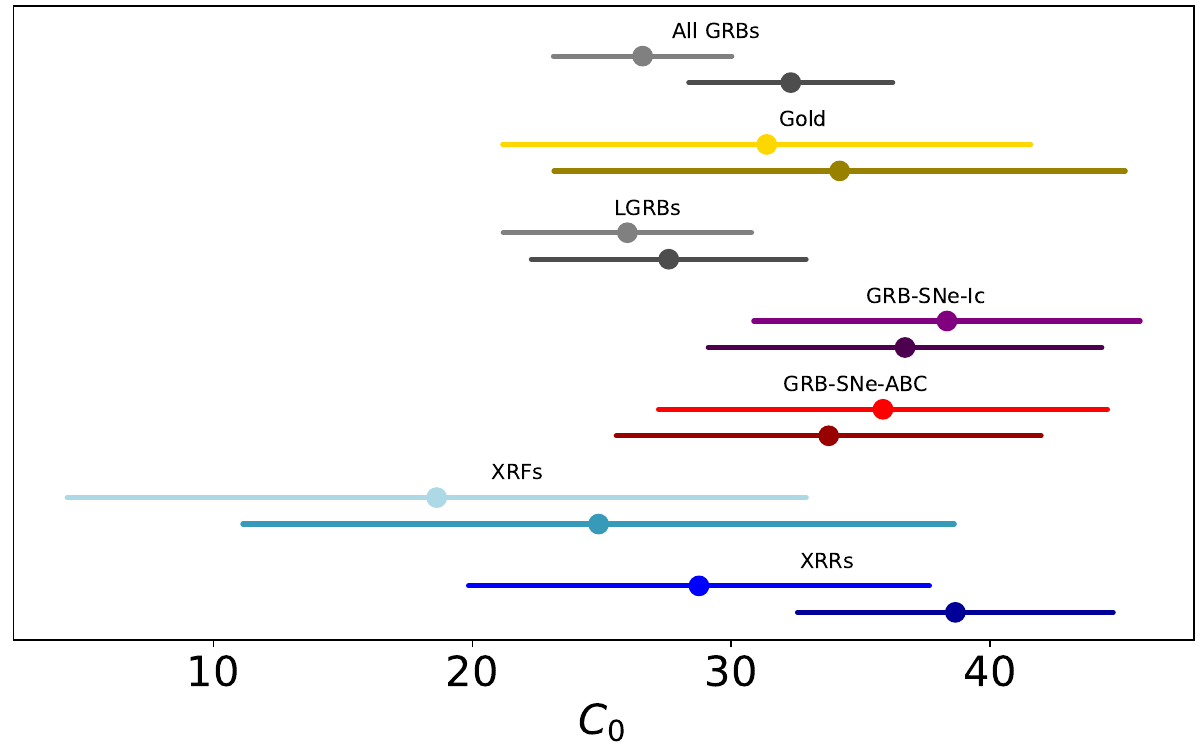}
\includegraphics[width=0.49\textwidth,angle=0,clip]{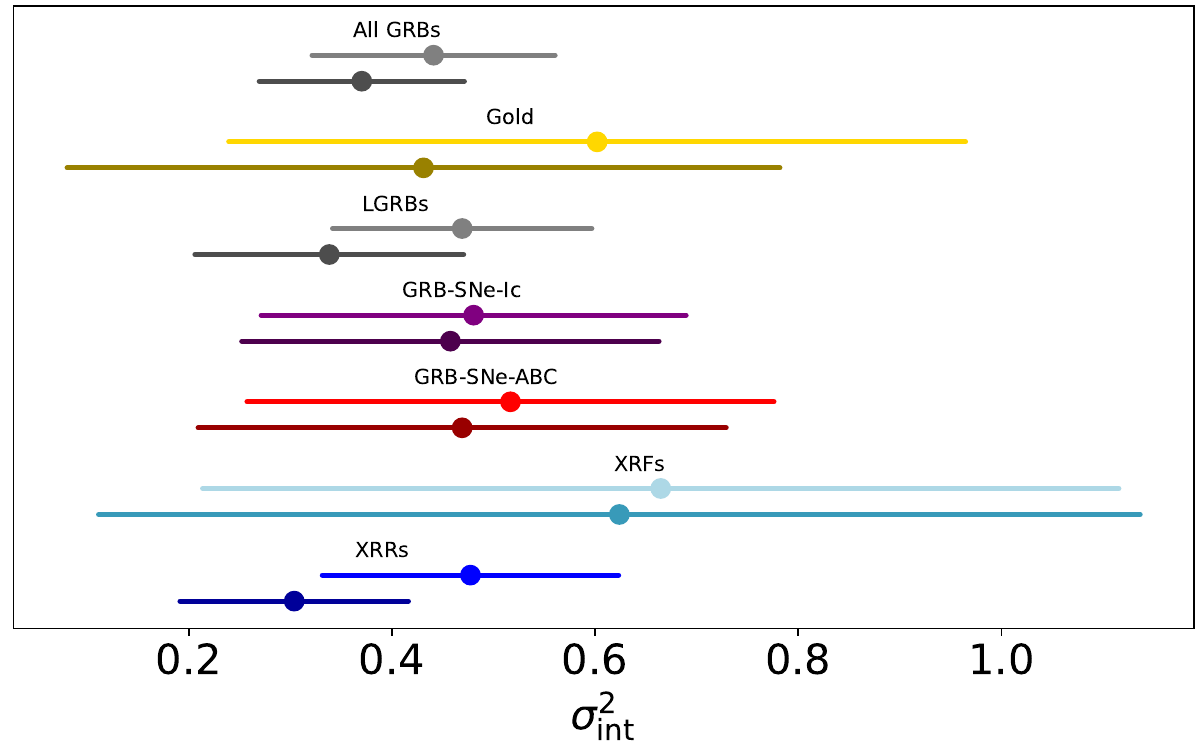}
\caption{An overview of 3D fundamental plane best fit parameters, where $a$, $b$ and $C_0$ are color coded according to the GRB class, with darker colors denoting correction. The x-axis displays the $1\sigma$ intervals for each parameter. The color-coding is as follows: grey for LGRBs, gold for the Gold sample, light blue for XRFs, blue for XRRs, bright green for UL, purple for SNe-Ic, red for SNe-ABC, cyan for short and type I (since both samples are identical) and orange for type II. The same colors, but darker are for the corrected class. Exact values for each parameter can be found in Table \ref{table:corr_paramsWL}.}
\label{fig:3Dquickres}
\end{figure*}

The derived parameters for each class can be found in Fig. \ref{fig:3Dquickres} and Table \ref{table:corr_paramsWL}. The $a$ parameter between the Gold and all samples are compatible with $1\sigma$ both for the corrected and uncorrected case.
The $a$ parameter of the uncorrected and corrected correlation in all other classes are compatible with one another within $1\sigma$.
For the uncorrected $b$ parameter is compatible with all classes with the exception of a few cases. The GRB-SNe-Ic is compatible with XRR with $2\sigma$ and GRB-SNe-Ic and LGRBs are compatible within 2 $\sigma$.
The corrected $b$ parameter is compatible with all classes with the exception of only LGRBs and XRR which are compatible within $2\sigma$. 
The $C_0$ uncorrected and corrected are  all compatible within $1\sigma$ except with the uncorrected case in which GRB-SNe-Ic and LGRBs are compatible within $2\sigma$.

The last column of Table \ref{table:corr_paramsWL} shows the percentage decrease for all samples compared between the uncorrected and corrected correlation in 3D. The intrinsic scatter of the corrected variables as compared to those uncorrected for evolution is smaller by at least $4\%$ in the case of the GRB-SNe-Ic and up to $36\%$ for the XRR subsample. This illustrates the importance of accounting for selection biases.

\begin{sidewaystable}
\centering
\hspace*{-3.5cm}
\begin{tabular}{@{}llccccccccccc@{}}
\toprule
GRB & Author & z & T90 & Class & logFa & logFaErr  & logTa & logTaErr  & beta & betaErr & logL(Ta) & logL(Ta)Err  \\
&  &  &  (s) &  & ($\mathrm{erg~cm^{-2} s^{-1}}$) &  ($\mathrm{erg~cm^{-2} ~s^{-1}}$)& (s)  & (s) & (s) & (s) & ($\mathrm{erg~~s^{-1}}$)  & ($\mathrm{erg~s^{-1}}$) \\ \midrule
970508A	&	Kann{[}1{]}, Kann{[}2{]}	&	0.83	&	13.2	&	XRF	&	-13.88	&	0.07	&	6.25	&	0.07	&	0.32	&	0.15	&	43.46	&	0.08	\\
980326A	&	GCN{[}3{]}	&	1	&	5	&	L	&	-14.54	&	1.12	&	5.31	&	0.60	&	0.66	&	0.70	&	43.07	&	1.14	\\
990510A	&	Kann{[}2{]}	&	1.62	&	67.58	&	L	&	-12.30	&	0.02	&	4.83	&	0.01	&	0.17	&	0.15	&	45.59	&	0.07	\\
000301C	&	Si{[}4{]}	&	2.03	&	2	&	IS	&	-13.83	&	0.12	&	5.88	&	0.05	&	0.59	&	0.12	&	44.45	&	0.13	\\
000926A	&	Kann{[}1{]}	&	2.07	&	25	&	L	&	-12.89	&	0.03	&	5.12	&	0.02	&	1.01	&	0.16	&	45.61	&	0.08	\\
010222A	&	Li{[}5{]}, Watanabe{[}6{]}	&	1.48	&	2	&	L	&	-12.61	&	0.02	&	4.64	&	0.02	&	0.76	&	0.22	&	45.44	&	0.09	\\
011211A	&	Kann{[}2{]}	&	2.14	&	270	&	L	&	-13.52	&	0.06	&	5.23	&	0.04	&	0.41	&	0.15	&	44.72	&	0.10	\\
020124A	&	GCN{[}7{]}	&	3.198	&	45.91	&	L	&	-12.77	&	0.05	&	4.45	&	0.07	&	1.32	&	0.25	&	46.38	&	0.17	\\
021004A	&	Li{[}5{]}	&	2.34	&	100	&	L	&	-12.92	&	0.02	&	5.40	&	0.02	&	0.67	&	0.14	&	45.53	&	0.08	\\
030226A	&	Kann{[}1{]}	&	1.99	&	22.09	&	L	&	-13.14	&	0.06	&	4.97	&	0.03	&	0.57	&	0.12	&	45.11	&	0.09	\\
030328A	&	Kann{[}1{]}	&	1.52	&	199.2	&	L	&	-12.70	&	0.02	&	4.38	&	0.02	&	0.36	&	0.45	&	45.21	&	0.18	\\
030329A	&	Si{[}4{]}	&	0.17	&	62.9	&	SN-A	&	-11.76	&	0.09	&	5.50	&	0.05	&	0.41	&	0.17	&	44.10	&	0.09	\\
030429A	&	Li{[}5{]},Leven{[}8{]}	&	2.658	&	10.3	&	XRF	&	-14.08	&	0.82	&	5.46	&	0.28	&	0.22	&	0.24	&	44.25	&	0.83	\\
040924A	&	Kann{[}1{]}	&	0.86	&	2.39	&	SN-C	&	-12.20	&	0.04	&	3.50	&	0.04	&	0.63	&	0.48	&	45.26	&	0.14	\\
041006A	&	Si{[}4{]}	&	0.72	&	17.4	&	SN-C	&	-12.45	&	0.03	&	4.08	&	0.03	&	0.36	&	0.27	&	44.77	&	0.07	\\
050319A	&	Zaninoni{[}9{]}	&	3.24	&	152.54	&	XRR	&	-12.83	&	0.02	&	4.44	&	0.03	&	0.74	&	0.42	&	45.98	&	0.26	\\
050401A	&	Li{[}5{]}, Kamble{[}10{]}	&	2.90	&	32.09	&	L	&	-13.60	&	0.12	&	4.14	&	0.16	&	0.39	&	0.05	&	44.89	&	0.12	\\
050408A	&	Si{[}4{]}	&	1.24	&	34	&	L	&	-13.25	&	0.03	&	4.36	&	0.05	&	0.28	&	0.27	&	44.45	&	0.10	\\
050416A	&	Li{[}5{]}	&	0.65	&	2.49	&	XRF-D-IS-SN	&	-13.54	&	0.05	&	4.15	&	0.06	&	0.92	&	0.30	&	43.70	&	0.08	\\
050502A	&	GCN{[}11{]}	&	3.79	&	20	&	L	&	-12.60	&	0.04	&	3.72	&	0.03	&	0.76	&	0.16	&	46.37	&	0.12	\\
050502B	&	GCN{[}12{]}	&	5.2	&	17.5	&	L	&	-12.45	&	0.12	&	3.58	&	0.14	&	0.90	&	0.06	&	46.93	&	0.13	\\
050525A	&	Kann{[}2{]}	&	0.61	&	8.83	&	SN-B-XRR	&	-11.70	&	0.07	&	3.22	&	0.10	&	0.52	&	0.08	&	45.39	&	0.07	\\
050603A	&	Kann{[}2{]}	&	2.82	&	21	&	L	&	-11.88	&	0.13	&	4.45	&	0.08	&	0.95	&	0.08	&	46.91	&	0.14	\\
050730A	&	Kann{[}2{]}	&	3.97	&	156.5	&	L	&	-12.37	&	0.04	&	4.34	&	0.05	&	0.52	&	0.05	&	46.48	&	0.06	\\
050801A	&	Kann{[}2{]}	&	1.38	&	19.4	&	XRR	&	-10.98	&	0.02	&	2.64	&	0.02	&	0.69	&	0.34	&	46.97	&	0.13	\\
050802A	&	Kann{[}2{]}	&	1.71	&	30	&	L	&	-11.61	&	0.08	&	2.91	&	0.09	&	0.72	&	0.03	&	46.56	&	0.08	\\
050820A	&	Kann{[}2{]}, Zaninoni{[}9{]}  	&	2.61	&	244.69	&	L	&	-11.97	&	0.01	&	4.46	&	0.02	&	0.72	&	0.03	&	46.62	&	0.02	\\
050824A	&	Kann{[}2{]}	&	0.83	&	22.58	&	XRF-SN-E	&	-12.50	&	0.03	&	3.65	&	0.06	&	0.45	&	0.18	&	44.87	&	0.06	\\
050904A	&	Li{[}5{]}, Yost{[}13{]}	&	6.295	&	181.57	&	UL	&	-14.18	&	0.13	&	5.54	&	0.07	&	1.00	&	0.09	&	45.47	&	0.15	\\
050908A	&	Zaninoni{[}9{]}	&	3.34	&	17.37	&	XRR	&	-12.61	&	0.08	&	3.26	&	0.13	&	1.80	&	0.09	&	46.90	&	0.10	\\
050922C	&	Kann{[}2{]}, Zaninoni{[}9{]}	&	2.2	&	4.54	&	IS	&	-11.65	&	0.01	&	3.77	&	0.01	&	0.56	&	0.01	&	46.69	&	0.01	\\
051028A	&	Li{[}5{]}, Castro-Tirado{[}15{]}	&	3.6	&	16	&	SN-B	&	-13.07	&	0.23	&	4.04	&	0.25	&	1.10	&	0.18	&	46.07	&	0.26	\\
051109A	&	Zaninoni{[}9{]}	&	2.35	&	37.23	&	L	&	-12.14	&	0.03	&	3.74	&	0.04	&	0.96	&	0.04	&	46.47	&	0.04	\\
051111A	&	Si{[}4{]}	&	1.55	&	59.78	&	L	&	-10.91	&	0.03	&	2.77	&	0.04	&	0.76	&	0.07	&	47.18	&	0.04	\\
060124A	&	Zaninoni{[}9{]}	&	2.3	&	13.63	&	XRR	&	-11.66	&	0.03	&	3.63	&	0.04	&	0.73	&	0.08	&	46.81	&	0.05	\\
060206A	&	Zaninoni{[}9{]}	&	4.05	&	7.59	&	XRR-IS	&	-12.01	&	0.00	&	4.29	&	0.00	&	0.77	&	0.01	&	47.03	&	0.01	\\
060210A	&	Kann{[}2{]}	&	3.91	&	255	&	L	&	-12.45	&	0.04	&	3.24	&	0.04	&	0.99	&	0.03	&	46.70	&	0.04	\\
060313A	&	Kann{[}16{]}	&	1.1	&	2	&	S	&	-12.89	&	0.13	&	3.58	&	0.39	&	0.85	&	0.06	&	44.88	&	0.13	\\
060418A	&	Kann{[}2{]}	&	1.49	&	144	&	XRR	&	-10.43	&	0.02	&	2.58	&	0.02	&	0.69	&	0.11	&	47.59	&	0.05	\\
  \bottomrule
\end{tabular}
\caption{The columns show in order the name of the GRB, the GCN information, the redshift, the $T_{90}$, the GRB class, the log of the flux at the end of the plateau emission, its error, the time at the end of the plateau emission, its error, the spectral index of the plateau and its error, the logarithm of the luminosity at the end of the plateau emission and its error. Errors are quoted at the $1\sigma$ level.}
\label{table:finaltable}
\end{sidewaystable}

\begin{sidewaystable*}
\centering
\hspace*{-30mm}
\begin{tabular}{lcCCCCCCCCCCCCC}
\multicolumn{13}{c}{Best Fit Parameters for the Subsamples in 2D with the W07 Model} \\ \hline
 \multicolumn{2}{l}{} & \multicolumn{5}{c}{Uncorrected for Evolution} & \multicolumn{5}{c}{Corrected for Evolution} & \multicolumn{1}{l}{} \\ \hline
Class & N & $a_\text{opt}$ & $C_0$ & $\sigma_\text{int}^2$ & $\text{z}$ & $\text{Chow}$ & $a'_\text{opt}$ & $C'_0$ & $\sigma_\text{int}^{\prime2}$ & $\text{z}^\prime$ & $\text{Chow}$ & $\Delta \sigma_\text{int}^2$ \\
All GRBs	&	179	&	-0.91	\pm	0.06	&	49.03	\pm	0.23	&	0.73	\pm	0.04	&	-1.65	&	0.87	&	-0.71	\pm	0.06	&	47.27	\pm	0.26	&	0.58	\pm	0.03	&	-0.95	&	0.92	&	-20$\%$ \\
Gold	&	12	&	-0.85	\pm	0.21	&	48.69	\pm	0.76	&	0.54	\pm	0.14	&	-	&	1.00	&	-0.66	\pm	0.17	&	46.95	\pm	0.78	&	0.34	\pm	0.10	&	-	&	1.00	&	-37$\%$ \\
LGRBs	&	102	&	-0.84	\pm	0.08	&	48.87	\pm	0.31	&	0.70	\pm	0.05	&	-2.39	&	0.69	&	-0.68	\pm	0.07	&	47.18	\pm	0.34	&	0.60	\pm	0.05	&	-1.75	&	0.42	&	-14$\%$ \\
SGRBs	&	13	&	-0.79	\pm	0.35	&	48.26	\pm	1.37	&	1.32	\pm	0.36	&	0.76	&	0.84	&	-0.55	\pm	0.25	&	45.92	\pm	1.22	&	0.91	\pm	0.25	&	2.17	&	0.67	&	-31$\%$ \\
GRB-SNe-Ic	&	26	&	-0.79	\pm	0.09	&	47.93	\pm	0.39	&	0.52	\pm	0.07	&	5.23	&	0.05	&	-0.65	\pm	0.12	&	46.83	\pm	0.57	&	0.54	\pm	0.09	&	1.41	&	0.54	&	+3$\%$ \\
GRB-SNe-ABC	&	19	&	-0.86	\pm	0.10	&	48.33	\pm	0.43	&	0.45	\pm	0.08	&	4.34	&	0.16	&	-0.71	\pm	0.11	&	47.23	\pm	0.55	&	0.44	\pm	0.10	&	0.26	&	0.63	&	+0$\%$ \\
ULGRBs	&	7	&	-1.31	\pm	0.45	&	50.77	\pm	2.19	&	1.00	\pm	0.45	&	0.54	&	0.74	&	-1.20	\pm	0.34	&	49.74	\pm	1.90	&	0.72	\pm	0.34	&	0.80	&	0.49	&	-28$\%$ \\
XRFs	&	16	&	-1.04	\pm	0.16	&	49.02	\pm	0.66	&	0.70	\pm	0.15	&	2.75	&	0.22	&	-0.69	\pm	0.15	&	46.76	\pm	0.70	&	0.49	\pm	0.12	&	2.92	&	0.18	&	-30$\%$ \\
XRRs	&	44	&	-1.07	\pm	0.12	&	49.67	\pm	0.41	&	0.69	\pm	0.08	&	-1.91	&	0.48	&	-0.85	\pm	0.11	&	47.89	\pm	0.46	&	0.49	\pm	0.06	&	-1.39	&	0.58	&	-28$\%$ \\
Type I GRBs	&	13	&	-0.79	\pm	0.35	&	48.26	\pm	1.37	&	1.32	\pm	0.36	&	0.76	&	0.84	&	-0.55	\pm	0.25	&	45.92	\pm	1.22	&	0.91	\pm	0.25	&	2.17	&	0.67	&	-31$\%$ \\
Type II GRBs	&	171	&	-0.93	\pm	0.06	&	49.12	\pm	0.21	&	0.72	\pm	0.04	&	-1.86	&	0.81	&	-0.73	\pm	0.05	&	47.38	\pm	0.23	&	0.56	\pm	0.03	&	-1.39	&	0.86	&	-21$\%$ \\
\hline
\multicolumn{13}{c}{Best Fit Parameters for the Subsamples in 3D with the W07 Model} \\ \hline
 \multicolumn{2}{l}{} & \multicolumn{5}{c}{Uncorrected for Evolution} & \multicolumn{5}{c}{Corrected for Evolution} & \multicolumn{1}{l}{} \\ \hline
Class & N & $a_\text{opt}$ & $b_\text{opt}$ & $C_0$ & $\sigma_\text{int}^2$ & $\text{z}$ & $a'_\text{opt}$ & $b'_\text{opt}$ & $C'_0$ & $\sigma_\text{int}^{\prime2}$ & $\text{z}$ &$\Delta \sigma_\text{int}^{\prime2}$ \\
All GRBs	&	58	&	-0.87	\pm	0.09	&	0.48	\pm	0.07	&	26.57	\pm	3.44	&	0.44	\pm	0.12	&	-0.97	&	-0.82	\pm	0.10	&	0.34	\pm	0.08	&	32.30	\pm	3.94	&	0.37	\pm	0.10	&	-0.94	&	-16$\%$ \\
Gold	&	6	&	-1.02	\pm	0.39	&	0.38	\pm	0.22	&	31.37	\pm	10.19	&	0.60	\pm	0.36	&	0.0	&	-0.96	\pm	0.40	&	0.31	\pm	0.26	&	34.18	\pm	11.03	&	0.43	\pm	0.35	&	0.0	&	-28$\%$ \\
LGRBs	&	31	&	-0.93	\pm	0.13	&	0.49	\pm	0.10	&	25.99	\pm	4.79	&	0.47	\pm	0.13	&	-0.98	&	-0.92	\pm	0.13	&	0.45	\pm	0.11	&	27.58	\pm	5.30	&	0.34	\pm	0.13	&	-0.94	&	-27$\%$ \\
GRB-SNe-Ic	&	9	&	-0.69	\pm	0.24	&	0.20	\pm	0.16	&	38.33	\pm	7.46	&	0.48	\pm	0.21	&	0.20	&	-0.66	\pm	0.22	&	0.23	\pm	0.16	&	36.71	\pm	7.60	&	0.46	\pm	0.21	&	-0.27	&	-4$\%$ \\
GRB-SNe-ABC	&	7	&	-0.81	\pm	0.29	&	0.26	\pm	0.18	&	35.86	\pm	8.67	&	0.52	\pm	0.26	&	0.19	&	-0.78	\pm	0.25	&	0.30	\pm	0.17	&	33.76	\pm	8.20	&	0.47	\pm	0.26	&	-0.26	&	-9$\%$ \\
XRFs	&	4	&	-0.94	\pm	0.67	&	0.65	\pm	0.28	&	18.62	\pm	14.27	&	0.66	\pm	0.45	&	-0.02	&	-1.34	\pm	1.24	&	0.55	\pm	0.29	&	24.87	\pm	13.74	&	0.62	\pm	0.51	&	0.10	&	-6$\%$ \\
XRRs	&	19	&	-0.77	\pm	0.17	&	0.42	\pm	0.18	&	28.75	\pm	8.90	&	0.48	\pm	0.15	&	-1.04	&	-0.65	\pm	0.14	&	0.19	\pm	0.13	&	38.65	\pm	6.10	&	0.30	\pm	0.11	&	-0.94	&	-36$\%$ \\
\hline
\end{tabular}
\caption{The best fit parameters of the 2D and 3D correlations with the W07 model. The first vertical half of the table (left) shows the GRB classes, the normalization, the slope with their respective errors and the intrinsic scatter $\sigma_int^{2}$ for each class for the 2D correlation uncorrected for selection biases. The second vertical half of the table (right) show the same variables, but corrected for selection biases (denoted by "$\prime$").
The final (13th) column shows the fractional change in $\sigma_\text{int}^2$ among the classes after correcting for redshift evolution and selection biases}.
In the 2D subtable, the 6th and 7th (11th and 12th) columns show the z-score and Chow test $p$-value, representing two statistical tests comparing subclasses with the Gold fundamental plane without (with) the correction for redshift evolution. Similarly, z-scores for the 3D case are found in columns 7 and 12 of the lower subtable.
\label{table:corr_paramsWL}
\end{sidewaystable*}

% UL GRBs are skipped here (only 2) 
\begin{sidewaystable*}
\centering
\hspace*{-25mm}
\begin{tabular}{lcCCCCCCCCCCC}
\multicolumn{13}{c}{Best Fit Parameters for the Subsamples in 2D with the Simple BPL Model} \\ \hline
 \multicolumn{2}{l}{} & \multicolumn{5}{c}{Uncorrected for Evolution} & \multicolumn{5}{c}{Corrected for Evolution} & \multicolumn{1}{l}{} \\ \hline
Class & N & $a_\text{opt}$ & $C_0$ & $\sigma_\text{int}^2$ & $\text{z}$ & $\text{Chow}$ & $a'_\text{opt}$ & $C'_0$ & $\sigma_\text{int}^{\prime2}$ & $\text{z}^\prime$ & $\text{Chow}$ &$\Delta \sigma_\text{int}^{\prime2}$ \\
All GRBs & 99	&	-0.98	\pm 	0.09	& 	49.27	\pm 	0.32	& 	0.74	\pm 	0.06	& 	3.26	&	0.35	&	-0.72	\pm 	0.08	& 	47.31	\pm 	0.36	& 	0.57	\pm 	0.05	& 	4.80	&	0.03 & -23\%	\\
Gold & 10	&	-1.24	\pm 	0.27	& 	50.37	\pm 	0.93	& 	0.68	\pm 	0.21	& 	-	&	1.00	&	-1.16	\pm 	0.27	& 	49.49	\pm 	1.13	& 	0.55	\pm 	0.21	& 	-	&	1.00 & -19\%	\\
LGRBs & 50	&	-0.91 \pm 	0.12	& 	49.14	\pm 	0.42	& 	0.68	\pm 	0.07	& 	1.18	&	0.26	&	-0.72	\pm 	0.12	& 	47.41	\pm 	0.53	& 	0.56	\pm 	0.07	& 	2.15	&	0.03 & -18\%	\\
SGRBs & 9	&	-1.00	\pm 	0.51	& 	48.92	\pm 	1.71	& 	1.35	\pm 	0.38	& 	1.78	&	0.30	&	-0.72	\pm 	0.34	& 	46.71	\pm 	1.54	& 	0.92	\pm 	0.32	& 	3.31	&	0.02 & -32\%	\\
GRB-SNe-Ic & 14	&	-0.75	\pm 	0.15	& 	47.65	\pm 	0.64	& 	0.57	\pm 	0.14	& 	4.31	&	0.001	&	-0.67	\pm 	0.18	& 	46.87	\pm 	0.85	& 	0.58	\pm 	0.15	& 	2.90	&	0.03 & +2\%\\
GRB-SNe-ABC & 9	&	-0.80	\pm 	0.13	& 	47.91	\pm 	0.50	& 	0.40	\pm 	0.13	& 	3.45	&	0.003	&	-0.63	\pm 	0.14	& 	46.80	\pm 	0.61	& 	0.35	\pm 	0.13	& 	2.32	&	0.03 & -13\% \\
% ULGRBs & 2	&	-1.55	\pm 	0.50	& 	51.48	\pm 	2.38	& 	1.14	\pm 	0.51	& 	-0.18	&	0.06	&	-1.50	\pm 	0.54	& 	51.00	\pm 	2.75	& 	1.10	\pm 	0.52	& 	0.39	&	0.11& -4$\%$	\\
XRFs & 10	&	-1.06	\pm 	0.24	& 	49.07	\pm 	0.96	& 	0.71	\pm 	0.20	& 	3.47	&	0.08	&	-0.67	\pm 	0.25	& 	46.78	\pm 	1.12	& 	0.56	\pm 	0.22	& 	3.14	&	0.02 & -21\%	\\
XRRs & 31	&	-1.05	\pm 	0.17	& 	49.62	\pm 	0.54	& 	0.77	\pm 	0.11	& 	1.48	&	0.75	&	-0.75	\pm 	0.14	& 	47.53	\pm 	0.59	& 	0.56	\pm 	0.10	& 	2.79	&	0.24 & -27\%	\\
Type I GRBs & 9	&	-1.00	\pm 	0.51	& 	48.92	\pm 	1.71	& 	1.35	\pm 	0.38	& 	1.78	&	0.30	&	-0.72	\pm 	0.34	& 	46.71	\pm 	1.54	& 	0.92	\pm 	0.32	& 	3.31	&	0.02 & -32\%	\\
Type II GRBs & 94	&	-0.99	\pm 	0.09	& 	49.32	\pm 	0.31	& 	0.72	\pm 	0.06	& 	2.91	&	0.40	&	-0.70	\pm 	0.08	& 	47.27	\pm 	0.34	& 	0.55	\pm 	0.05	& 	4.31	&	0.04 & -24\%	\\
\hline
\multicolumn{13}{c}{Best Fit Parameters for the Subsamples in 3D with Simple BPL Model} \\ \hline
 \multicolumn{2}{l}{} & \multicolumn{5}{c}{Uncorrected for Evolution} & \multicolumn{5}{c}{Corrected for Evolution} & \multicolumn{1}{l}{} \\ \hline
Class & N & $a_\text{opt}$ & $b_\text{opt}$ & $C_0$ & $\sigma_\text{int}^2$ & $\text{z}$ & $a'_\text{opt}$ & $b'_\text{opt}$ & $C'_0$ & $\sigma_\text{int}^{\prime2}$ & $\text{z}$ &$\Delta \sigma_\text{int}^{\prime2}$ \\
%skipping Gold, UL, SGRBs, SNe, SN-ABC, XRF, type I
All GRBs & 19	&	-0.84	\pm 	0.18	&	0.40	\pm 	0.13	&	30.07	\pm 	6.33	&	0.52	\pm 	0.12	& - &	-0.72	\pm 	0.19	&	0.29	\pm 	0.13	&	33.98	\pm 	5.88	&	0.45	\pm 	0.12 &  -	& -13\% \\
% Gold & 5	&	-1.20	\pm 	0.63	&	0.47	\pm 	0.24	&	27.51	\pm 	11.16	&	0.96	\pm 	0.60	& - &	-1.16	\pm 	0.58	&	0.47	\pm 	0.23	&	27.60	\pm 	10.76	&	0.84	\pm 	0.39 & -& -13$\%$	\\
LGRBs & 12	&	-0.95	\pm 	0.28	&	0.34	\pm 	0.15	&	33.36	\pm 	7.11	&	0.54	\pm 	0.16	& - &	-1.00	\pm 	0.30	&	0.34	\pm 	0.17	&	33.15 \pm 	7.75	&	0.51	\pm 	0.20 & - & -2\% \\
% GRB-SNe-Ic & 6	&	-0.82	\pm 	0.14	&	0.12	\pm 	0.11	&	42.69	\pm 	5.17	&	0.21	\pm 	0.13	& 0.88 &	-0.81	\pm 	0.16	&	0.19	\pm 	0.12	&	39.17	\pm 	5.68	&	0.21	\pm 	0.15 & 0.39	& 0$\%$ \\
% GRB-SNe-ABC & 5	&	-0.85	\pm 	0.18	&	0.14	\pm 	0.12	&	42.11	\pm 	5.73	&	0.26	\pm 	0.18	& 0.83 &	-0.86	\pm 	0.19	&	0.19	\pm 	0.12	&	39.50	\pm 	5.69	&	0.23	\pm 	0.19  & 0.45& -12$\%$ 	\\
XRRs & 5	&	-0.71	\pm 	0.36 &	0.39	\pm 	0.25	&	19.18	\pm 	12.33	&	0.83	\pm 	0.39	& - &	-0.37	\pm 	0.25	&	0.46	\pm 	0.22	&	25.02	\pm 	10.03	&	0.33	\pm 	0.32 & - & -60\% \\
Type II & 19 & -0.76 \pm 0.19 & 0.31 \pm 0.13 & 33.52 \pm 6.01 & 0.44 \pm 0.13 & - & -0.87 \pm 0.19 & 0.42 \pm 0.13 & 29.41 \pm 6.34 & 0.53 \pm 0.13 & - & -17$\%$ \\
\hline
\end{tabular}
\caption{The best fit parameters of the 2D and 3D correlations with the simple BPL model. The first vertical half of the table (left) shows the GRB classes, the normalization, the slope with their respective errors and the intrinsic scatter $\sigma_int^{2}$ for each class for the 2D correlation uncorrected for selection biases. The second vertical half of the table (right) show the same variables, but corrected for selection biases (denoted by "$\prime$").
The final (13th) column shows the fractional change in $\sigma_\text{int}^2$ among the classes after correcting for redshift evolution and selection biases}.
In the 2D subtable, the 6th and 7th (11th and 12th) columns show the z-score and Chow test $p$-value, representing two statistical tests comparing subclasses with the Gold fundamental plane without (with) the correction for redshift evolution. Z-scores could not be computed for the 3D case due to the lack of an appreciable Gold sample.
\label{table:corr_params_simpleBPL}
\end{sidewaystable*}

\begin{sidewaystable*}
\centering
\hspace*{-3cm}
\begin{tabular}{lcCCCCCCCCCCC}
\multicolumn{13}{c}{Best Fit Parameters for the Subsamples in 2D with the smoothly BPL Model} \\ \hline
 \multicolumn{2}{l}{} & \multicolumn{5}{c}{Uncorrected for Evolution} & \multicolumn{5}{c}{Corrected for Evolution} & \multicolumn{1}{l}{} \\ \hline
Class & N & $a_\text{opt}$ & $C_0$ & $\sigma_\text{int}^2$ & $\text{z}$ & $\text{Chow}$ & $a'_\text{opt}$ & $C'_0$ & $\sigma_\text{int}^{\prime2}$ & $\text{z}^\prime$ & $\text{Chow}$ &$\Delta \sigma_\text{int}^{\prime2}$ \\
All GRBs & 45	&	-0.97	\pm 	0.10	& 	49.48 \pm 	0.37	& 	0.74	\pm 	0.08	& 	0.27	&	0.49	&	-0.80	\pm 	0.11	& 	47.99	\pm 	0.47	& 	0.62	\pm 	0.08	& 	1.08	&	0.68 & -16\%	\\
Gold & 5	&	-0.57	\pm 	0.21	& 	48.14	\pm 	0.86	& 	0.31	\pm 	0.21	& 	0.00	&	1.00	&	-0.65	\pm 	0.26	& 	47.43	\pm 	1.24	& 	0.42	\pm 	0.25	& 	0.00	&	1.00 & +35\%	\\
LGRBs & 21	&	-1.04	\pm 	0.15	& 	49.97	\pm 	0.58	& 	0.65	\pm 	0.12	& 	-0.52	&	0.36	&	-0.98	\pm 	0.16	& 	49.01	\pm 	0.77	& 	0.60	\pm 	0.14	& 	-0.29	&	0.38 & -8\%	\\
SGRBs & 4	&	-1.25	\pm 	0.72	& 	50.37	\pm 	2.94	& 	2.23	\pm 	0.90	& 	0.49	&	0.72	&	-0.75	\pm 	0.46	& 	47.37	\pm 	2.31	& 	1.34	\pm 	0.61	&  1.57	&	0.59 & -40\%	\\
GRB-SNe-Ic & 5 &	-0.40	\pm 	0.23	& 	46.75	\pm 	0.85	& 	0.44	\pm 	0.22	& 	5.49	&	0.01	&	-0.33	\pm 	0.32	& 	45.82	\pm 	1.36	& 	0.56	\pm 	0.33	& 	1.37	&	0.26 & +27\%\\
GRB-SNe-ABC & 4	&	-0.69	\pm 	0.30	& 	47.99	\pm 	1.22	& 	0.38	\pm 	0.22	& 	4.91	&	0.01	&	-0.66	\pm 	0.45	& 	47.41	\pm 	1.91 & 	0.43	\pm 	0.25	& 	1.23	&	0.95 & +13\%	\\
% ULGRBs & 4	&	-1.55	\pm 	0.50	& 	51.48	\pm 	2.38	& 	1.14	\pm 	0.51	& 	-0.18	&	0.06	&	-1.50	\pm 	0.54	& 	51.00	\pm 	2.75	& 	1.10	\pm 	0.52	& 	0.39	&	0.11& -4$\%$	\\
XRFs & 4	&	-0.72	\pm 	0.29	& 	47.97	\pm 	1.05	& 	0.77	\pm 	0.32	& 	3.09	&	0.12	&	-0.59	\pm 	0.32	& 	46.38	\pm 	1.42	& 	0.81	\pm 	0.76	& 	4.70	&	0.09 & +5\%	\\
XRRs & 17	&	-1.27	\pm 	0.21	& 	50.27	\pm 	0.71	& 	0.81	\pm 	0.15	& 	0.11	&	0.15	&	-0.94	\pm 	0.21	& 	48.38	\pm 	0.85	& 	0.63	\pm 	0.14	& 	1.28	&	0.34 & -22\%	\\
Type I GRBs & 4	&	-1.25	\pm 	0.72	& 	50.37	\pm 	2.94	& 	2.23	\pm 	0.90	& 	0.49	&	0.72	&	-0.75	\pm 	0.46	& 	47.37	\pm 	2.31	& 	1.34	\pm 	0.61	&  1.57	&	0.59 & -40\%	\\
Type II GRBs & 44	&	-0.99	\pm 	0.11	& 	49.55	\pm 	0.41	& 	0.75	\pm 	0.09	& 	0.25	&	0.45	&	-0.82	\pm 	0.11	& 	48.09	\pm 	0.47	& 	0.63	\pm 	0.07	& 	1.02	&	0.67 & -16\%	\\
\hline
\multicolumn{13}{c}{Best Fit Parameters for the Subsamples in 3D with Smoothly BPL Model} \\ \hline
 \multicolumn{2}{l}{} & \multicolumn{5}{c}{Uncorrected for Evolution} & \multicolumn{5}{c}{Corrected for Evolution} & \multicolumn{1}{l}{} \\ \hline
Class & N & $a_\text{opt}$ & $b_\text{opt}$ & $C_0$ & $\sigma_\text{int}^2$ & $\text{z}$ & $a'_\text{opt}$ & $b'_\text{opt}$ & $C'_0$ & $\sigma_\text{int}^{\prime2}$ & $\text{z}$ &$\Delta \sigma_\text{int}^{\prime2}$ \\
All GRBs & 8 	&	-1.01	\pm 	0.31	&	0.31	\pm 	0.18	&	34.88	\pm 	8.30	&	0.60	\pm 	0.27	& - &	-0.71	\pm 	0.20	&	0.23	\pm 	0.11	&	36.90	\pm 	5.23	&	0.34	\pm 	0.17 &-	& -38\% \\
Type II & 8 	&	-1.01	\pm 	0.31	&	0.31	\pm 	0.18	&	34.88	\pm 	8.30	&	0.60	\pm 	0.27	& - &	-0.71	\pm 	0.20	&	0.23	\pm 	0.11	&	36.90	\pm 	5.23	&	0.34	\pm 	0.17 &-	& -38\% \\
% type II and all GRBs are the same sample 
% Gold & 5	&	-1.20	\pm 	0.63	&	0.47	\pm 	0.24	&	27.51	\pm 	11.16	&	0.96	\pm 	0.60	& - &	-1.16	\pm 	0.58	&	0.47	\pm 	0.23	&	27.60	\pm 	10.76	&	0.84	\pm 	0.39 & -& -13$\%$	\\
% LGRBs & 24	&	-0.76	\pm 	0.16	&	0.20	\pm 	0.10	&	39.24	\pm 	5.01	&	0.61	\pm 	0.14	& 0.72 &	-0.62	\pm 	0.15	&	0.19	\pm 	0.09	&	38.38	\pm 	3.91	&	0.44	\pm 	0.11 & 0.54& -28$\%$ \\
% GRB-SNe-Ic & 6	&	-0.82	\pm 	0.14	&	0.12	\pm 	0.11	&	42.69	\pm 	5.17	&	0.21	\pm 	0.13	& 0.88 &	-0.81	\pm 	0.16	&	0.19	\pm 	0.12	&	39.17	\pm 	5.68	&	0.21	\pm 	0.15 & 0.39	& 0$\%$ \\
% GRB-SNe-ABC & 5	&	-0.85	\pm 	0.18	&	0.14	\pm 	0.12	&	42.11	\pm 	5.73	&	0.26	\pm 	0.18	& 0.83 &	-0.86	\pm 	0.19	&	0.19	\pm 	0.12	&	39.50	\pm 	5.69	&	0.23	\pm 	0.19  & 0.45& -12$\%$ 	\\
% XRRs & 9	&	-0.84	\pm 	0.23	&	0.58	\pm 	0.25	&	21.40	\pm 	12.48	&	0.61	\pm 	0.22	& 0.53 &	-0.72	\pm 	0.19	&	0.33	\pm 	0.23	&	32.45	\pm 	10.94	&	0.38	\pm 	0.19 & 0.08& -38$\%$ \\
\hline
\end{tabular}
\caption{The best fit parameters of the 2D and 3D correlations with the smoothly BPL model. The first vertical half of the table (left) shows the GRB classes, the normalization, the slope with their respective errors and the intrinsic scatter $\sigma_{int}^{2}$ for each class for the 2D correlation uncorrected for selection biases. The second vertical half of the table (right) show the same variables, but corrected for selection biases (denoted by "$\prime$").
The final (13th) column shows the fractional change in $\sigma_\text{int}^2$ among the classes after correcting for redshift evolution and selection biases}.
In the 2D subtable, the 6th and 7th (11th and 12th) columns show the z-score and Chow test $p$-value, representing two statistical tests comparing sub-classes with the Gold fundamental plane without (with) the correction for redshift evolution. Z-scores could not be computed for the 3D case due to the lack of a sufficiently large Gold sample.
\label{table:corr_params_smoothBPL}
\end{sidewaystable*}

\subsection{The 2D optical relation with the Willingale (2007) model: $L_{\rm opt}^{(\prime)}$-$T_{\rm opt}^{(*,\prime)}$}

\begin{figure*}[!t]
\centering
\includegraphics[width=0.32\textwidth,angle=0,clip]{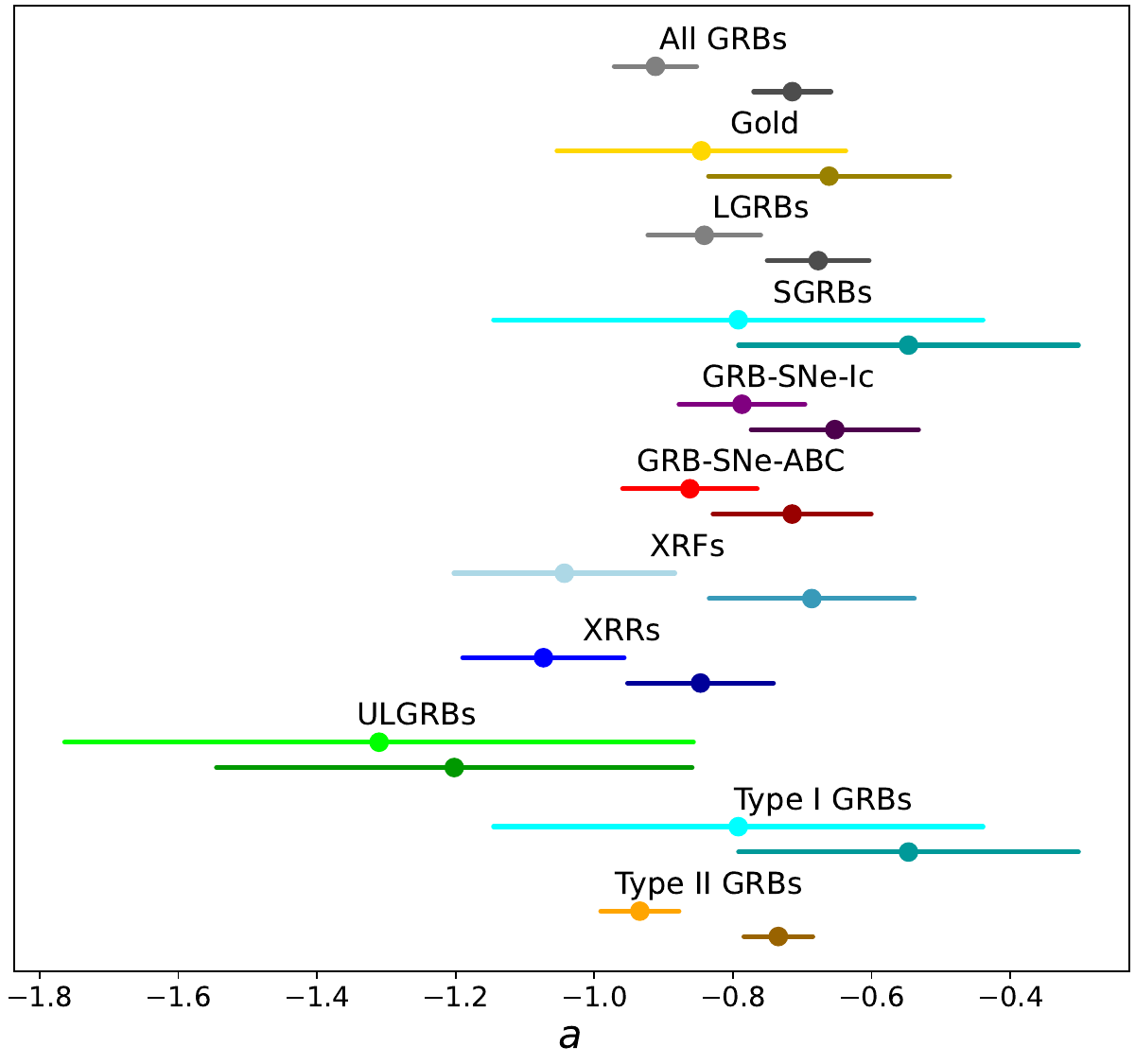}
\includegraphics[width=0.32\textwidth,angle=0,clip]{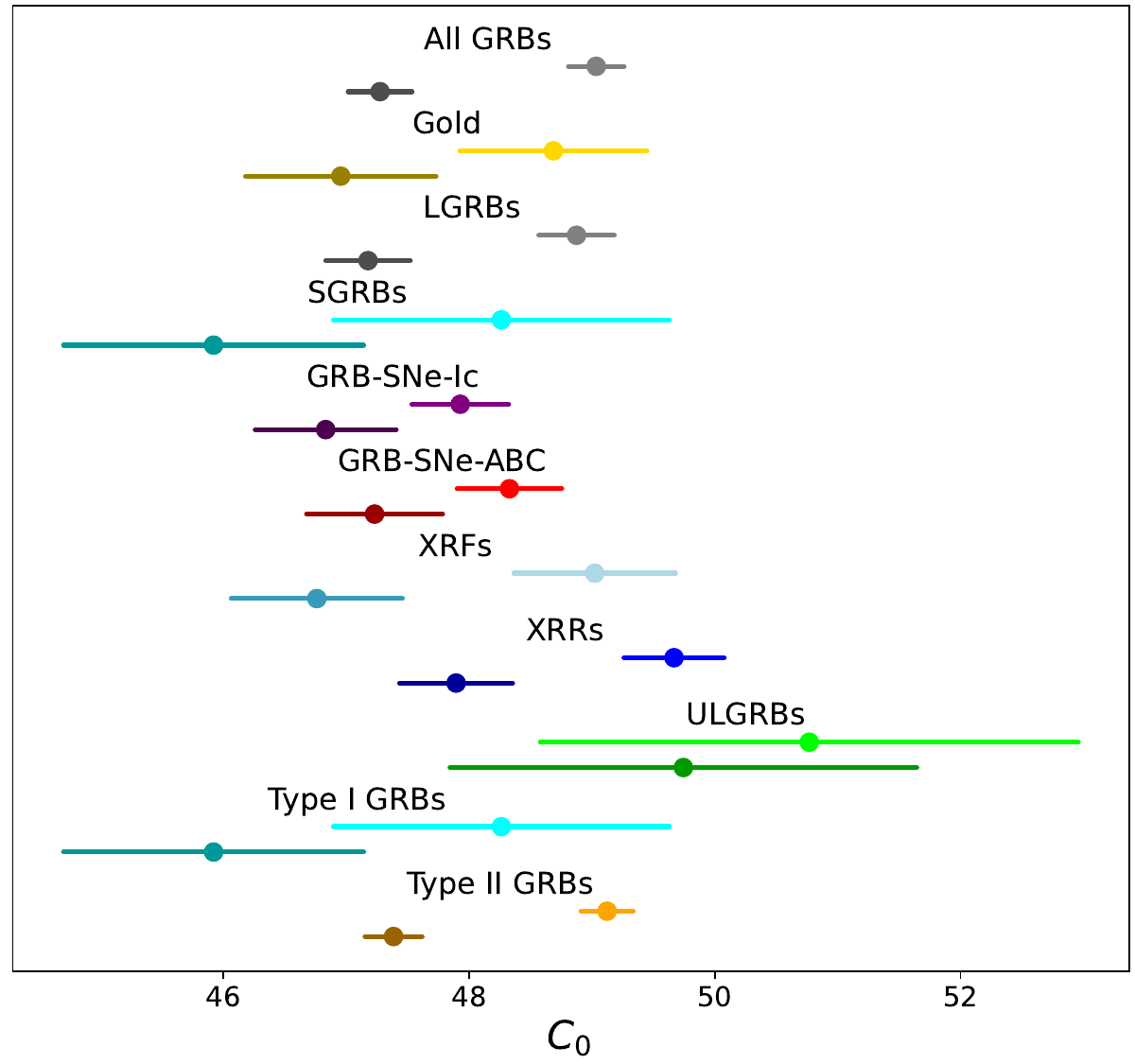}
\includegraphics[width=0.32\textwidth,angle=0,clip]{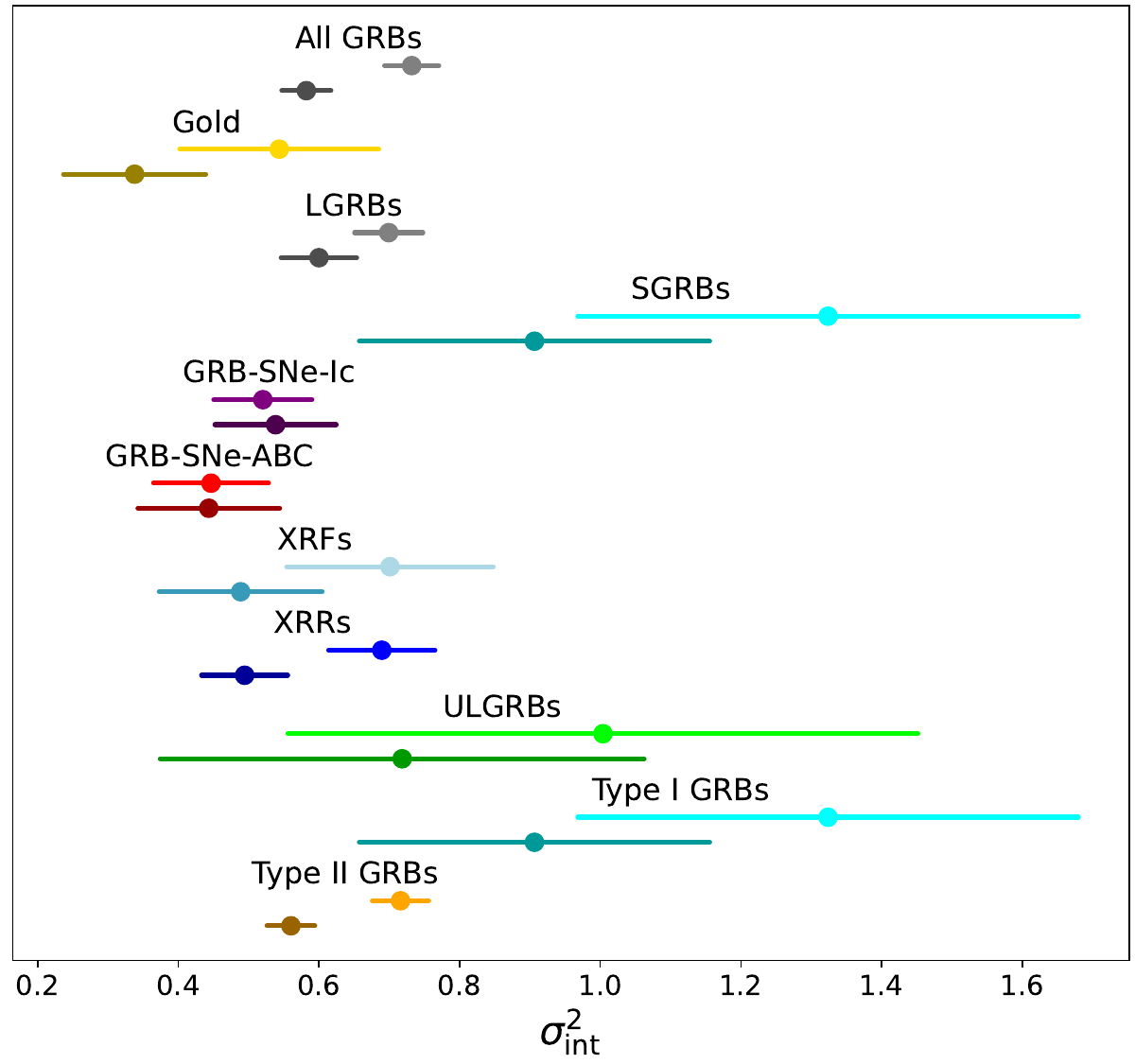}
\caption{An overview of 2D fundamental plane best fit parameters, where $a$ and $C_0$ are color coded according to the GRB class, with darker colors denoting correction. The x-axis displays the $1\sigma$ parameter intervals. The color-coding is as follows: grey for LGRBs, gold for the Gold sample, light blue for XRFs, blue for XRR, bright green for UL, purple for SNe-Ic, red for SNe-ABC, and cyan for both short and Type I (since both samples are identical) and Type II is orange. Exact values for each parameter can be found in Table \ref{table:corr_paramsWL}.}
\label{fig:2Dquickres}
\end{figure*}

\begin{figure*}[!t]
\centering
\includegraphics[width=0.49\textwidth,angle=0,clip]{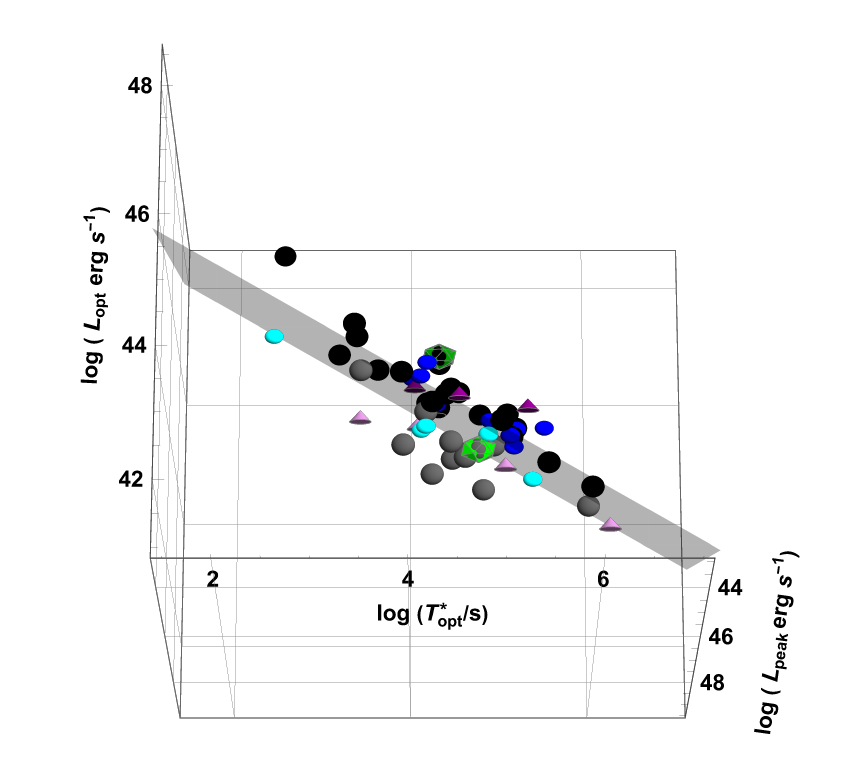}
\includegraphics[width=0.49\textwidth,angle=0,clip]{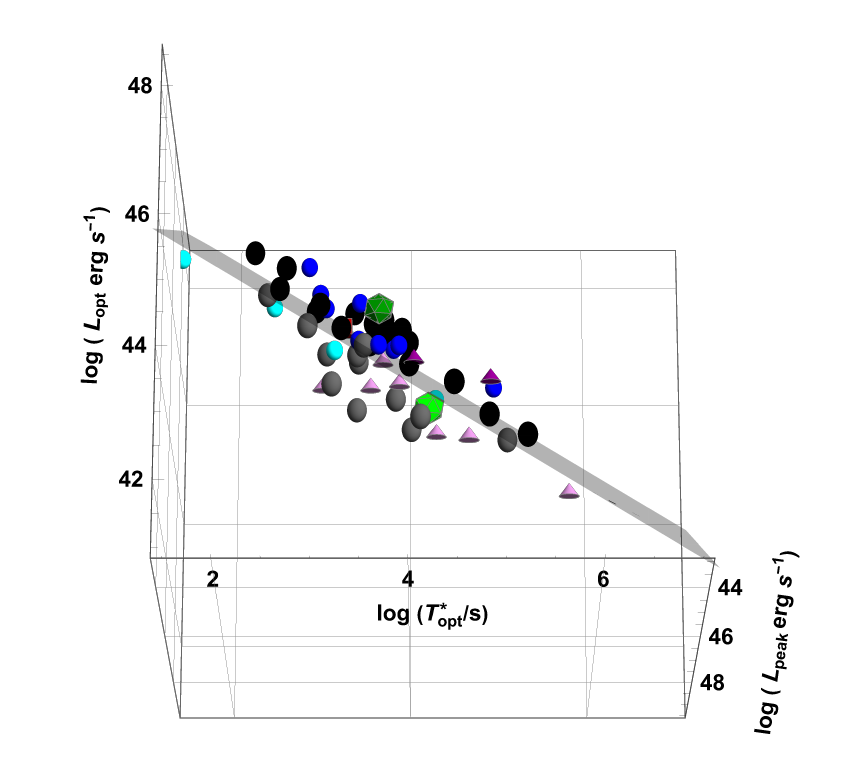}
\includegraphics[width=0.48\textwidth]{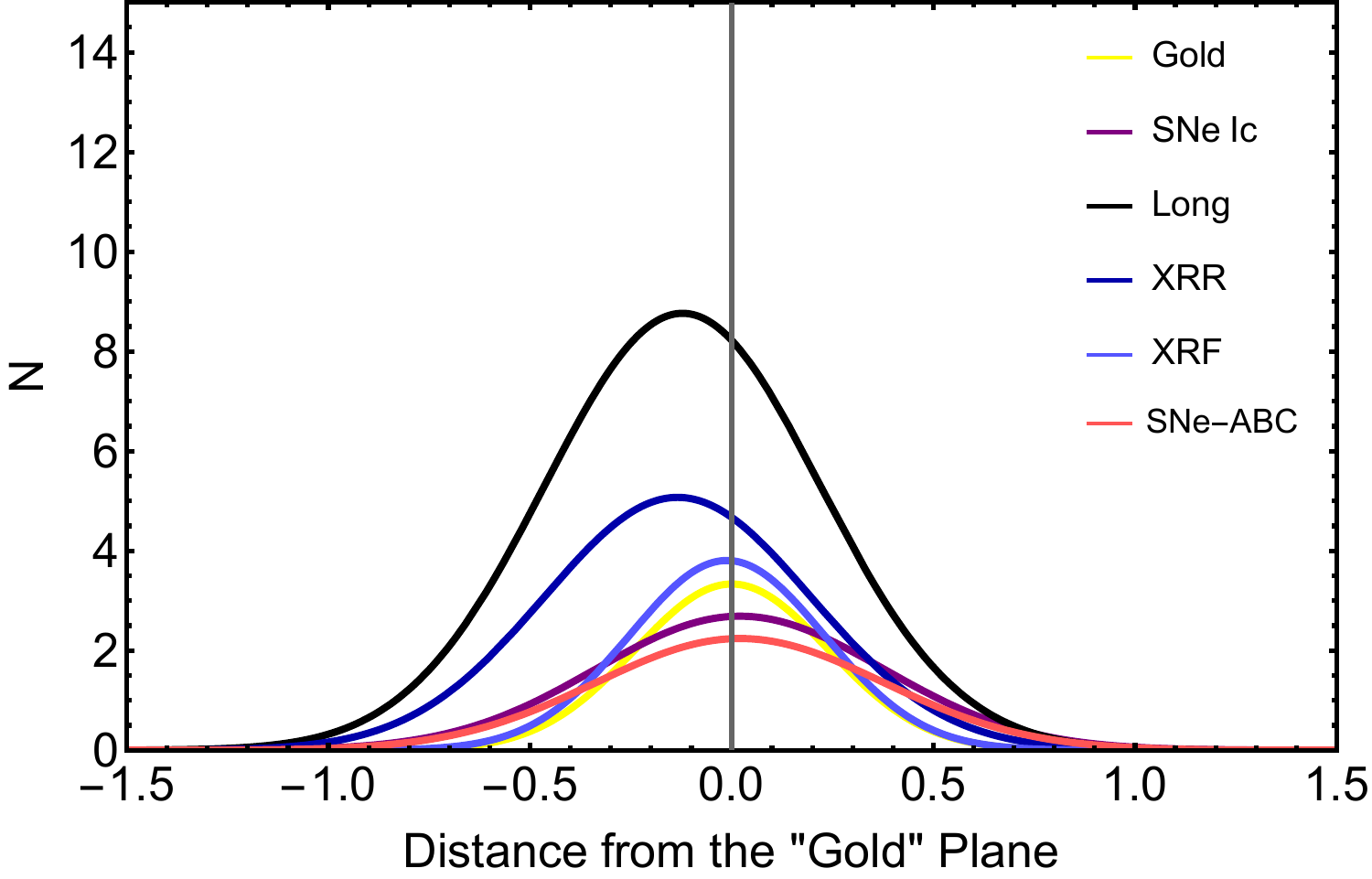}
\includegraphics[width=0.48\textwidth]{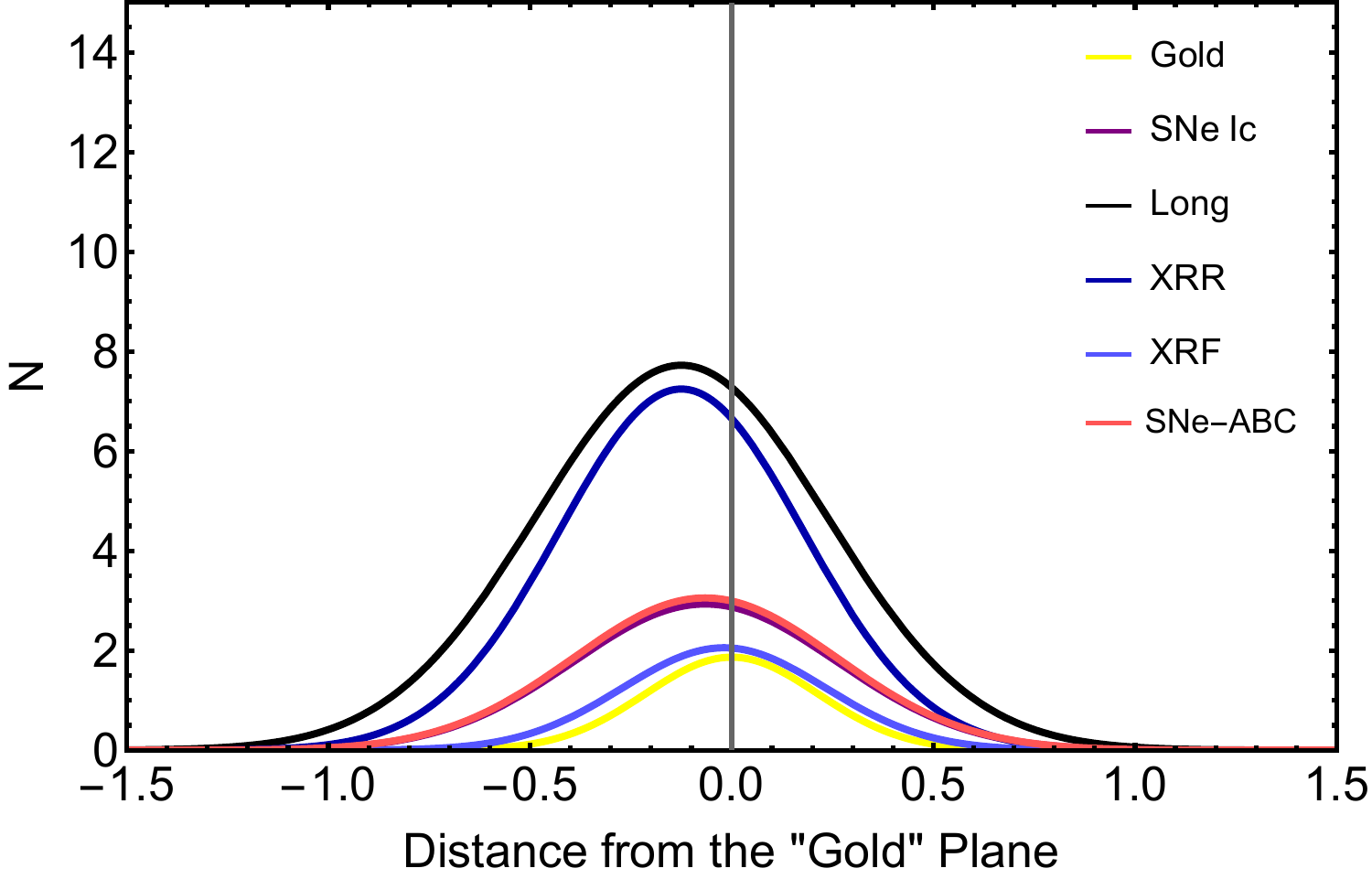}
\caption{Upper panels: 58 GRBs in the $L^{(\prime)}_\text{opt}-T_\text{opt}^{(*/\prime)}-L^{(\prime)}_\text{opt}$ parameter space with the fitted plane parameters in Table \ref{table:corr_paramsWL}, including LGRBs (black circles), SGRBs (red cuboids), GRB-SNe Ic (purple cones), XRFs and XRRs (blue spheres), and ULGRBs (green icosahedrons). The left and right panels display the 3D correlation with and without any correction for both redshift evolution and selection biases, respectively. Lower panels: the distances of the GRB of each class indicated with different colors from the Gold fundamental plane, which is taken as a reference, with and without correction for redshift evolution and selection biases, respectively.}
\label{fig:3D}
\end{figure*}

From Fig. \ref{fig:2Dquickres} and Table \ref{table:corr_paramsWL}, we see that the slope of the 2D correlation uncorrected and corrected for selection biases is compatible within $1\sigma$ for all classes with the exception of a few cases. For the uncorrected case XRR and GRB SNe-Ic are compatible within $2\sigma$.
For the corrected case the UL are compatible with the XRFs, SNe-Ic, SNe-ABC, SGRBs and the Gold within $2\sigma$.
The normalization constants ($C_0$) are all compatible within $1\sigma$ with the exception of a few cases. 
In the case of the uncorrected $C_0$ SNe-Ic are compatible within 2 sigma with XRRs, XRFs, LGRBs and UL. For the corrected case the UL are compatible within 2 $\sigma$ with SGRBs, Gold, SNe-Ic, SNe-ABC, LGRBs and XRFs. Additionally, SGRBs are compatible within $2\sigma$ with XRRs.

We assume that all classes undergo a similar redshift evolution for simplicity because, besides LGRBs, all other classes have too few GRBs to be considered alone to reliably apply the \cite{Efron1992} method. After the correction is applied, all slopes between classes except the ULGRB subsample are compatible within $1\sigma$. The ULGRB sample is compatible with the XRR subsample within $1\sigma$, and $2\sigma$ for all other subsamples. The normalization  constants are compatible within $1\sigma$ except for the ULGRBs, which are compatible with the XRRs within $1\sigma$ and with all other classes within $2\sigma$.

% ========================== 3D FIGURES HERE

%%%%%%%%%%%%%%%%%%%%%% ADD THE 2D LT correlation
\begin{figure*}[!t]
\centering
\includegraphics[width=0.46\textwidth,angle=0,clip]{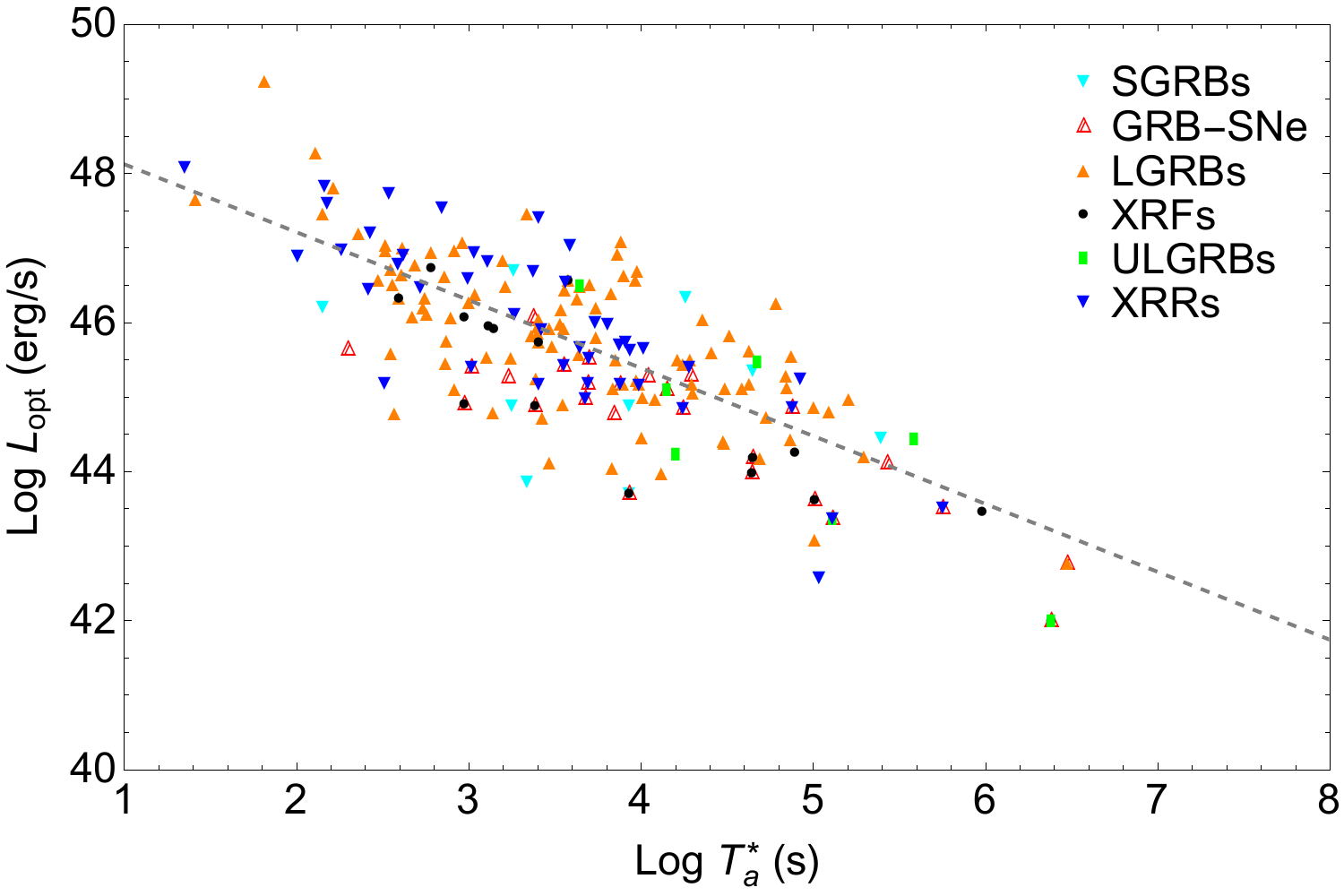}
\includegraphics[width=0.46\textwidth,angle=0,clip]{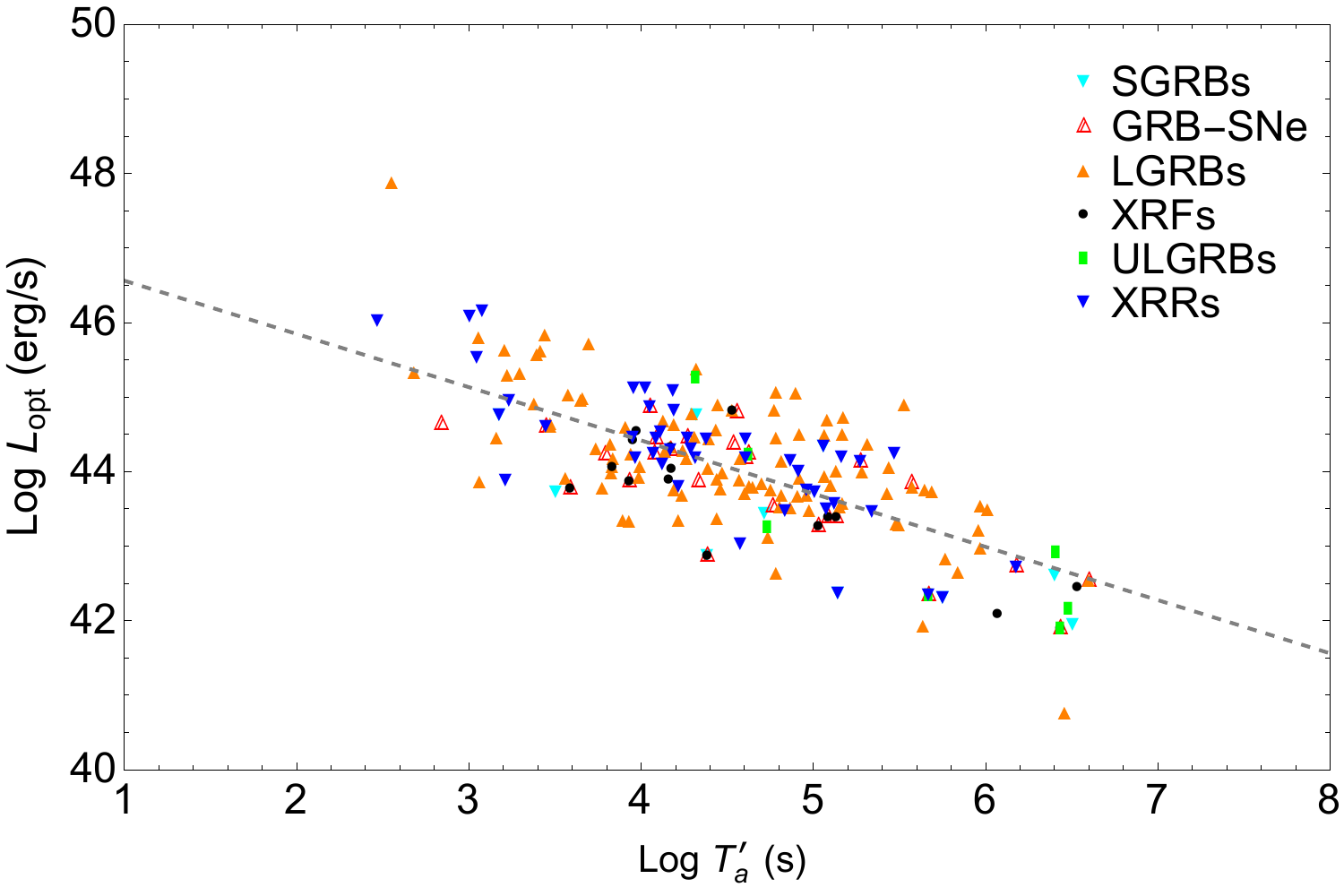}
\includegraphics[width=0.46\textwidth,angle=0,clip]{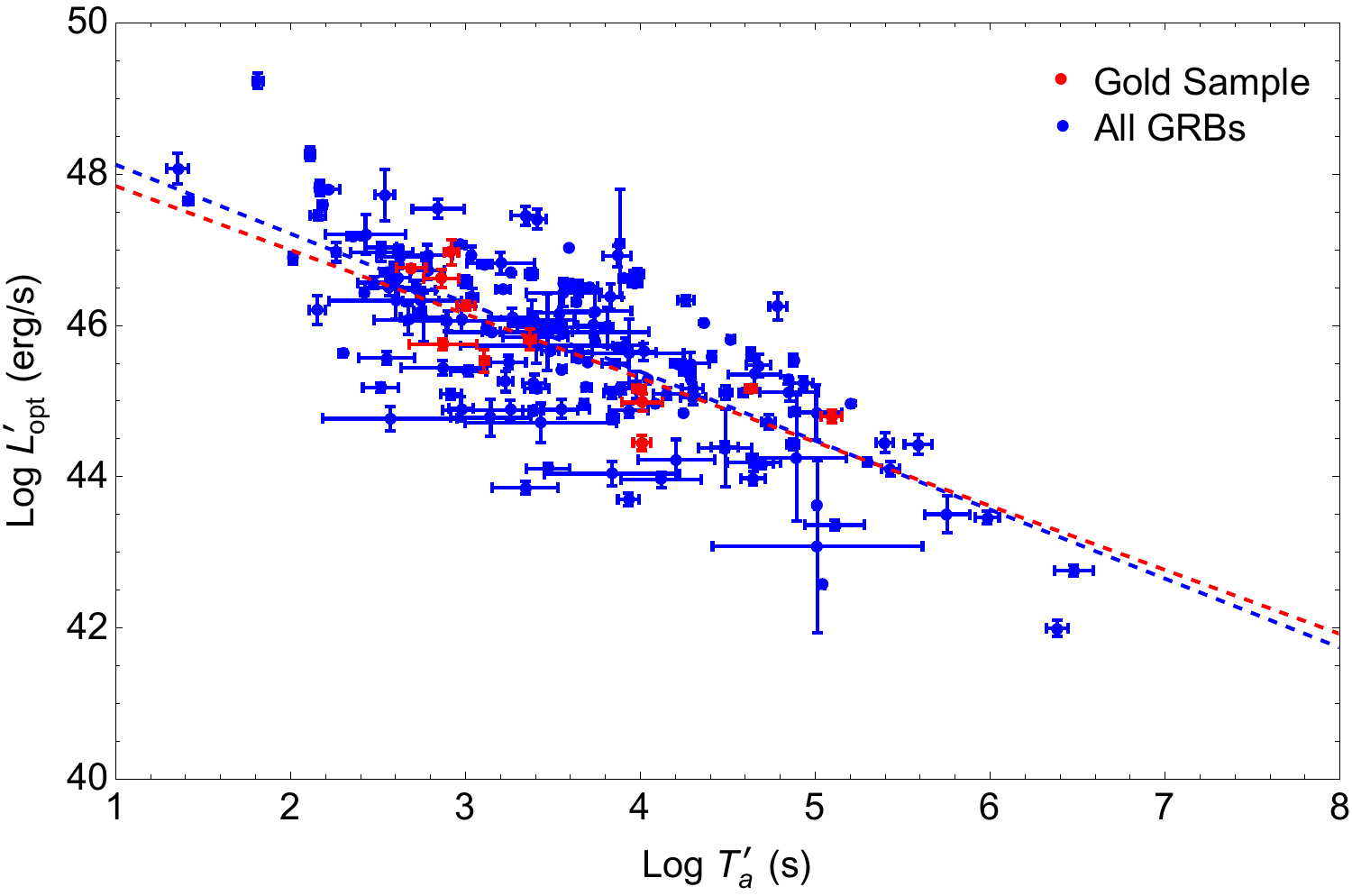}
\includegraphics[width=0.46\textwidth,angle=0,clip]{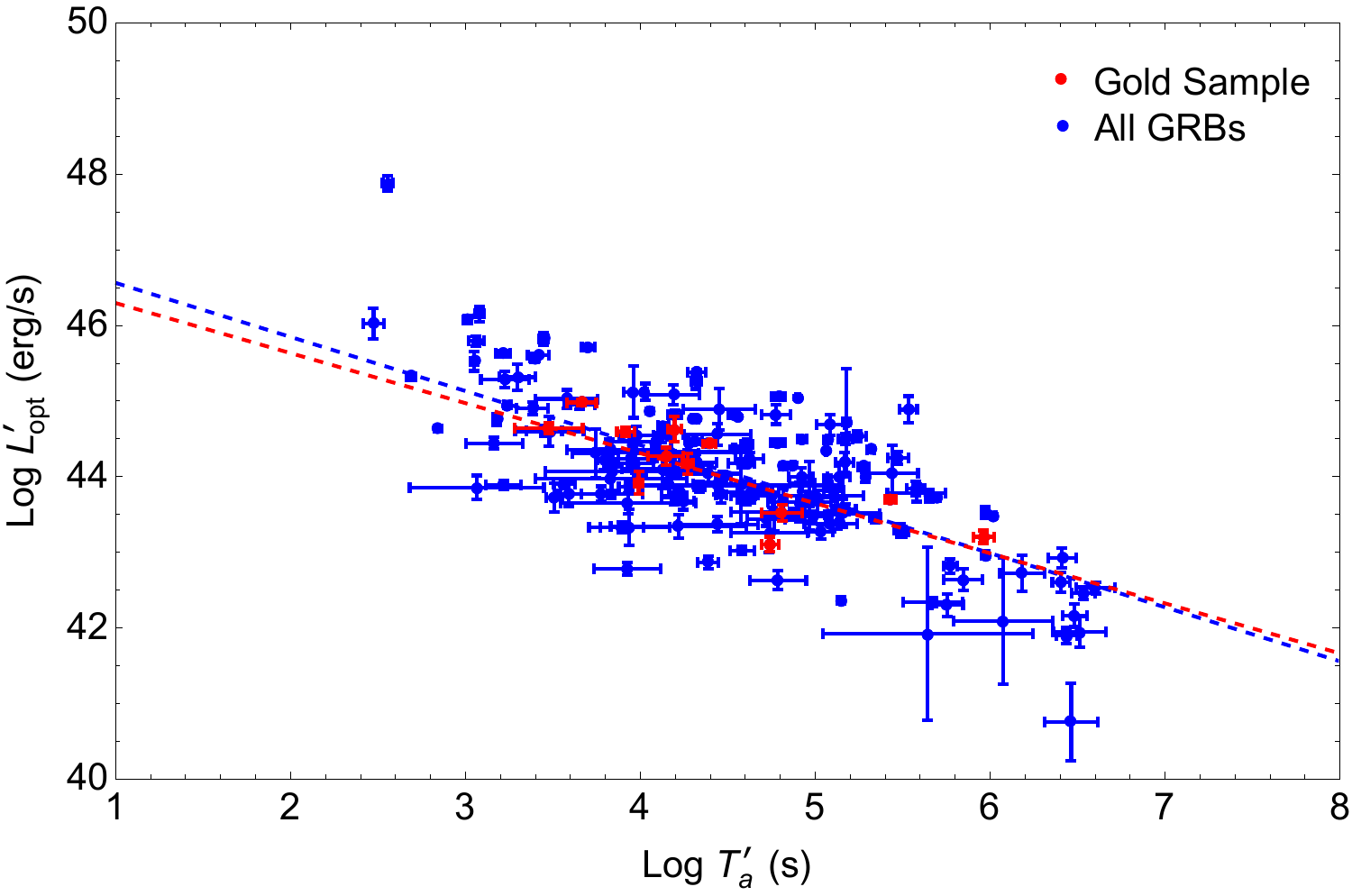}
\includegraphics[width=0.46\textwidth,angle=0,clip]{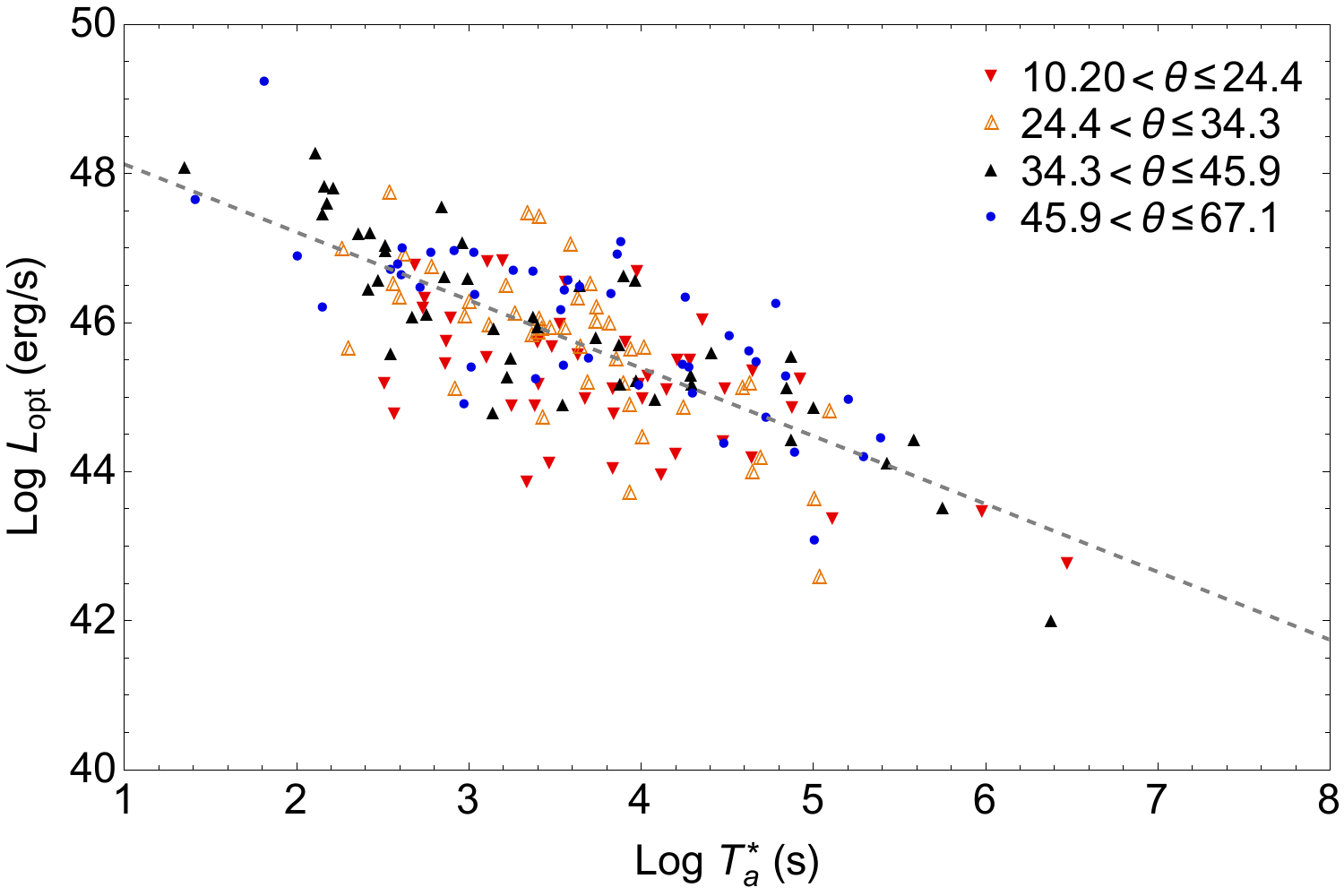}
\includegraphics[width=0.46\textwidth,angle=0,clip]{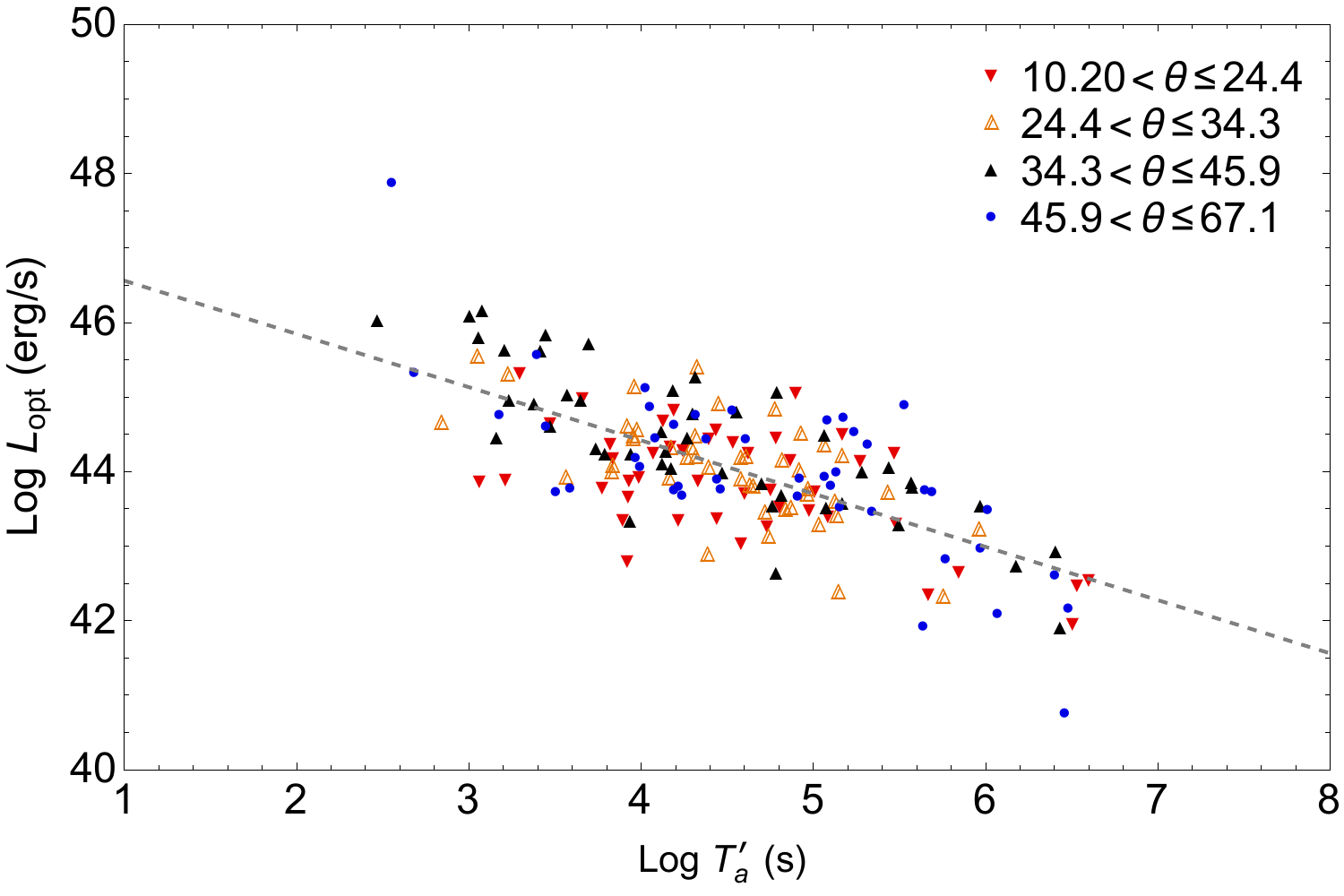}
\includegraphics[width=0.46\textwidth,angle=0,clip]{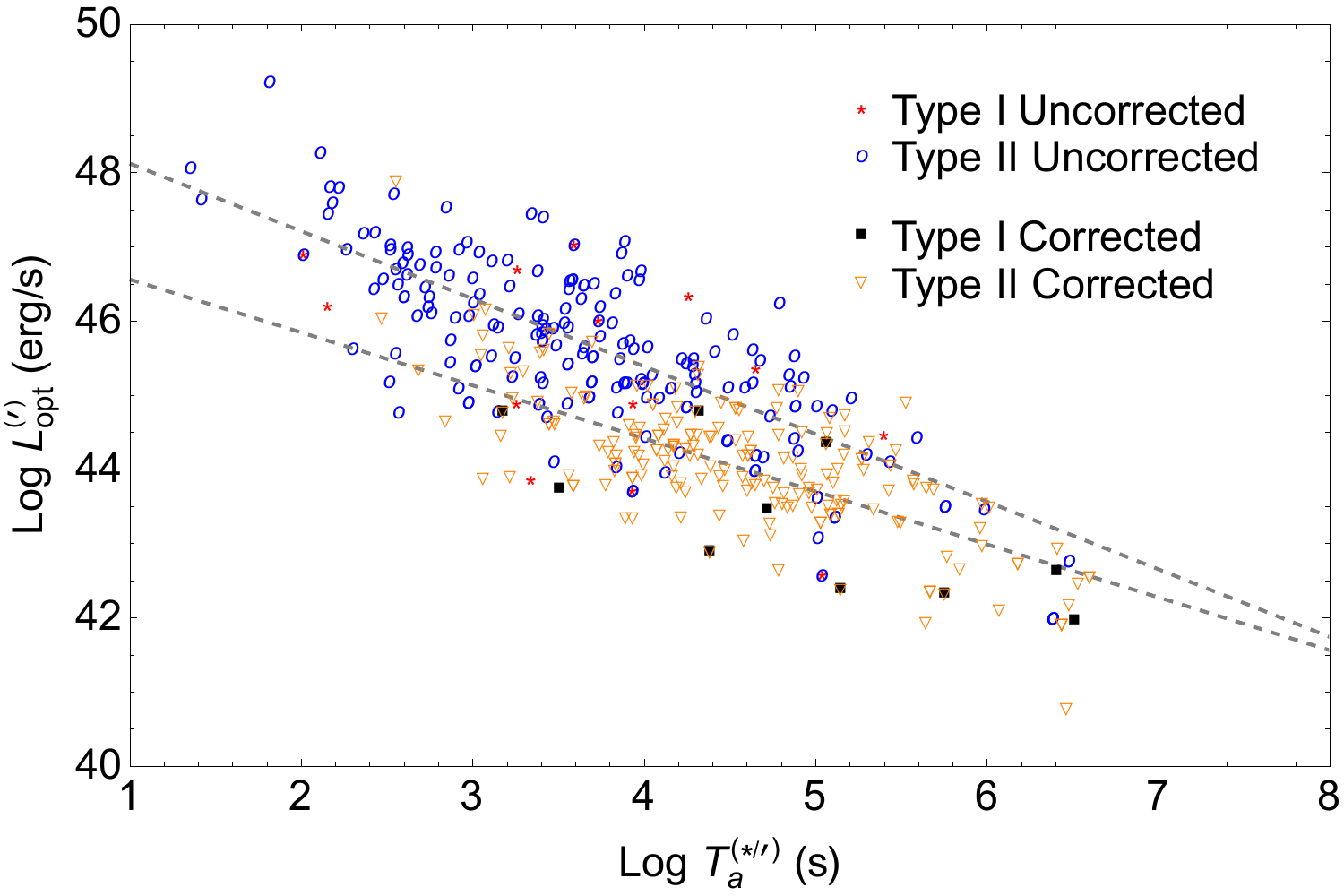}
\includegraphics[width=0.46\textwidth,angle=0,clip]{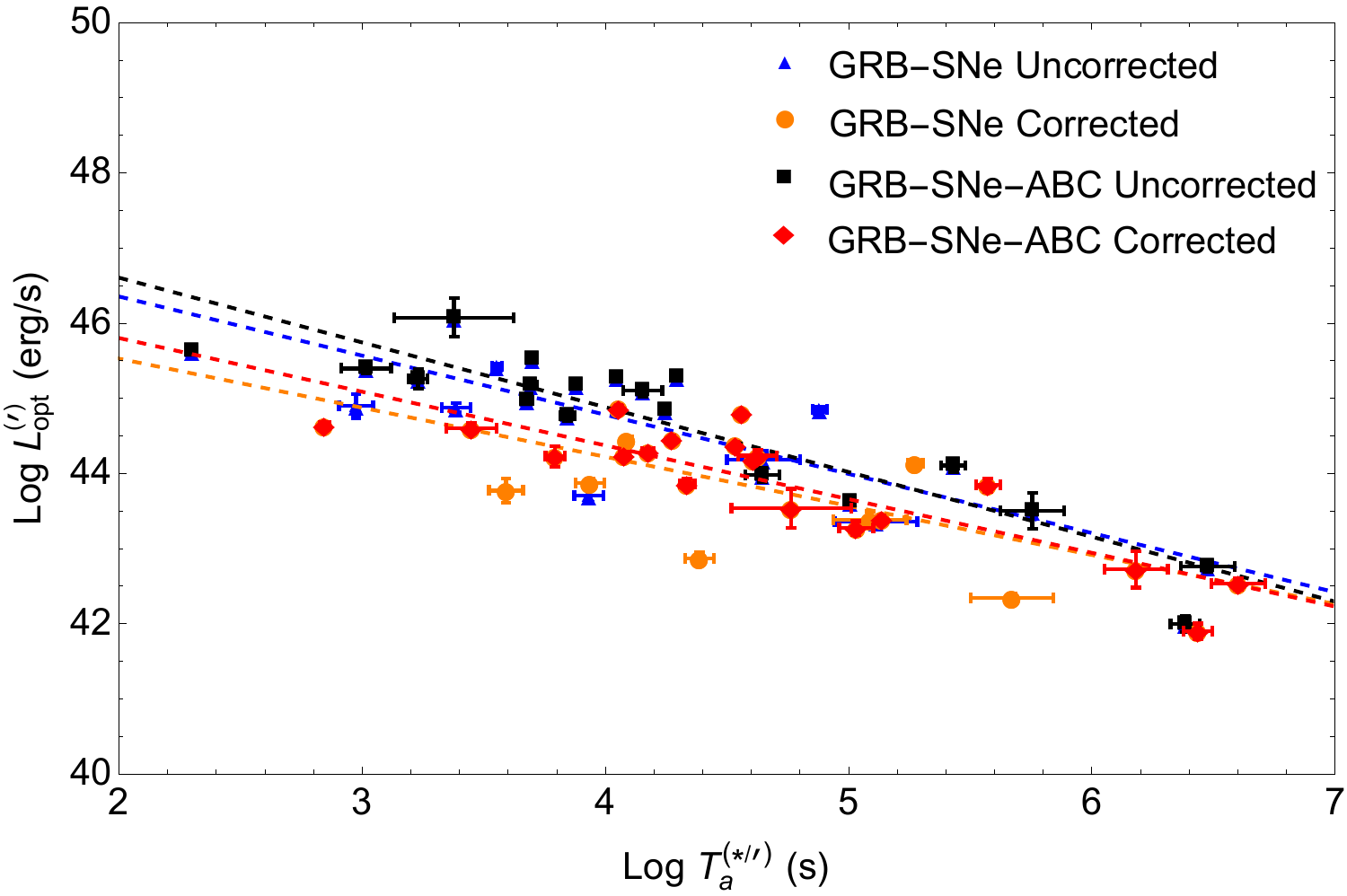}
\caption{The three upper left panels show the observed $L^{(\prime)}_\text{opt}$-$T^{(*/\prime)}_\text{opt}$ correlations uncorrected for biases and evolution with all classes (first row), the total and Gold samples (second row), and plateau angles (third row), respectively. Panels on the left for the first three rows show the correlation without any correction, on the right with correction. The bottom left panel shows Type I and Type II GRBs. The bottom right shows the GRB-SNe-Ic and GRB-SNe-ABC classes before and after correction.}
\label{fig:LT}
\end{figure*}
 
This highlights an important finding: the GRB classes are not distinguishable in terms of the parameters of the 2D optical correlation both with and without correction for selection biases and redshift evolution. The fact that the classes are indistinguishable is checked with the z-score. This trend is different in the 2D and 3D X-ray correlation, where these classes are distinguishable \citep{dainotti2010,dainotti2017a,dainotti2017b,dainotti2020arXiv}.

\subsection{The 2D and 3D correlation with the BPL functions}

Here, we repeat the fitting of the selected 179 GRBs with the BPL with the requirement that the angle of the plateau should be less than $\ang{41}$, corresponding to $|\alpha_1|<0.8$, and found that 99 GRBs can be appropriately fitted with a simple BPL (Eq. \ref{eq:simpleBPL}) while 45 can be fitted with a smoothly BPL (Eq. \ref{eq:smoothBPL}).
The results of the fitting the simple and smoothly BPL models in 2D and in 3D are shown in Table \ref{table:corr_params_simpleBPL} and Table \ref{table:corr_params_smoothBPL}.

Checking the agreement of the parameters for the simple BPL correlation in 2D, we find the slope of the uncorrected correlation $a_{opt}$ of the Gold sample agrees with all other classes within $1\sigma$ except for the GRB-SNe-Ic and GRB-SNe-ABC which agree within $2\sigma$. For the corrected correlation, the slope of the Gold sample is compatible with SGRBs, XRF, and XRR within $1\sigma$ - all other classes (LGRBs, GRB-SNe-Ic, GRB-SNe-ABC) are compatible within $2\sigma$.
The normalization constant $C_0$ of the uncorrected correlation for the Gold sample is compatible with all classes within $1\sigma$ except for the GRB-SNe-Ic, and GRB-SNe-ABC for which it is compatible within $2\sigma$. For the corrected correlation, the $C_0$ of the Gold sample is compatible with all classes within $2\sigma$.

For the 3D simple BPL correlation, we instead compare the classes to the total sample, as there are too few data points to define a Gold sample. The slope $a_{opt}$, $b_{opt}$ parameter of both the corrected and uncorrected correlations are compatible with all classes within $1\sigma$. The $C_0$ parameter is similarly compatible with all classes in all cases except for the corrected LGRBs, which are compatible with the corrected total sample within $2\sigma$.

For the smoothly BPL in 2D, the slope and normalization parameter of the uncorrected correlation in the Gold sample are compatible with all classes within $1\sigma$ except the LGRBs and XRR which are compatible within $2\sigma$. The slope and $C_0$ of the corrected Gold sample are compatible with all other classes within $1\sigma$.

The 3D correlation sample only contains 8 GRBs, which are all classified as type II GRBs, so we do not conduct the same analysis.

\subsection{The distance from the Gold fundamental plane}

Following \cite{dainotti2021b}, we test if the fundamental plane can be a discriminant between classes within the W07 samples. In each 2D and 3D sample, we compute the distance of any GRB class from the plane identified by the Gold sample for both corrected and uncorrected GRBs via calculating a z-score, which is defined as 

\begin{equation}
\label{eq:zscore}
z = \frac{\mu-\mu_{\rm Gold}}{\sqrt{(\sigma/\sqrt{N})^2 + (\sigma_{\rm Gold}/\sqrt{N_{\rm Gold}})^2}},
\end{equation}

\noindent where $\mu$ denotes the mean, $\sigma$ the standard deviation, and N the number of points in a given sample (see bottom panels of Fig. \ref{fig:3D}). 
We find that in the uncorrected 2D W07 correlation, there is a significant indication ($|z^{(\prime)}|>3$) of differences between classes and the Gold plane with the GRB-SNe-Ic, GRB-SNe-ABC, at $z=5.23$ and $4.34$, respectively. This result is expected because indeed in X-rays, the 2D correlation \citep{dainotti2017a} for the GRB-SNe-ABC subsample of 7 GRBs has a slope of -1.9 and a slope of -1.5 with 19 GRBs associated with all GRB-SNe-Ic types. The sample size in the current investigation for GRB-SNe-Ic is 26 (27\% increase) and 19 (92\% increase) for the GRB-SNe-ABC. This result then shows that with a much larger sample that the GRB-SNe-Ic may behave differently from the regular LGRBs in terms of the 2D correlation and the energy reservoir may not be constant any longer for these classes. However, this result becomes insignificant even at the $2\sigma$ level, so it may possible that this is due to redshift evolution. 

Interestingly, after correcting the sample for redshift evolution and selection biases, the magnitude of the $z$ scores are reduced, but still indicate some difference ($z\approx 2-3$) between the Gold sample and the type I/SGRB and XRF subsamples.  Exact z-values can be found in Table \ref{table:corr_paramsWL}. The uncorrected 2D z-scores range from -2.39 to 5.23 for the W07, from 1.18 to 4.31 for the simple BPL, and from -0.52 to 5.49 for the smoothly BPL. For the corrected 2D correlation, the z-scores range from -1.75 to 2.17 for the W07, from 2.15 to 4.80 for the simple BPL, and from -0.29 to 4.70 for the smoothly BPL.

Here, we notice that the range of variations of the z-scores for the 2D uncorrected sample are largest for the W07, while the variation of the z-scores for the corrected 2D sample is largest for the smoothly BPL. The simple BPL carries the minimum change of variation in both the corrected and uncorrected case.

The uncorrected 3D z-scores range from -1.04 to 0.19 for W07, while the corrected 3D z-scores range from -0.94 0.10. We find no significant indication of differences between classes from the Gold plane (see Fig. \ref{fig:3D}).

This analysis would benefit from an increased sample, as z-scores are being penalized for large $\sigma$s and small sample sizes. With a larger sample, we may be able to use the 3D optical fundamental plane as a class discriminator, similarly to the X-ray fundamental plane.

We also test if this total sample of 179 GRBs still holds the same features in terms of each class (upper panels of Fig. \ref{fig:LT}), the presence of the Gold sample (second row), groupings of plateau angles (third row of panels). The left panels of these three rows show the correlations without correction for selection biases and redshift evolution, while the right panels show the correlations after correction. We also investigate if there is clustering in terms of type I and II GRBs in the bottom left panel and in the case of GRB-SNe Ic and GRB-SNe-ABC (bottom right panel). Similar to \cite{dainotti2021b}, there is no particular clustering of GRBs around any given class, plateau angle, nor within the Gold sample. This occurs within the total sample both with and without the correction for redshift evolution and selection biases.
These GRB features are less distinguishable compared to those in X-rays \citep{dainotti2021b}.

\subsection{The Anderson-Darling, Anderson, and Chow tests}
We also test if the functional relationship is valid with the Anderson-Darling test by checking the null hypothesis that $L_{a,opt}$ is drawn by the same population as $L_{opt,theor}$ distribution as determined by the 2D and 3D correlations of each total sample.
In 3D, if we consider the W07 function and the BPL functions for both the corrected and uncorrected correlation, we find that for all functions, the null hypothesis is heavily favored with $p \geq 0.25$ for both the uncorrected and corrected correlations.

In 2D correlations, we find that the uncorrected correlation in the simple BPL fitting again favors the null hypothesis at $p = 0.25$. In contrast, the corrected correlation accepts the null hypothesis at 12$\%$. The smoothly BPL fitting favors the null hypothesis at 25\% for both the corrected and uncorrected correlation. When we apply the test W07 function, we obtain that the null hypothesis is favored at the 9\% level for the uncorrected sample and the corrected one.
 
We also applied the Anderson test for normality (or Gaussianity)\footnote{The Anderson-Darling and Anderson tests have been performed with the \texttt{scipy} Python package.} of best-fit residuals ($L_{opt,observed}- L_{opt,theor}$). For both the corrected and uncorrected 3D correlation in all models,and  regardless of the fitting procedure, the null hypothesis (that the distribution is drawn from a Gaussian) is accepted at $p > 15\%$, with the exception of the corrected W07 sample, which is still accepted at a $p$-value of $0.12$. 
%^ still true for both BPL ; changed W07
For the 2D correlations, both the smoothly and simple BPL, as well as the corrected W07 correlation are accepted at the $p > 15\%$ level.
%^ still true for both BPL & W07
The uncorrected W07 fitting is accepted  at the $p > 10\%$ level.
Thus, we can safely state that in all cases, the 2D and 3D correlations do fulfill both the test of Gaussianity and the Anderson-Darling test for verifying the true nature of the correlation.

Although we have presented the values of the z-score and have shown the compatibility among all classes in the 2D relations, it may be valuable to consider additional statistical tests so that we can draw more reliable conclusions. 
We stress that parameter confidence intervals calculations are based on asymptotic theory. For discussion of multiple regression with small samples, see discussion and references in \cite{Kelly2003}. We assume the regression models in Equation \ref{eq:w07}, \ref{eq:simpleBPL}, and \ref{eq:smoothBPL} are correct.

We apply the Chow test \citep{chow_test} to verify whether the true coefficients in two linear regressions between different subsamples and the Gold sample are equal. In parallel with the z-score test, we take the Gold sample as a reference. We compute the Chow test $p$-values for the several classes against the Gold sample both for the cases of the W07 and both BPL models in 2D, and for both the cases with and without correction for evolution and selection biases. Due to the lack of an appreciable Gold sample for the simple and smoothly BPLs, the Chow test was not computed in the 3D case.

We first discuss the 2D relation without the correction for redshift evolution and sample bias. From the Chow test results for the W07 sample with no evolution, we can assess that the Gold and full sample have similar coefficients with the highest probability of 87\%; the Gold sample also has similar coefficients to the SGRBs/type I at 84\% and the type II GRBs at 81\%.
The classes of GRBs associated with GRB-SNe-Ic, GRB-SNe-ABC, and XRFs have coefficients which are the least similar to the Gold sample at 5\%, 16\%, and 22\% respectively. 
The GRB-SNe Ic and GRB-SNe-ABC show a smaller percentage of agreement with the Gold sample, and this is aligned with the idea explained in the introduction that the GRB-SNe may not be following the same trend as the regular GRBs for which the SNe has not been seen. 
The rest of the classes yield similar parameters with $p\geq 60\%$ (see Table \ref{table:corr_paramsWL}). Interestingly and unexpectedly, the SGRBs have the highest probability of the parameters being similar after the total sample. When we consider the simple BPL sample for the case without correction for evolution, the results are, as expected, comparable with the one with W07, though the most similar class to the Gold sample is the XRRs, with a probability 75\%). The Chow $p$-value for all other classes is $<60\%$, the smallest being the GRB-SNe-Ic and GRB-SNe-ABC subsamples with $p$-values of 0.1\% and 0.3\%, as seen in the W07. The low value of probability for these classes further strengthen the idea that the GRB-SNe population may follow a different trend compared to the regular LGRBs. 

Additionally, in considering the smoothly BPL sample, we see that the $p$-values show the same trend as the W07 and simple BPL samples. In this case the Chow $p$-value for the short GRB sample is the largest (72\%) and it is followed by the type II GRBs (45\%), as in the W07. Nevertheless, in all cases, the GRB-SNe-Ic, GRB-SNe-ABC, and XRF subsamples represent the lowest $p$-values of all tests within the context of samples uncorrected for evolution.

After correcting for evolution in all three samples, $p$-values tend to have less spread among each class.  For the W07 fitting, the full sample still has the highest probability at 92\%, and it is still followed by the type II at 86\% and SGRBs/type I at 67\%. The XRF now has the lowest similarity, with a probability of 18\%, while the other classes range from 42\% to 63\% probability. Considering the simple BPL fitting, the spread of values lessened considerably - the XRR has the greatest similarity to the Gold sample with 24\% probability, while the XRF and SGRB/type I have the lowest similarity at 2\%. The other classes range from 3\% to 4\% probability. For the smoothly BPL, the GRB-SNe-ABC now has the highest probability (95\%), while the XRF has the lowest probability (9\%), as in the W07. The type II are the second highest at 67\% probability, and the other classes range from 26\% to 59\% probability (See Table \ref{table:corr_paramsWL}, \ref{table:corr_params_simpleBPL}, and \ref{table:corr_params_smoothBPL} for exact values.)

%the ones without correction follow the same trend, but with correction really don't, should we fix this
It is interesting to note that although the $p$-values vary from the BPL and the W07 function, the conclusion of the analysis remains the same within the uncorrected fittings and corrected fittings, thus supporting the reliability of these conclusions which are independent from the functional forms used for fitting the LCs.

%%%%%%%%%%%%%%%%%%%%%%%%%%%% ADD THE HISTOGRAMS

\section{Comparison between optical and X-rays}\label{sec:comparison}
\subsection{Discussion of the current status in the literature}
\begin{figure*}[!t]
\centering
\includegraphics[width=0.32\textwidth,angle=0,clip]{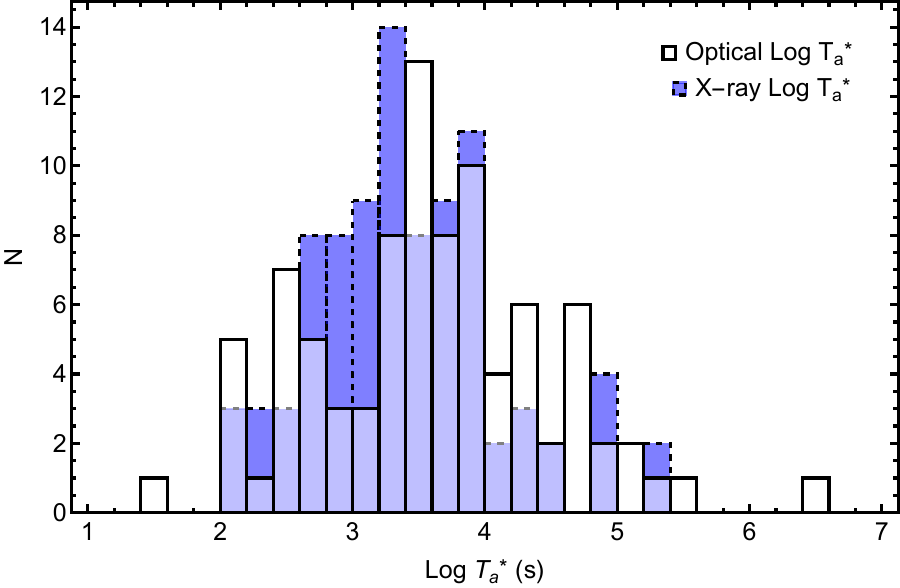}
\includegraphics[width=0.32\textwidth,angle=0,clip]{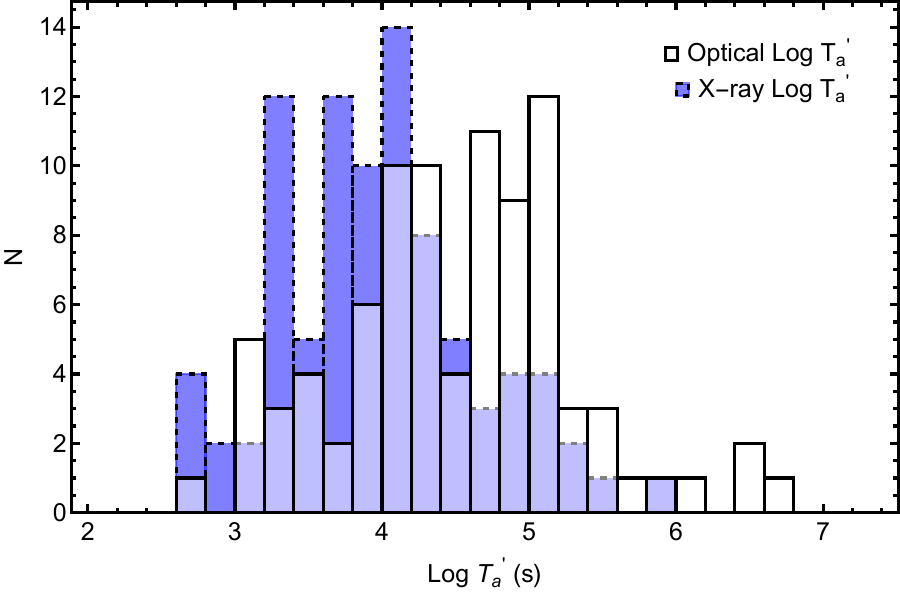}
\includegraphics[width=0.325\textwidth,angle=0,clip]{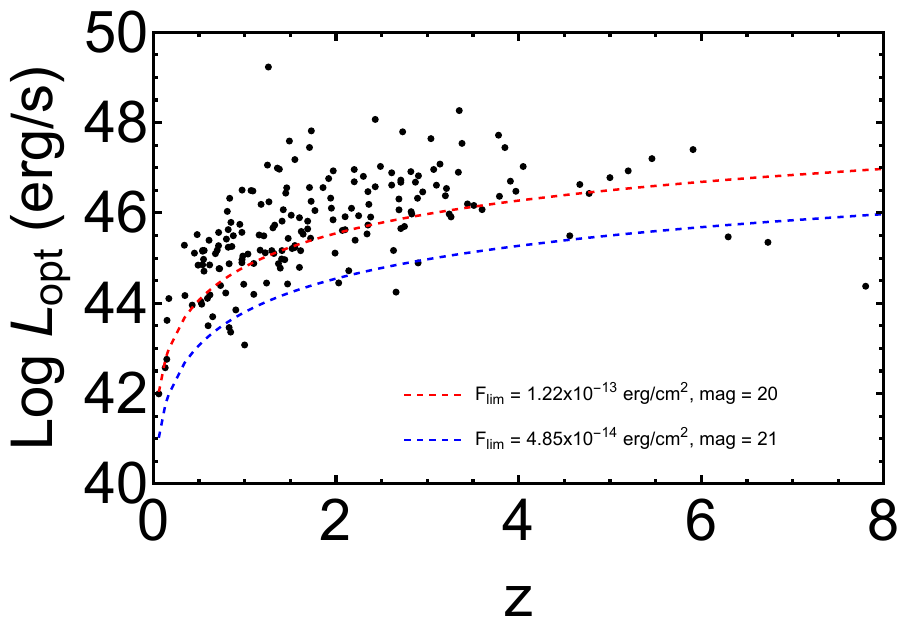}
\caption{Left panel: The differential distribution of $T^{*}_{X,a}$ (white solid line) and $T^{*}_{opt,a}$ (blue dashed line). Middle panel: The same as the left panel, but corrected for evolution. Right panel: The observed luminosity vs. redshift with limiting luminosity curves given by the observability with Tomo-e Gozen.}
\label{fig:XvO}
\end{figure*}

GRB collimation has been inferred with the observations of achromatic steepening in GRB LCs, such steepenings are called jet breaks. Pinpointing a jet break from afterglow LCs enables us to measure the jet opening angle and consequently the GRBs’ energy. Investigating which GRBs are chromatic and achromatic is crucial tackling this issue. 
The topic of the achromaticity vs. chromaticity of the LCs in X-rays and optical has been extensively studied in the literature \citep{Panaitescu2006,Huang2006,Liang2007,2007A&A...469L..13M}. To be more specific in \cite{Panaitescu2006} several GRBs such as 050319, 050401, 050607, 050713A, 050802, and 050922C exhibit a steepening at 1–4 hours in X-rays after the burst, which, surprisingly, is not accompanied by a break in the optical emission. The reason for this behavior is still a puzzling issue. Out of the several GRBs presented in  \cite{Panaitescu2006} we have 4 GRBs which are also common to our sample: 050401, 050319, 050802, and 050922C. Most likely, the behavior of not being accompanied by the spectral break does not originate from the outflow collimation.  If the optical and X-ray observations stem from the same synchrotron forward-shock model, we identify the temporal breaks with the passage of the synchrotron cooling break through the X-ray band. That evolution of the synchrotron cooling break depends on the equivalent kinetic energy, circumburst density,  microphysical parameters, and the electron population’s spectral index. It is worth noting that a temporal break in the optical observations without observing a temporal break in the X-rays could also be explained by the passage of the spectral break in optical bands.
\cite{Liang2007} analyzed the origin of the shallow X-ray phase in a sample of 53 long bursts detected by \emph{Swift}. Among the 13 bursts with well-sampled optical LCs, six had an optical break, $t_b$, consistent with being achromatic. However, the remaining cases either did not show an optical break or had a break at an epoch different from $t_b$. This observational result poses challenges for the synchrotron forward-shock scenario with energy injection, opening up to the possibility that the optical and X-ray emission may not be emitted by the same mechanism, at least for some bursts. There are four significant outliers in the sample, GRBs 060413, 060522, 060607A, and 070110. The last two bursts are also present in our sample.   A very steep decay immediately follows the shallow decay phase exhibited in these X-ray lightcurves after $t_b$, which is inconsistent with any external shock model. 
The optical and X-ray observations show that these bursts evolve independently, indicating  these X-ray plateaus may have an internal origin \citep{troja2007,depasquale2016}.
 
However, a one to one comparison among the GRB lightcurves in X-rays and optical is needed to determine more realistically whether or not we have an achromatic or chromatic plateau. 
 
In \cite{Liang2007} at least some X-ray breaks are chromatic. In the current paper we check consistency of the chromaticity versus achromaticity scenario for 89 GRBs which have both X-ray and optical lightcurves. As a result, we have 10 cases of achromatic emission when we do not correct for selection biases and redshift evolution and 13 cases when we apply these corrections. The achromatic cases, which are a fraction of 11.2\% in case of no evolution and 14.6\% in case of evolution may reflect that these are associated with an external origin (e.g. refreshed shocks).  However, the remaining cases cannot be explained within this scenario. %the number with ev did not change
 
Invoking different emission regions (e.g., \citealt{2002ZhangMeszaros}) may solve the problem, although more detailed modeling is needed. Crossing a cooling break would also result in a temporal break, but it would also be accompanied by a spectral index variation by $\sim 0.5$. \cite{Liang2007} found that the changes in the X-ray spectral indices across the breaks of GRB 050318, 050319, 050802, 050401 are between $0.01$ and $0.12$, and therefore much smaller than $0.5$. Thus, the suggestion of a cooling spectral break is ruled out. \cite{Genet2007} accounted for these chromatic breaks as being due to a long-lived reverse shock in which only a small fraction of the electrons are accelerated. The main issue for such an interpretation is how to ‘‘hide’’ the emission from the forward shock, which carries most of the energy.
 
\cite{2007A&A...469L..13M} analyzed the cases of GRB 060418 and GRB 060607A and concluded regarding GRB 060418 that due to the difference between the spectral indices in X-rays, $\beta_X$, and Near Infrared, $\beta_{NIR}$ a different origin for those wavelengths may occur. For the case of GRB 060607A a definite conclusion cannot be reached due to the presence of flares.
\cite{Wang2015, Wang2018} used a large sample of GRBs that have an optical break consistent with being achromatic in the X-ray band. Their sample includes 99 GRBs from 1997 February to 2015 March that have optical and X-ray LCs for \emph{Swift} GRBs. These X-ray LCs are consistent with the jet break interpretation. Out of these 99 GRBs, 55 GRBs have temporal and spectral behaviors both before and after the break, consistent with the theoretical predictions of the jet break models, respectively. These include 53 long/soft (Type II) and 2 short/hard (Type I) GRBs.

\subsection{The analysis in this paper}
\begin{figure*}[!t]
\includegraphics[width=0.5\textwidth,angle=0,clip]{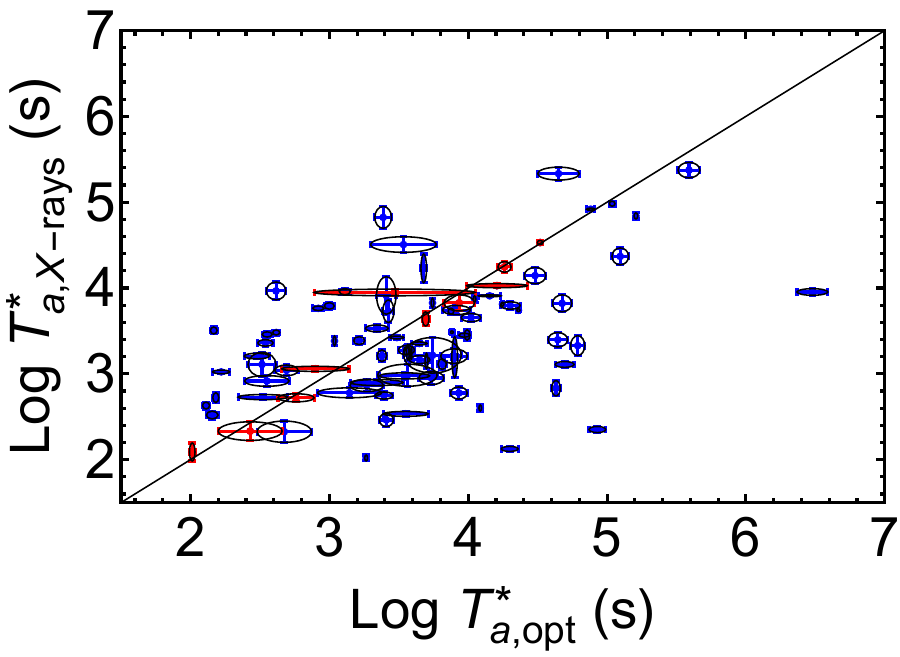}
\includegraphics[width=0.5\textwidth,angle=0,clip]{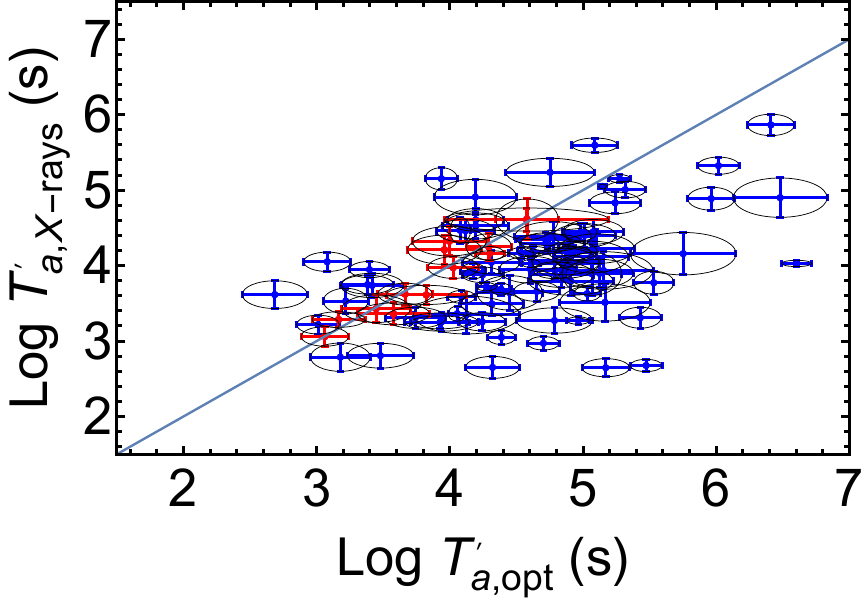}
\includegraphics[width=0.5\textwidth,angle=0,clip]{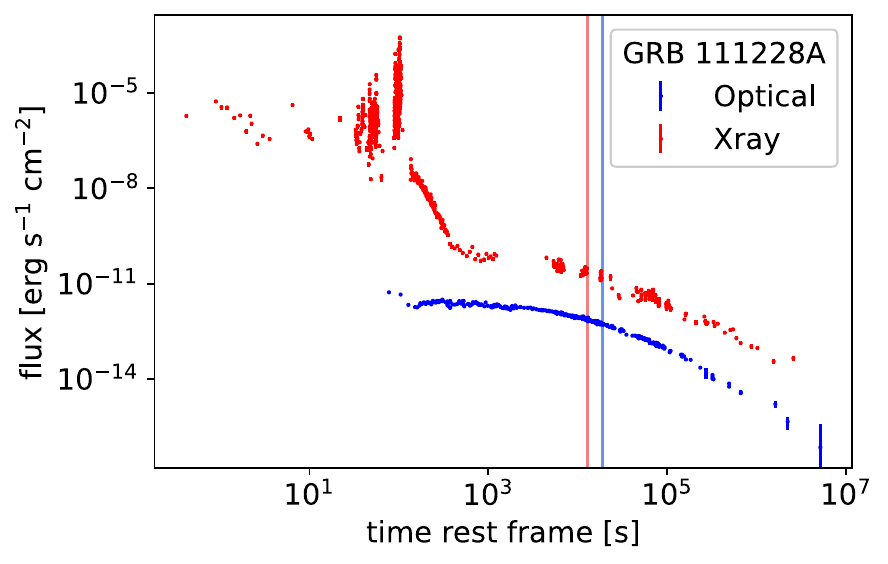}
\includegraphics[width=0.5\textwidth,angle=0,clip]{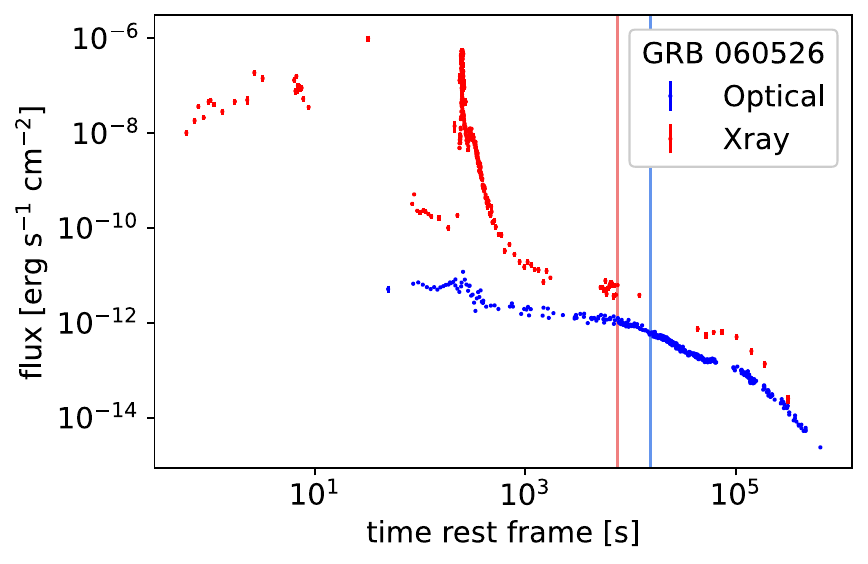}
\includegraphics[width=0.5\textwidth,angle=0,clip]{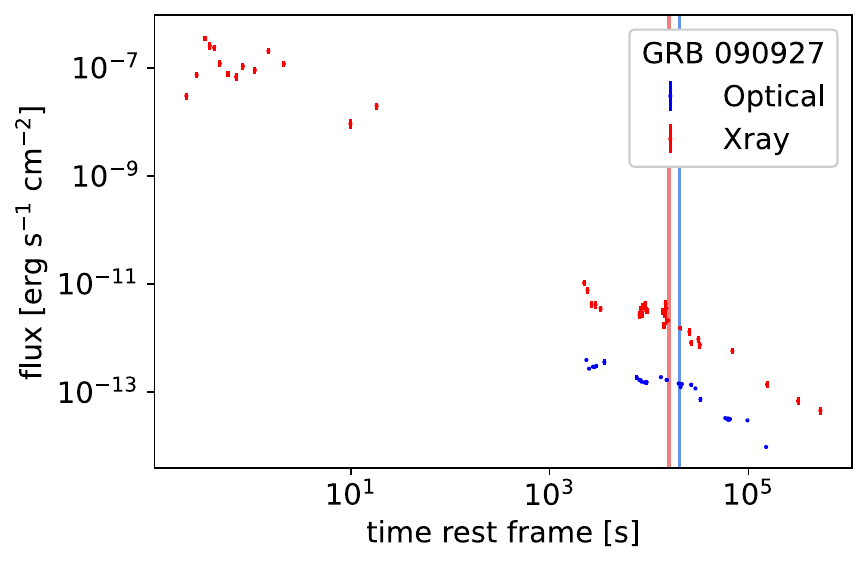}
\includegraphics[width=0.5\textwidth,angle=0,clip]{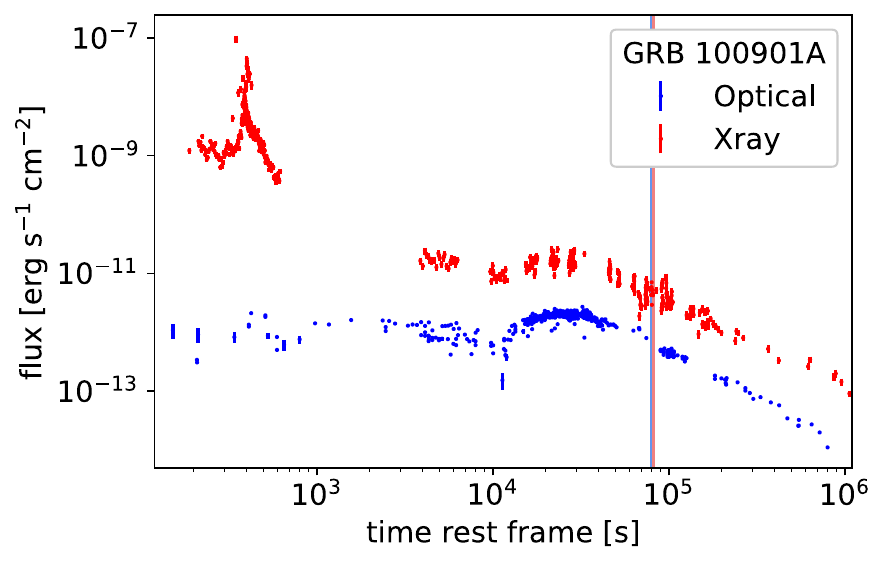}
\caption{Left top panel: plot of $T^*_{a_X}$ vs. $T^*_{a_\text{opt}}$} for uncorrected data. Right top panel: the same plot, but for corrected data. Middle panels: example of GRBs in which the plateau is chromatic in more than $1\sigma$ between the X-ray and optical LCs. Lower panels: example of GRBs in which the plateau is achromatic between the X-ray and optical LCs within $1\sigma$. $T_a^*$ from W07 fitting appears as a vertical line, denoted in bright red for X-ray and bright blue for the optical both in the middle and lower panels. \label{Tacomparison}
\end{figure*}

Because the above discussions are based on a one-to-one comparison, we follow two approaches in this work: the first one is to show the distribution of the sample to study the population as a whole, shown in the left and middle panels of Fig. \ref{fig:XvO}, and the second approach is to perform a one-to-one analysis in which we show the plot of $T_{X}$ vs $T_\text{opt}$ to highlight the coincidence of the breaks in Fig. \ref{Tacomparison}.

Following \cite{dainotti2021b}, we test if the rest-frame end times ($T_a^{(*/\prime)}$) of the plateau are achromatic in X-rays and optical.

The left panel of Fig. \ref{Tacomparison} shows the cases without evolution, while the right-hand panel shows the cases with evolution. We have shown the uncertainties as ellipses because they are not independent. We also show two examples of the LC comparison between X-ray and optical in that figure. We mark $T_a$ as vertical lines, denoted in red for X-ray and blue for the optical.
Specifically, we show two cases in which the plateau seems to show chromaticity (the middle panel of Fig. \ref{Tacomparison}) and two cases of achromaticity (lower panel of Fig. \ref{Tacomparison}) within $1\sigma$. 
We then perform the Kolmogorov-Smirnov (KS) Test for the uncorrected $T^{*}_{X,a}$ and $T^{*}_{opt,a}$ and the corrected $T^{(*/\prime)}_{X,a}$ and $T^{(*/\prime)}_{opt,a}$ distributions. We find that, for times not corrected for evolution or selection biases, the distance among these distributions, $D$, is $D=0.20$ with a probability that they are drawn by the same parent population $P=0.05$.
However, when we correct for selection biases and redshift evolution the null hypothesis is clearly rejected with $D=0.36$ and $P=\num{1.66e-5}$, showing that $T_a^\prime$ is chromatic across X-ray and optical bands, with the observed X-ray breaks happening earlier than those in the optical. This points toward the hypothesis that the end of the plateaus may be associated with an external origin, indicating that the continuous energy injection has finished \citep[e.g., see][]{Li2014}.

\begin{figure*}[!t]
\centering
\includegraphics[width=0.45\textwidth,angle=0,clip]{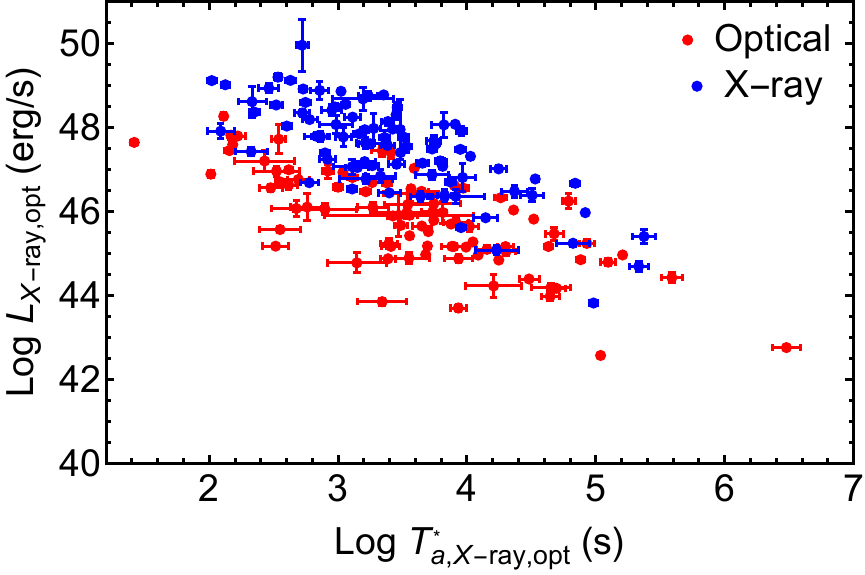}
\includegraphics[width=0.45\textwidth,angle=0,clip]{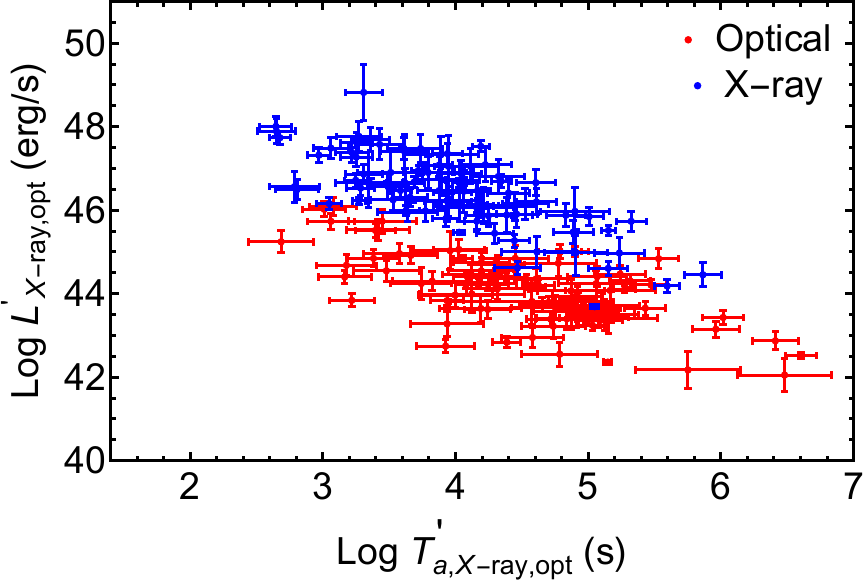}
\caption{Left panel: plot of $L_{X,a}$ and $L_{opt,a}$ vs. $T^{*}_{X,a}$ and $T^{*}_{opt,a}$ in X-ray (blue) and optical (red). Right panel: same plot corrected for evolution.}
\label{fig:LaTaevo}
\end{figure*} 

We compute the 2D correlation in X-rays for 89 GRBs in common between the X-rays and optical sample, see the left-hand  panel of Fig. \ref{fig:LaTaevo}, obtaining $a_{X} = -1.17 \pm 0.10$ and $a_\text{opt} = -0.93 \pm 0.09$. After correcting for observational biases and redshift evolution, we find $a'_X = -0.94 \pm 0.13$ and $a'_\text{opt} = -0.67 \pm 0.08$, see the right-hand panel of Fig. \ref{fig:LaTaevo}. In both the corrected and uncorrected case, the slope agrees within $2\sigma$.

It is clear from this analysis that we need to increase the number of observations to be able to tell whether the uncorrected end-time distributions are chromatic across X-ray and optical wavelengths. Thus, we will soon start an observational campaign for optical GRB follow-up with Tomo-e Gozen, a wide-field CMOS camera mounted on the 1.05-m KISO Schmidt telescope located in Nagano, Japan \citep{Sako2018a}.
We show the limiting luminosities of Tomo-e Gozen at each redshift, assuming we can detect at $20$ mag with an exposure time of 100 s, corresponding to $F_\text{lim} =\num{1.22e-13}$ erg cm$^{-2}$ s$^{-1}$ (red dashed line in Fig. \ref{fig:XvO}) and with a $21$ mag afterglow and an exposure time of 1000 s, corresponding to $F_\text{lim} = \num{4.85e-14}$ erg cm$^{-2}$s$^{-1}$ (blue dashed line in Fig. \ref{fig:XvO}). If we consider our sample with redshift with an exposure of 1000 s or more, we can observe the majority of bursts with plateau emission.
We have also initiated an international collaboration with DDOTI \citep{2016SPIE.9910E..0GW} where we have similar limiting luminosities as the ones from the KISO and the Telescope in Krakow CDK 500 within the Skynet network for joint observations. The limiting magnitude of the telescope in Krakow is 18 mag (corresponding to $F_\text{lim}=1.31 \cdot 10^{-12}$ in the $V$ band) when the condition of observations are favorable. Based on the current sample of 179 GRBs, we estimate that with the CDK 500 we will still be able to catch a good fraction of the high luminosity plateaus ($36\%$). The great advantage of this synergy is that the three telescopes are located in three different parts of the world (Europe, Mexico and Japan) thus allowing a good coverage of the GRB afterglows if they are observable in three locations.

\section{\label{conclusion} Discussion and Conclusions}\label{sec:discussion}

We have gathered the largest compilation of optical plateaus to date (179 GRBs, 76\% larger than the previous sample presented in \citealt{dainotti2020b}) and show that the $L_{\rm opt}^{(\prime)}-T^{(*/\prime)}_{\rm opt}$ correlation holds, and is compatible with the previous X-ray sample within $1\sigma$ both before and after correcting for redshift evolution and selection biases.
We also discover the existence of a 3D optical correlation, an extension of the 2D correlation by adding the peak prompt luminosity, $L_{peak,opt}$.

The optical correlation in 3D fitted with W07 model is:
\begin{equation}
\log L_{\rm opt} = (-0.87 \pm 0.09) \log T^{*}_{opt} + (0.48 \pm 0.07) \log L_{peak} +(26.57 \pm 3.44)
\end{equation}
\noindent with $\sigma_{\rm int}^{2} = 0.44 \pm 0.12$.
For the case of the simple BPL, $a=-0.84 \pm 0.18$, $b=0.40 \pm 0.13$ and $C_0=30.07 \pm 6.33$ and $\sigma=0.52 \pm 0.12$. For the case of the smoothly BPL, $a=-0.97 \pm 0.25$, $b=0.30 \pm 0.16$ and $C_0=35.45 \pm 7.68$ and $\sigma=0.55 \pm 0.25$.
The 3D correlation fitted with W07 after correcting for selection biases and redshift evolution is:
\begin{equation}
\log L '_{\rm opt} = (-0.82 \pm 0.10) \log T '_{opt} + (0.34 \pm 0.08) \log L '_{peak} +(32.30 \pm 3.94)
\end{equation}
\noindent with $\sigma_{\rm int}^{\prime 2}=0.37 \pm 0.10$.
Similarly, we find that the parameters for the corrected simple BPL sample are $a'=-0.72 \pm 0.19$, $b'=0.29 \pm 0.13$, $C_0=33.98\pm 5.88$, $\sigma'=0.45 \pm 0.12$. For the corrected smoothly BPL sample, we find that $a=-0.71 \pm 0.20$, $b=0.23 \pm 0.11$ and $C_0=36.90 \pm 5.23$ and $\sigma=0.34 \pm 0.17$.
The 3D fundamental plane fitted with W07 for the whole sample corrected for evolution has a $\sigma_{int}^{2}$ $16\%$ smaller than the 3D correlation for the uncorrected sample.
When we consider the simple and smoothly BPL models, the reduction of scatter is 13\% and 38\% smaller, respectively.
For the 3D Gold sample, fitted with the W07, the corrected $\sigma_{int}^{2}=0.43$, which is 28\% smaller than the uncorrected $\sigma_{int}^{2}=0.60$.

Regarding the 2D fitting, for the corrected Gold sample, $\sigma_{int}^{2}=0.34$ is $37\%$ smaller than the Gold sample without this correction ($\sigma_{int}^{2}=0.54$) for W07; for the case of the simple and smoothly BPL models, $\sigma_{\rm int}^{2}$ is 19\% and 16\% smaller, respectively. Similarly to what has been discussed in \citet{dainotti2020b}, the 2D correlation for the uncorrected and corrected Gold sample has a $\sigma_{int}^2$ $23\%$ and $36\%$ smaller than the corresponding $\sigma_{int}^2$ for the full sample, respectively. For the simple BPL, the uncorrected Gold sample has a $\sigma_{int}^2$ that is 8\% smaller than the full sample, while the corrected Gold sample has a $\sigma_{int}^2$ that is 4\% smaller than the corrected full sample. For the smoothly BPL, the uncorrected Gold sample has a $\sigma_{int}^2$ that is 58\% samller than the full sample, while the corrected Gold sample has a $\sigma_{int}^2$ that is 32\% smaller than the full sample.

Comparing between the 2D and 3D optical correlations for the W07 sample, the total sample uncorrected for biases for 3D has a $\sigma_{int}^2$ $40\%$ smaller than the corresponding 2D correlation. Similarly, the 3D correlation for the total sample corrected for selection biases has a $\sigma_{int}^2$ $36\%$ smaller than the corresponding corrected 2D correlation.

Thus, with our new definition, the Gold sample still reduces the scatter of 2D correlation both before and after correction for selection biases and redshift evolution.
%both demonstrate high linear anti-correlations.
Given that the slope of each 2D correlation is nearly $-1$, it is implied that the plateau has a fixed energy reservoir independent of a given class. This can be explained within the magnetar scenario. 
Additionally, we find that the $L_{\rm opt}^{(\prime)}-T^{(*/\prime)}_{\rm opt}-L_{peak}^{(\prime)}$ correlation holds regardless of GRB class and plateau angle (Table \ref{table:corr_paramsWL}, \ref{table:corr_params_simpleBPL}, and \ref{table:corr_params_smoothBPL}).

Furthermore, we find that $T^{*}_{a}$ is achromatic between X-ray and optical observations for a sub-sample of GRBs observed at both wavelengths (Fig. \ref{fig:XvO}) if we do not consider selection biases for 10 cases and 13 cases when we consider evolution.
An underlying chromatic behaviour between the X-rays and optical ($T^{\prime}_{a}$) is shown regardless of correction for selection biases and redshift evolution. This investigation casts a new light in the long-standing debate whether or not the plateau is achromatic in nature. The chromaticity of the plateau between X-rays and optical is aligned with the result of GRB 090510 in which the plateau is chromatic between the {\it Fermi}-LAT in high-energy $\gamma$-rays and the X-ray observations (\citealt{dainotti2021c}). 

%\section{\label{sec:acknowledgement} Acknowledgements}

%TC:ignore
\acknowledgements
This work made use of data supplied by the UK Swift Science Data Centre at the University of Leicester. We are particularly grateful to Y. Niino for his precious discussion on the observability and limiting luminosity of the Tomo-e Gozen.
We are particularly grateful to T. Sakamoto for the suggestions on the structure of the paper and to T. Moriya for the discussion on the spectral index and on the GRB-SNe classes distinction.
We thank G. Sarracino for his help in modifying our host extinction code in Python and R. Fatima for performing some of the fitting, J. Osęka, G. Krężel, Z. Kania, C. Wala, S. Gupta, N. Osborne and E. Johnson for data gathering, A. Rabęda, N. Osborne and D. Zhou for the fitting of some GRB LCs, and B. Simone for help with conversions. D.A.K. acknowledges support from Spanish National Research Project RTI2018-098104-J-I00 (GRBPhot).
SY and DL acknowledge the support by the United States Department of Energy in funding the Science Undergraduate Laboratory Internship (SULI) program. We thank E. Cuellar for his work in managing the SULI summer program.
SY gratefully acknowledges the support of the Vagelos Challenge Award at the University of Pennsylvania.
RLB acknowledges support from the DGAPA/UNAM IG100820 and the DGAPA/UNAM postdoctoral fellowship.
NF acknowledges support from the DGAPA/UNAM IN106521
Some of the data used in this paper were acquired with the RATIR instrument, funded by the University of California and NASA Goddard Space Flight Center, and the 1.5-meter Harold L. Johnson telescope at the Observatorio Astronómico Nacional on the Sierra de San Pedro Mártir, operated and maintained by the Observatorio Astronómico Nacional and the Instituto de Astronomía of the Universidad Nacional Autónoma de México. Operations are partially funded by the Universidad Nacional Autónoma de México (DGAPA/PAPIIT IG100414, IT102715, AG100317, IN109418, IG100820, IN106521 and IN105921). We acknowledge the contribution of Leonid Georgiev and Neil Gehrels to the development of RATIR. MF, LZ and AZ thank the support from the Scientific Caribbean Foundation. ES thanks the support from the Latino Education Advancement Foundation. 
\software{dustmaps \citep{dustmaps}, astroquery \citep{astroquery}, scipy \citep{scipy}}
%TC:endignore

% \nocite{*}
\bibliographystyle{gammaraygang_custombibliography.bst} %the reason for custom bibliography style is to truncate the number of authors displayed to 3
\bibliography{references.bib}

\end{document}